\documentclass[12pt,preprint]{aastex}
\usepackage[dvips]{color}
\usepackage{longtable,lscape}
\usepackage{amsmath}
\usepackage[usenames,dvipsnames]{xcolor}

\begin{document}
\newcommand{\wcen}  {$\omega$~Cen}
\newcommand{\boo}   {Bo\"{o}tes~I}
\newcommand{\boos}  {Boo-1137}
\newcommand{\kms}   {\rm km~s$^{-1}$} 
\newcommand{\teff}  {$T_{\rm eff}$} 
\newcommand{\logg}  {$\log g$} 
\newcommand{\loggf} {log~$gf$} 
\newcommand{\bvz}   {$(B-V)_{0}$} 
\newcommand{\feh}   {[Fe/H]}
\newcommand{\alphafe} {[$\alpha$/Fe]}
\newcommand{\vxi}   {$\xi_t$}
\newcommand{\uf}    {ultra-faint}
\newcommand{\mv}    {$M_{V}$}
\newcommand{\ebv}    {$E(B-V)$}
\newcommand{\ebi}    {$E(B-I)$}
\newcommand{\bi}    {$B-I$}
\newcommand{\bio}   {$(B-I)_{0}$}
\newcommand{\vhb}   {$V_{\rm HB}$}
\newcommand{\msun} {\rm M$_\odot$}
\newcommand{\ltsima} {$\; \buildrel < \over \sim \;$}
\newcommand{\simlt} {\lower.5ex\hbox{\ltsima}}
\newcommand{\gtsima} {$\; \buildrel > \over \sim \;$}
\newcommand{\simgt} {\lower.5ex\hbox{\gtsima}}

\title{THE POPULATIONS OF CARINA. II. CHEMICAL ENRICHMENT\footnote{Based on
    observations collected at the European Southern Observatory, Paranal, Chile
    (Proposal 180.B-0806(B), PI: G. Gilmore)}}

\author{
JOHN E. NORRIS\altaffilmark{1},
DAVID YONG\altaffilmark{1},
KIM A. VENN\altaffilmark{2},
GERARD GILMORE\altaffilmark{3},
LUCA CASAGRANDE\altaffilmark{1}, AND
AARON DOTTER\altaffilmark{1}
}

\altaffiltext{1}{Research School of Astronomy and Astrophysics, The
  Australian National University, Canberra, ACT 2611, Australia;
  jen@mso.anu.edu.au, yong@mso.anu.edu.au, luca@mso.anu.edu.au,
  aaron.dotter@gmail.com}

\altaffiltext{2}{Department of Physics and Astronomy, University of 
Victoria, 3800 Finnerty Road, Victoria, BC V8P 1A1, Canada; kvenn@uvic.ca}

\altaffiltext{3}{Institute of Astronomy, University of Cambridge,
  Madingley Road, Cambridge CB3 0HA, UK; gil@ast.cam.ac.uk}

\begin{abstract}

Chemical abundances are presented for 19 elements in a sample of 63
red giants in the Carina dwarf spheroidal galaxy (dSph), based on
homogeneous 1D/LTE model atmosphere analyses of our own observations
(32 stars) and data available in the literature (a further 31
independent stars).  The (Fe) metallicity and {\alphafe} distribution
functions have mean values and dispersions of --1.59 and 0.33~dex
([Fe/H] range: --2.68 to --0.64), and 0.07 and 0.13~dex ({\alphafe}
range: --0.27 to 0.25), respectively.  We confirm the finding of
\citet{venn12} that a small percentage (some 10\% in the present
investigation) of the sample show clear evidence for significant
enrichment by Type~Ia supernovae ejecta.  Calcium, with the most
accurately determined abundance of the $\alpha$-elements, shows an
asymmetric distribution towards smaller values of [Ca/Fe] at all [Fe/H], most
significantly over --2.0 $<$ [Fe/H] $<$ --1.0, suggestive of incomplete
mixing of the ejecta of Type~Ia SNe with the ambient medium of each of
Carina's generations.  Approximate color-magnitude-diagram age
estimates are presented for the sample and, together with our chemical
abundances, compared with the results of our previous synthetic CMD
analysis, which reported the details of Carina's four well-defined
populations.

We searched for the Na-O anti-correlation universally reported in the
Galaxy's globular clusters, and confirm that this phenomenon does not
exist in Carina.  We also found that one of the 32 stars in our sample
has an extremely enhanced lithium abundance -- A(Li)$_{\rm NLTE}$ =
+3.36, consistent with membership of the $\sim$1\% group of Li-rich
stars in dSph described by \citet{kirby12}.

\end{abstract}


\section{INTRODUCTION} \label{sec:intro}

This is the second of two papers concerning the temporal and chemical
evolution of the stellar populations of the Carina dwarf spheroidal
(dSph) galaxy.  In Paper I \citep{norris17a} we reported an analysis
of Carina's color magnitude diagram (CMD) to constrain the nature of
the four populations evident in the very high quality CMD data of
\citet{stetson11}.  As noted there, an understanding of the
populations turns on a determination of their ages and chemical
abundances; and while the age structure is best constrained by an
interpretation of the CMD, a more complete understanding of the
chemical evolution depends on an accurate knowledge of not only the
abundance of iron ([Fe/H]), but also of those of the some 20 elements
that can be determined only from model stellar atmosphere analysis of
high-spectral-resolution, high-$S/N$ spectroscopic data.  That is,
while the CMD is essential for the task of relatively accurate age
determination, it does not strongly constrain the details of the
system's chemical evolution, in particular the role of the
$\alpha$-elements.  For that, a large sample of homogeneously analyzed
high-resolution spectroscopic material is required to adequately
constrain the details of Carina's chemical enrichment.  To date the
major high-resolution investigations are those of \citet{shetrone03},
\citet{koch08a}, \citet{venn12}, \citet{lemasle12}, and
\citet{fabrizio12, fabrizio15}. To further address this problem we
report here observations and analysis of 32 Carina red giants,
together with re-analysis of the equivalent widths from
high-resolution spectra of a similar quality available in the
literature for stars that include 31 Carina red giants not in our
sample.  These results together provide chemical abundances for a
sample of 63 independent stars that have been analyzed homogeneously.
We use these to seek clearer insight into the chemical enrichment of
Carina's populations.

\subsection{The Role of Carina in Constraining Dwarf Galaxy Enrichment}

Carina is an interesting dwarf spheroidal galaxy, with a very complex
star formation history as seen in its CMD (\citealp{smecker94},
\citealp{smecker96}, \citealp{mighell97}, \citealp{hurley98},
\citealp{hernandez00}, \citealp{dolphin02}, \citealp{monelli03},
\citealp{bono10}, \citealp{stetson11}, \citealp{deboer14},
\citealp{weisz14}, \citealp{monelli14}, \citealp{kordopatis16}, and
\citealp{santana16}).  Standard broad-band $BVRI$ photometry reveals
several distinct sequences, clearly identifying a punctuated star
formation history, with individual star forming events of different
epochs and durations.  And yet, this dSph has one of the narrowest red
giant branches in the Local Group.  The degeneracy in age and
metallicity/chemical composition on its red giant branch (RGB) seems
to have conspired perfectly in these $BVRI$ colors.  It is clear that
Carina is dominated by an intermediate-aged population, with minor
contributions at both old and younger times.  In Paper I we reported
our investigation of the Carina CMD of \citet{stetson11} using
synthetic CMDs based on the isochrones of \citet{dotter08}, in terms
of the three basic population parameters [Fe/H], age, and {\alphafe},
for the cases when (i) {\alphafe} is held constant and (ii) {\alphafe}
is permitted to vary.  We found four epochs of star formation, well
described in terms of [Fe/H] = --1.85, --1.5, --1.2, and $\sim$--1.15
and ages $\sim$~13, 7, $\sim3.5$, and $\sim$1.5 Gyr, respectively (for
{\alphafe} = 0.1 (constant {\alphafe}) and {\alphafe} = 0.2, 0.1, 0.0,
--0.2 (variable {\alphafe})), with small spreads in [Fe/H] and age of
order 0.1 dex and 1 -- 3 Gyr\footnote{ The parameters of the youngest
  population are less certain than those of the other three, and given
  it is less centrally concentrated it may not be directly related to
  them. We conjectured that rather than having [Fe/H] $\sim$--1.15 it
  might be more metal-poor by $\Delta$[Fe/H] $\sim$~0.3 dex.  We also
  noted that more work is needed to determine whether Carina's
  horizontal branch, its blue stragglers, and its stellar rotation
  might play a role in determining the position of this population in
  the CMD.}. (In Paper I we referred to these four groups
chronologically, as the ``first'', ``second'', ``third'', and
``fourth'' populations, respectively.  We shall adopt this
nomenclature later in the present work.)  These parameters reproduce
five basic observed features in Carina's CMD (two distinct subgiant
branches of old and intermediate-age populations, two younger,
main-sequence components and the small color dispersion observed on
its RGB).

These complexities nothwithstanding, dwarf galaxies such as Carina are
simpler systems than spiral galaxies, having lower masses, and
undergone fewer accretion episodes, fewer star formation events, and
less chemical evolution (see e.g., \citealp{mateo98},
\citealp{tolstoy09}).  \citet{weisz14} have shown significant scatter
in the star formation history of the Local Group dSph galaxies, even
among those with similar masses, indicating the importance of additional
factors such as their local environment, the effects of
stellar/supernova (SN) feedback, and variations in the metallicity
yields and timescales for SN and asymptotic giant branch (AGB) events.

One aspect of current interest in dwarf galaxy research is the
importance of inhomogeneous mixing of the interstellar medium (ISM).
A poorly mixed ISM aids in the removal of low angular momentum gas in
numerical simulations of dwarf galaxies, avoiding the formation of
bulges \citep{governato10}.  Inhomogeneous mixing has also been
invoked to explain the large range in metallicities found in the
ultrafaint galaxies, presumed to be due to a single core collapse SN
event (\citealp{simon11}, \citealp{tominaga14}, and
\citealp{frebel15}).  We note that the metallicity dispersions
observed in the ultrafaint systems are nevertheless in excellent
agreement with the inhomogeneous and binomial model for the chemical
evolution of galaxies presented by \citet{leaman12}.

The mixing efficiency and timescale for Type Ia SN events is less
clear.  The usual assumption is that Type Ia SN (and AGB)
contributions lag those of Type II SNe by $\sim$1 Gyr (e.g.,
\citealp{argast02}, \citealp{revaz12}), due to the lower mass of their
progenitor stars.  The location of a knee in the {\alphafe} vs. [Fe/H]
abundances for stars in dwarf galaxies compared with similar
metallicity stars in the Milky Way is one piece of evidence for the
late contributions of iron from SN Ia -- though that trend has also
been attributed to metallicity dependent SN Ia yields
(\citealp{kobayashi98} and \citealp{kobayashi09}), or a truncated upper IMF
(\citealp{tolstoy03}, \citealp{mcwilliam13}).  It could be that
several mechanisms are in play.

Direct evidence for inhomogeneous mixing of Type Ia SN material in a dwarf
galaxy was presented for stars in Carina from ESO/FLAMES-UVES and
Magellan/MIKE spectra by \citet{venn12}.  Two (of nine) stars showed
an enhancement of the iron-group elements (Cr, Mn, Fe) by factors of 3
-- 7 relative to the other elements, which suggested those stars
formed in a pocket of iron-enriched material; one of these (Car-612)
is amongst the most metal-rich stars in Carina, suggesting late-time
inhomogeneous mixing.  Additional stars observed with
ESO VLT FLAMES-GIRAFFE by \citet{lemasle12} also showed a wide dispersion
in [X/Fe] values; but only three wavelength settings were used
(covering $\sim$1000\,{\AA}) providing fewer elements to clearly identify
the signatures of inhomogeneous mixing vs. smooth/punctuated chemical
evolution.  Previous analyses of a few stars in Carina did not clearly
identify these features as the result of too few stars
\citep{shetrone03} and elements \citep{koch08a} being analyzed.

\subsection{Outline of the Present Work}

We have undertaken a kinematic and chemical analysis of the 63 red
giants in Carina in an effort to better understand the formation and
evolution of this system.  In Section~\ref{sec:obsmat} we present our
observational material for 32 Carina red giants, based on
high-resolution, moderate $S/N$, spectra obtained with the
ESO/FLAMES-UVES combination during ESO Proposal 180.B-0806(B) (PI
Gilmore)\footnote{These spectra have also been analyzed by
  \citet{fabrizio12, fabrizio15}.}, which we use in
Section~\ref{sec:analysis} to determine chemical abundances for 19
elements for these stars, using model atmosphere techniques.  In
Section~\ref{sec:litsam} we augment this data set with spectroscopic
material of comparable quality for other Carina red giants in the
literature.  To produce a set of homogeneously determined abundances,
we analyze the literature equivalent width values using the same
techniques as adopted for the primary sample.  The total sample
comprises 63 independent Carina red giants.  Their ([Fe/H])
metallicity distribution function (MDF) is presented in
Section~\ref{sec:mdf}, while Section~\ref{sec:relabund} gives an
overview of relative elemental abundances in the [X/Fe] vs. [Fe/H]
planes.  In Section~\ref{sec:alphas} we discuss the abundances of the
$\alpha$-elements and what they have to tell us about Carina's
chemical evolution, while Section~\ref{sec:sne} examines the role
played by Type Ia SNe in its chemical enrichment.  In
Sections~\ref{sec:age} and Section~\ref{sec:distributions} we present
rough age estimates, and compare them and our observed MDF and
{\alphafe} distribution with those obtained using the synthetic CMD
predictions in Paper I.  Section~\ref{sec:nao} presents the result of
our search for the ubiquitous anti-correlation between sodium and
oxygen abundances that exists within individual Galactic globular
clusters, while in Section~\ref{sec:li} we present the discovery of an
extremely lithium-rich Carina red giant. We summarize our results in
Section~\ref{sec:summary}.

\section{OBSERVATIONAL MATERIAL}\label{sec:obsmat}

The present high-resolution spectroscopic investigation of Carina is
part of a larger program involving the kinematic and abundance
analysis of a much larger sample of Carina giants observed at lower
resolution.  It utilizes the power of the ESO VLT FLAMES-UVES system in
UVES-Fiber mode \citep{pasquini02}, which permits simultaneous
observation of a large number of stars at intermediate resolution (R
$\sim$ 6500), via a system of 130 fibres, together with a smaller
number at high resolution (R $\sim$ 47000), via eight fibres.  The
analysis of the lower resolution sample has been reported elsewhere
\citep{kordopatis16}.  The obvious advantage of the higher resolution
capability is that it enables insight into the chemical abundances of
considerably more elements ($\sim$ 20) than is possible at the lower
resolution.  We refer the reader to our previous work on the Bootes I
ultrafaint galaxy for an example of the synergy of the approach we
adopt in the analysis of FLAMES material (see \citealp{koposov11} and
\citealp{gilmore13}).

\subsection{Sample Selection}\label{sec:sample}

\begin{figure}[htbp]
\begin{center}

\includegraphics[width=10.0cm,angle=0]{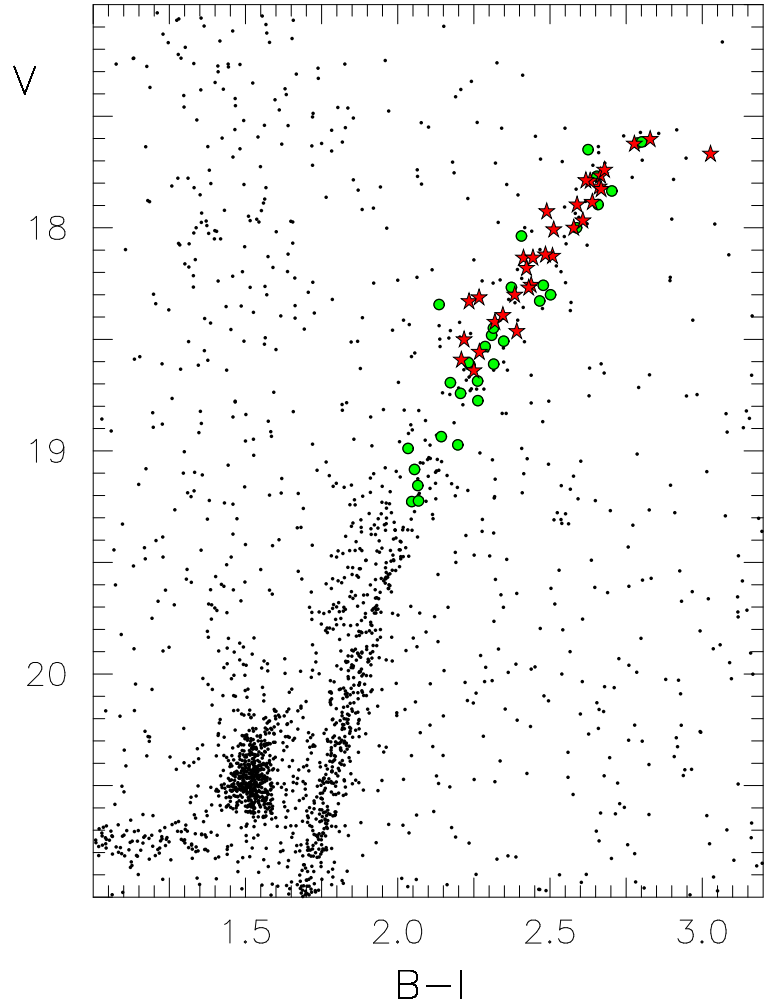}
  \caption{\label{fig:cmd}\scriptsize The Carina ($V,~B-I$)
    color$-$magnitude diagram for stars observed in our initial
    photometric survey.  The high quality photometry presented here
    has been made available by P.B. Stetson.  The small black symbols
    present results for the general survey; large red star symbols
    represent objects observed spectroscopically at high-resolution in
    the present work; and large green circles stand for independent
    stars in the literature having spectroscopy of a similar
    quality.}

\end{center}
\end{figure}

Our selection of objects is based on unpublished CCD $V,I$
observations that we have made of the Carina galaxy.  The only
criterion we adopted in the selection was that a star should lie close
to the well-defined RGB of the system\footnote {We note here for
  completeness that a comparison of the final 63 spectroscopic red
  giant sample presented in this paper with the radial-velocity
  selected samples of \citet{koch06} and \citet{walker09} shows that
  our sample is incomplete insofar as it is narrower in the ({\mv},
  {\bio}) CMD than theirs, by some 3 -- 4$\sigma$. See Paper I for
  further discussion of this point.}.  In Figure~\ref{fig:cmd} we show
the positions of these stars in the ($V,~B-I$) color-magnitude
diagram, together with that of the larger sample surveyed, for the
sample described by \citet{stetson11} and made available by
P.B. Stetson (2014, private communication).  The large red star
symbols represent objects for which we have obtained the
high-resolution spectra discussed above.  The large green circles
stand for stars having high-resolution spectroscopic data in the
literature, to which we shall return in Section~\ref{sec:litsam} and
determine chemical abundances, using the techniques we adopt in the
analysis of our program stars.

\subsection{High-resolution Spectroscopy}\label{sec:obsspec}

High-resolution, moderate-$S/N$, spectra were obtained of 39 Carina
red giants, during 2007 November -- 2008 March, with the FLAMES system
at the 8.2 m Kueyen (VLT/UT2) telescope at Cerro Paranal, in
UVES-Fiber mode. Of the eight UVES fibers, $\sim$5 were allocated to
Carina candidate members and $\sim$3 to nearby sky positions to enable
background measurement.  We obtained 9 individual exposures in Service
Mode, 8 of duration 60\,min, leading to an effective total integration
time of 8.8 hrs.  The spectra were obtained using the 580nm setting,
and cover the wavelength ranges 4800 -- 5750\,{\AA} and
5840 -- 6800\,{\AA}.  The resolving power was R = 47000.

The spectra of the individual exposures of the program stars were
reduced by using the FLAMES-UVES
pipeline\footnote{{http://www.eso.org/sci/software/pipelines/}}.
Following this, the individual spectra were cross-correlated to
determine relative wavelength shifts between them in order to
compensate for the Earth's motion during the data-taking interval.
After sky-subtracting and shifting the individual spectra to the rest
frame, the spectra were co-added to produce the summed spectrum of
each star.  Our subsequent abundance analysis of these data indicated
that the signal-to-noise ratio ($S/N$) of the spectra of 7 objects was
too low to permit reliable abundance results, and these will not be
considered further here.  

Details of the remaining 32 program stars are presented in
Table~\ref{tab:obsdat}, where Columns 1 and 2 contain their
identification and coordinates. Alternative nomenclatures of these
stars, together with the names of others that will be the subject of
re-analysis of equivalent data available in the literature, using
techniques described below in Section~\ref{sec:analysis}, are
presented in Table~\ref{tab:appendix} of the Appendix.

\subsection{Photometry}\label{sec:photometry}

Photometry has been obtained from several sources: P.B. Stetson
provided us with homogenized $BVI$, M.J. Irwin furnished $JHK$ from
ESO VISTA survey photometry, and M. Gullieuszik supplied
$BVIJHK_s$. Here all $BVI$ magnitudes are corrected to the
Johnson-Kron-Cousins system, following \citet{stetson05}, while
$JHK_s$ are corrected to be on the 2MASS system.  A comparison of the
photometric dataset per star between the three sources shows excellent
agreement.  We estimate from intercomparison of the photometry that
the $B$, $V$, $I$, $J$, $H$, and $K$ values have errors of 0.006,
0.006, 0.011, 0.009, 0.009, and 0.016, respectively.  We have
therefore averaged the magnitudes, and present the results in Table 1,
which we use in our determination of the atmospheric parameters in
Section~\ref{sec:atpar}.

\subsection{Radial Velocities}\label{sec:radvel}

Radial velocities were measured for each of the individual spectra for
each star described above.  To do this we used the Fourier
cross-correlation techniques described by \citet[Section
  2.4]{norris10} and \citet[Section 2.2.1]{gilmore13}, to which we
refer the reader for details.  We note here, for completeness, that
there were two minor differences in the present work.  First, we
cross-correlated our spectra against the Arcturus high resolution
spectrum of
\citet{hinkle00}\footnote{ftp://ftp.noao.edu/catalogs/arcturusatlas/}
(rebinned to have the same pixel size as our data). Second, for each
spectrum we cross-correlated over the three wavelength regions 5160 --
5190\,{\AA}, 5400 -- 5498\,{\AA}, and 6502 -- 6598\,{\AA} to produce
three velocity estimates, which we averaged to determine the velocity
for that spectrum.  Then, for each star we averaged the velocities
available for the $\sim$8 exposures of each star and (having applied
appropriate heliocentric corrections) we averaged these values,
weighted by the inverse square of their errors, to produce the final
radial velocity for each object.  Our velocities and their internal
errors are presented in the final two columns of
Table~\ref{tab:obsdat}.  The data are of high precision, with the mean
of the errors for the 32 stars being 0.16~{\kms}.  Velocities for
these spectra have also been determined by \citet{fabrizio12}:
comparison between our velocities and theirs shows that the mean
velocity difference is 0.03 ~{\kms} with a dispersion of 0.76~{\kms}.
The mean radial velocity of our sample of 32 stars (all of which are
radial-velocity members) is 224.8 $\pm$ 1.0 {\kms}, with dispersion
5.65 $\pm$ 0.71 {\kms}.  We note that the dispersion agrees well with
those of \citet{walker09b}, within the limits, but is somewhat lower
than those of 10.4 $\pm$ 1, 7.6 $\pm$ 0.5, and 8.5 $\pm$ 0.8 {\kms},
reported by \citet{kordopatis16} for their metal-poor,
intermediate-metallicity, and metal-richer RGB stars.  For
completeness, we also note that we find no dependence of velocity
dispersion on metallicity in our relatively small sample of stars.

\subsection{Equivalent Widths}\label{sec:ew} 

\begin{figure}[htbp!]
\begin{center}

\includegraphics[width=10.0cm,angle=-90]{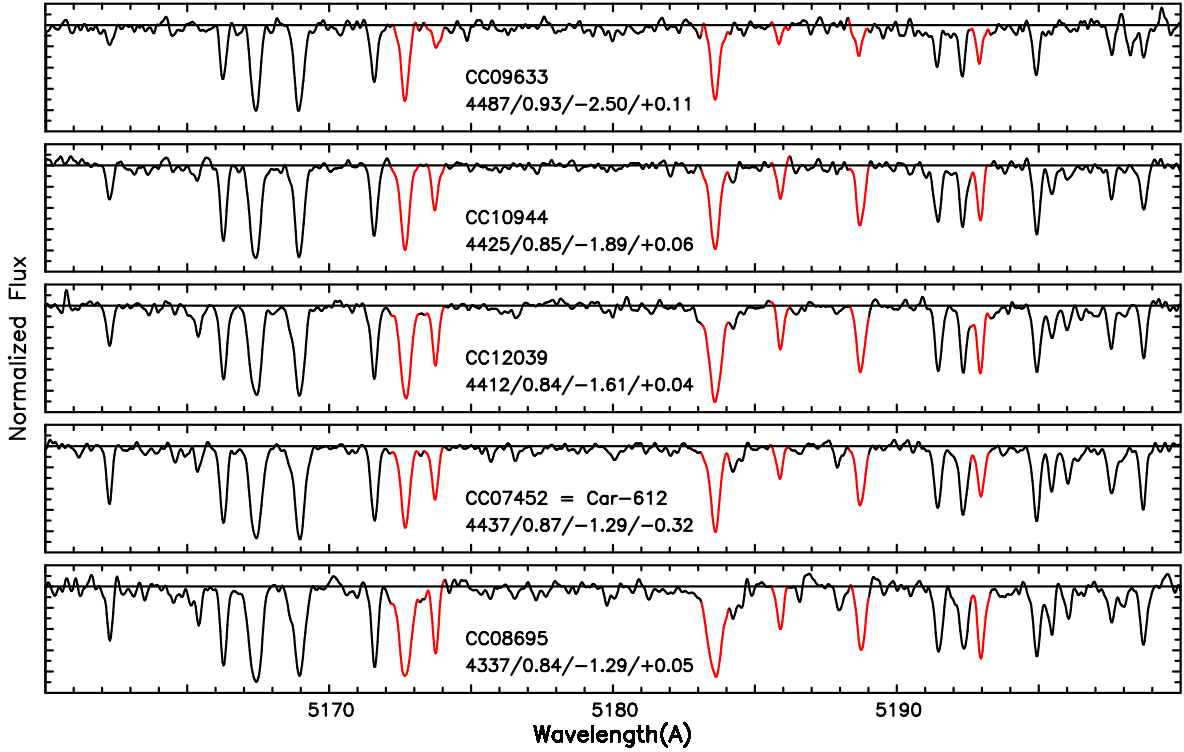}
  \caption{\label{fig:spectra} \scriptsize Representative spectra of
    the program stars in the wavelength range 5160 -- 5200\,{\AA}.
    Also shown are atmospheric parameters and abundances
    {\teff}/{\logg}/[Fe/H]/[Ca/Fe], where the values are those
    determined in the present work, as will be described below.
    [Fe/H] increases from top to bottom, and the accompanying increase
    in line strength is clear.  We note that all of the stars, except
    CC07452 (= Car-612), have [Ca/Fe] $\sim$0.05 -- 0.10.
    CC07452, with [Ca/Fe] = --0.32, is ``$\alpha$-challenged'', an
    effect first established by \citet{koch08a} and \citet{venn12},
    and described in the Introduction.  Inspection of the red-colored
    parts of the spectra, which cover unblended features of the
    $\alpha$-elements Mg I, Ti I and Ti II, shows the important result
    (which we shall revisit in Section~\ref{sec:sne}) that not only
    Ca, but also Mg and Ti have lower abundances relative to Fe in
    CC07452, than in the majority of the stars.}
\end{center}
\end{figure}

Before attempting to determine equivalent widths from the co-added
pipelined spectra described in Section~\ref{sec:obsspec}, we undertook
three further steps.  First, given the somewhat low $S/N$ of a
significant number of spectra, and the line crowding on some of the
relatively high-abundance stars in our sample, we resolved to measure
equivalent widths only redward of 5300\,{\AA}.  Inspection of the
degree of line blending in the high-resolution spectrum of Arcturus
\citep{hinkle00} supports this decision.  We also note the previous
decision of \citet{norris95a} in their abundance analysis of the red
giants in the globular cluster $\omega$ Centauri (of spectra having R
$\sim$ 38,000), who stated: ``Initially we sought to measure ... all
lines in the wavelength range 5050 -- 6810\,{\AA}, but in view of the
difficulty of continuum placement in the coolest stars, decided to
reduce this to 5285 -- 6810\,{\AA}''.  Second, given the high
resolution (R = 47000) of the present spectra, we double-binned the
data to facilitate reliable continuum placement (yielding a pixel size
of $\sim$0.028\,{\AA})\footnote{A referee has suggested that
  double-binning of the data will affect the measurement of equivalent
  widths.  This is not the case.  In the FLAMES manual
  http://www.eso.org/sci/facilities/paranal/instruments/flames/doc/VLT-MAN-ESO-13700-2994\_p99.pdf
  Section 2.1 notes that there are five pixels per FLAMES fibre.
  Double-binning means that there are 2.5 pixels per resolution
  element, which is not undersampled according to Nyquist sampling and
  thus the resolution remains at R=47,000.  We also note that the
  spectral lines we measured have a median FWHM of $\sim$~0.28{\AA}
  which corresponds to some 10 double-binned pixels.}.  We measured
$S/N$ per double-binned pixel in several intervals of width 2.0 --
5.5\,{\AA} in the range 5310 -- 5723\,{\AA}, in which the continuum is
clearly seen in the Arcturus spectrum of \citet{hinkle00}.  The $S/N$
values (averaged over the several wavelength intervals) for the 32
Carina giants for which we analyze equivalent widths lie in the range
8 -- 22, with median value 12, and are presented for the individual
stars in the first rows of Tables~\ref{tab:ew1} -- \ref{tab:ew3}.
Third, and finally, we removed the inter-order undulations in the
pipelined data and re-normalized the spectra by determining the
position of the ``continuum'' using five-pixel Gaussian smoothing and
fitting a low order Legendre polynomial, followed by k-sigma clipping
using a scale length of 10\,{\AA}, following \citet{venn12}.  Examples
of the resulting spectra in the wavelength range 5160 -- 5200\,{\AA},
including for interest that of the $\alpha$-challenged star Car-612
discussed in Section~\ref{sec:intro} (and designated here CC07452) are
presented in Figure~\ref{fig:spectra}.  (We note that these spectra
lie outside the region of our abundance analysis; they are presented
here for heuristic purposes only.)

\begin{figure}[htbp]
\begin{center}

\includegraphics[width=4.5cm,angle=0]{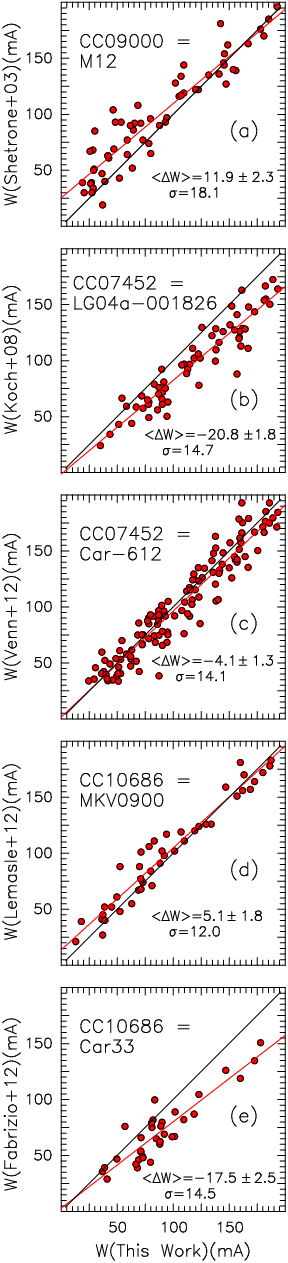}
  \caption{\label{fig:ew} \scriptsize Comparison of the equivalent
    widths of the present work with those of (a) \citet{shetrone03},
    (b) \citet{koch08a}, (c) \citet{venn12}, (d) \citet{lemasle12},
    and (e) \citet{fabrizio12}.  In the legend of each panel the upper
    number presents the star identification we use, while the next
    line contains the name used by the other authors.  The black line
    is the 1-1 relationship, while the red line represents the linear
    least-squares linear best fit. The mean difference in equivalent
    widths, $\langle$$\Delta$W$\rangle$ (in the sense other work -- this
    work), its standard error, and dispersion ($\sigma$) are also
    shown.}
\end{center}
\end{figure}

Equivalent widths were measured independently for all stars by both
J.E.N. and D.Y. for the set of unblended lines formed by the merging
of the linelists of \citet{venn12} and \citet{yong13a}. Three
independent equivalent measures were obtained: J.E.N. used the
equivalent width measurement techniques described by \citet{norris01},
while D.Y. utilized IRAF software as described by \citet{yong08} as
well as the automated package DAOSPEC (see
\citealp{stetson08}{\footnote{http://www1.cadc-ccda.hia-iha.nrc-cnrc.gc.ca/community/STETSON/daospec}). As
  a further check, K.A,V. used IRAF and DAOSPEC for a small subset of
  the spectra, which confirms the results of J.E.N. and D.Y.
  
Before accepting an equivalent width measurement, both J.E.N. and D.Y.
inspected the Gaussian fits to each spectrum line to ensure an
acceptable representation.  For each line in a given star, the
measurements of J.E.N. and D.Y. produced three independent estimates
that were combined as follows: (i) we required that at least two
estimates were available; (ii) when three measurements were recorded,
we rejected one of them if it clearly disagreed with the average of
the other two; and (iii) of the cases where two estimates were
averaged, we rejected some 30 lines for which large differences of
order greater than 3$\sigma$} existed between them.  We rejected a
line from the abundance analysis if an equivalent width was measured
for it in only one star, or the majority of stars in the sample had an
equivalent width greater than 200~m{\AA}.  Finally, in four cases we
rejected a line that was the only representative of an atomic species
and yielded an abundance that was clearly in error.
Figure~\ref{fig:ew} presents a comparison of the equivalent widths of
the present work with those of \citet{shetrone03},
\citet{koch08a},\citet{venn12}, \citet{lemasle12}, and
\citet{fabrizio12, fabrizio15}.  In these comparisons, there is good
agreement between the present work and the results of
\citet{shetrone03}, \citet{venn12}, and \citet{lemasle12}.

We find a systematic difference, however, between our data and those
of \citet{koch08a} between their LG04a\_01826 and our CC07452, in the
sense that their equivalent widths become systematically smaller than
ours with increasing line strength. (Note, however, that Figure~\ref{fig:ew}
shows good agreement between our work and that of \citet{venn12} for
this object.)  We have five other stars (CC06122, CC08469, CC10194,
CC11217, CC12039) in common with the work of \citet{koch08a}, and find
similar differences in all of these objects as well.  In
Section~\ref{sec:litsam} below, when we shall re-analyze equivalent
width data from the literature, we shall not include the work of
\citet{koch08a} in view of these systematic equivalent width
differences.

\begin{figure}[htbp]
\begin{center}

\includegraphics[width=10.0cm,angle=0]{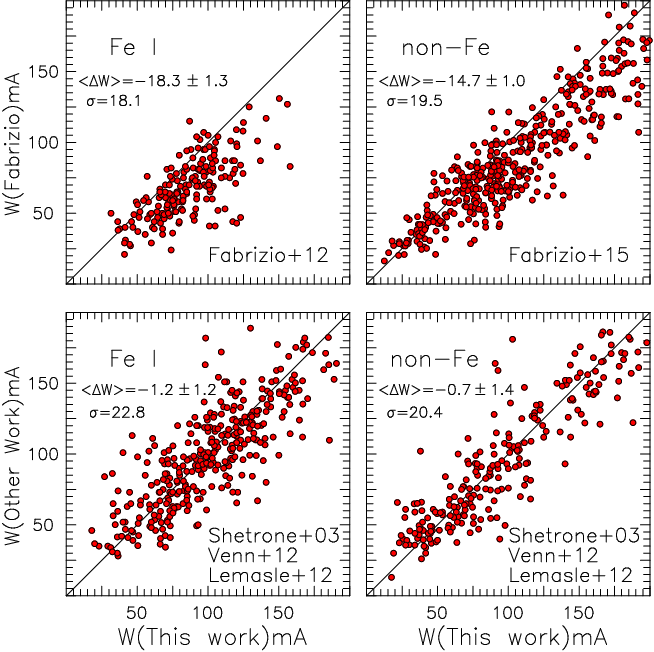}
  \caption{\label{fig:gof} \scriptsize {Comparison of the Fe I
      equivalent widths, W, of stars in common between the present
      work and those of \citet{fabrizio12} (top left panel) and those
      of \citet{shetrone03}, \citet{venn12} and \citet{lemasle12}
      (bottom left panel); and between the equivalent widths of non-Fe
      lines of the present work and \citet{fabrizio15} (top right
      panel) and those of \citet{shetrone03}, \citet{venn12} and
      \citet{lemasle12} (bottom right panel). The mean difference in
      equivalent widths, $\langle$$\Delta$W$\rangle$ (in the sense
      other work -- this work), its standard error, and dispersion
      ($\sigma$) are also shown.}}
\end{center}
\end{figure}

We obtain relatively poor agreement between our results and those of
\citet{fabrizio12, fabrizio15} for CC10686, as may be seen in the
bottom panel of Figure~\ref{fig:ew}.  To investigate this further, in
the upper left panel of Figure~\ref{fig:gof} we present a comparison
between the equivalent widths of Fe I lines in the 31 stars in common
between the samples of \citet{fabrizio12} and the present
work\footnote{The sample comprises the stars in our
  Table~\ref{tab:obsdat}, excluding CC06975.}, and in the lower left
panel we compare the collective Fe I data of \citet{shetrone03},
\citet{venn12}, and \citet{lemasle12} on the one hand, and the present
work, on the other.  In the right panel of the figure we present a
similar comparison for the non-Fe lines, using the equivalent widths
of \citet{fabrizio15}.  From these panels we conclude that the line
strengths of \citet{fabrizio12, fabrizio15} are on average smaller
than those of both the present investigation and those of the other
workers.  As seen in Figure~\ref{fig:gof}, the equivalent widths of
  \citet{fabrizio12} and \citet{fabrizio15} are smaller by 18 and 15
  m{\AA} for Fe~I and non-Fe lines, respectively, than those of the
  present work.  Offsets of this order in comparison with the work of
  \citet{venn12} and \citet{lemasle12} were also reported by
  \citet{fabrizio15}.  We shall return to the significance of these
differences in Section~\ref{sec:finabs}.

As part of our quality control, we estimated the smallest equivalent
width, W(min), that we could reliably measure in each star.  To
achieve this, for each spectrum we examined the number of lines we
measured as a function of decreasing equivalent width, and on the
conservative assumption that at lowest line strength our estimates
might not be reliable we adopted the fifth smallest equivalent width
we measured as a conservative estimate of W(min).  These values lie in
the range 14 -- 34~m{\AA}, with median 24~m{\AA}, and are presented in
the second row of Tables~\ref{tab:ew1} -- ~\ref{tab:ew3}. If we use
the \citet{cayrel88} formula, with the correction from
\citet{battaglia08}, to calculate the minimum equivalent width, we
find a range from 4~m{\AA} ($S/N$ = 22) to 11~m{\AA} ($S/N$ = 8) for
our high resolution data.  These values are a factor of three smaller than our
conservative estimates above, and therefore consistent with a 3
$\sigma$ certainty for these estimates.

In Section~\ref{sec:nao} we shall investigate the existence or
otherwise of anti-correlation between the abundances of Na and O,
which is ubiquitous within individual Galactic globular clusters, and
which the work of \citet{shetrone03} suggests may not exist in Carina.
As part of that exercise we closely estimated the upper limits of the
equivalent widths of the lines of these elements (together with Al,
which also anti-correlates with O in the globular clusters) by visual
inspection of them and nearby lines of other elements.  These limits
are also presented, for future consideration, in Tables~\ref{tab:ew1}
-- \ref{tab:ew3}, specifically for O~I~6300.3\,{\AA},
Na~I~5688.2\,{\AA}, and Al~I~6696.0\,{\AA}.

To form our final equivalent width (EW) dataset we accepted lines with
wavelengths in the range 5300 -- 6780\,{\AA} and equivalent widths
less than 200 m{\AA}.  Our adopted equivalent widths for 211 unblended
lines are presented in Tables~\ref{tab:ew1} -- \ref{tab:ew3}.  Line
identifications, together with their lower excitation potentials,
$\chi$, and {\loggf} values are presented in Columns 1 -- 4,
respectively.  The equivalent widths populate the remainder of the
tables. These data are suitable for model atmosphere abundance
analysis (see Section~\ref{sec:analysis}).

In what follows we shall exclude lines weaker than the relevant W(min)
values described above, except for the O~I~6300.3\,{\AA},
Na~I~5688.2\,{\AA}, and Al~I~6696.0\,{\AA} transitions, discussed in
the previous paragraph.  We consider these weaker lines, but note that
their detections are at the 1 -- 2 $\sigma$ level, rather than the 3
$\sigma$ level we have adopted for the other lines.

\section{CHEMICAL ABUNDANCE ANALYSIS}\label{sec:analysis}

\subsection{Atmospheric Parameters}\label{sec:atpar} 

\subsubsection{Effective Temperature (\teff)}\label{sec:teff}

In most abundance analyses of cool giants, effective temperatures are
based on either spectroscopic or photometric techniques.  The former,
denoted ``excitation'' temperatures, are based on the requirement that
model atmosphere analysis of Fe~I lines should yield abundances
independent of lower excitation potential, while the latter, denoted
``photometric'' temperatures, rely on calibrations of the stellar
continuum energy distribution as a function of effective temperature.
Insofar as our spectra have relatively low $S/N$, on the one hand,
while we have access to accurate $BVIJHK$ photometry, on the other, we
have chosen to adopt photometric temperatures, which we expect to
have considerably higher accuracy in the present case.

We have determined {\teff} by using the InfraRed Flux Method (IRFM),
following \citet{casa06}, which solves for {\teff} in the basic
equation
\begin{equation}
\label{irfmeq}
\frac{\mathcal{F}_{Bol}(\textrm{Earth})}
{\mathcal{F}_{\lambda_{\textrm{\tiny{IR}}}}(\textrm{Earth})} = \frac{\sigma
T_{\mathrm eff}^4}{\mathcal{F}_{\lambda_{\textrm{\tiny{IR}}}}(\textrm{model})}
\end{equation}

where $\mathcal{F}_{Bol}(\textrm{Earth})$ and
$\mathcal{F}_{\lambda_{\textrm{\tiny{IR}}}}(\textrm{Earth})$ refer to
bolometric and monochromatic infrared fluxes at the Earth, while
$\mathcal{F}_{\lambda_{\textrm{\tiny{IR}}}}(\textrm{model})$ refers to
a corresponding model atmosphere synthetic spectrum.

The observational input data were the $BVIJHK$ values presented in
Table~\ref{tab:obsdat} and a distance modulus and reddening for Carina
of $(m-M)_{V}$ = 20.05 $\pm$ 0.11 and E($B-V$) = 0.06 $\pm$ 0.02
\citep[Section~3.4]{venn12}, while the adopted model synthetic spectra
were those of \citet{castelli03}.  As found by \citet{casa06}, these
temperatures tend to be slightly ($\le$100 K) hotter than those
determined with the \citet{ramirez05} $BVIJHK$ color-temperature
calibrations.  They tracked this effect primarily to the zero point
calibration of the 2MASS photometric system.  The resulting \teff,
which are the averages of the three independent estimates obtained by
applying the IRFM to the observed $J$, $H$, and $K$ magnitudes, are
presented in Column 2 of Table~\ref{tab:atpar}, while Column 3
contains their uncertainties, based on the spread in the $JHK$
estimates.  The average uncertainty in the {\teff} for the 32 stars in
the table is 76 $\pm$ 2 K, which we shall use in
Section~\ref{sec:errors}.

\subsubsection{Surface Gravity (\logg)}\label{sec:logg}

In many analyses surface gravities are determined by requiring that
abundances obtained by using Fe~I and Fe~II lines are the same.  There
are two potential problems with this method.  The first is that the
results of non-LTE (NLTE) calculations show that while the LTE assumption is
acceptable for the analysis of Fe~II lines, it leads to erroneous
results for Fe~I (e.g., \citealp{lind12}).  An additional
consideration is that at lowest abundances the number of Fe~II lines
decreases dramatically.  A second method of gravity determination is
to use theoretical isochrones in which the observed {\teff} is
interpolated in the theoretical ({\logg}, {\teff})--relationship, or
(with color as proxy for {\teff}) in a ({\logg}, color)--relationship.
This method is, however, critically dependent on the accuracy of the
model {\teff} values and the (color, {\teff}) dependence.

A third approach is to use the fundamental definitions of gravity and
{\teff} to express gravity as a function of mass, {\teff}, and
luminosity.  The basic challenge of this method is that one needs a
reliable distance in order to determine luminosity, which is not
generally possible for field stars.  For star clusters, however, this
is not an insuperable problem, insofar as their distances can be
obtained with reasonable accuracy.  We choose to follow this approach.

The surface gravity, \logg, of a star of mass M, effective temperature
\teff, and bolometric magnitude $M_{\rm Bol}$, may be written:

\logg~=~\logg${_\odot}$ + 0.4$\times$log(M/M${_\odot}$) +
4.$\times$log({\teff}/{\teff}${_\odot}$) + 0.4$\times(M_{\rm Bol}$ -
$M_{\rm Bol\sun}$).

To apply this relationship, we adopt for the Sun \logg$_\odot$ = 4.44,
T$_{\rm eff\odot}$ = 5777\,K, and $M_{\rm Bol_\odot}$ = 4.75; and for
the Carina giants M = 0.8~M$_{\odot}$, the distance modulus and
reddening of Section~\ref{sec:teff}, and A($V$)/E($B-V$) = 3.24
(following \citet{venn12} and \citet{schlegel98}).  We use the $V$
magnitudes in Table~\ref{tab:obsdat} together with $V$ bandpass
bolometric corrections (BC) from \citet{alonso99} to determine $M_{\rm
  Bol}$.  The resulting gravities are presented in Column 4 of
Table~\ref{tab:atpar}.

The error budget for {\logg} may be expressed as: 

$\sigma^{2}_{{\rm log}g}$~=~$(0.4)^{2}\times(\sigma_{\rm M}$/(2.302$\times$M)$^{2}$ +
                        $(4)^{2}\times(\sigma_{T_{\rm eff}}/(2.302{\times}T_{\rm eff})^{2}$ + 
$(0.4)^{2}\times(\sigma^{2}_{V} + \sigma^{2}_{(m-M)_{V}} + \sigma^{2}_{\rm BC}$)

We estimate the following representative errors: $\sigma_{M}$/M = 0.26
(assuming the bulk of the stars in our sample has ages in the range 5
-- 12 Gyr and {\alphafe} in the range --0.2 to 0.4 dex, together with
the isochrones of
\citet{dotter08}\footnote{http://stellar.dartmouth.edu/models/isolf\_new.html},
and note that this estimate is relatively insensitive to metal
abundance; $\sigma_{T_{\rm eff}}$/{\teff} = 0.019
(Section~\ref{sec:teff}); $\sigma_{V}$ = 0.020
(Section~\ref{sec:photometry}); $\sigma_{(m-M)_{V}}$ = 0.110
(Section~\ref{sec:teff}); and $\sigma$(BC) = 0.048 (following
\citealp{alonso99}), with $\sigma_{T_{\rm eff}}$ = 80\,K and
$\sigma_{\rm [Fe/H]}$ = 0.15 (Section~\ref{sec:errors}).  With these
values $\sigma_{{\rm log}g}$ = 0.07 dex; and we note in passing that
errors in {\teff} are the dominant contributor to the total error
budget.

\subsubsection{Metal Abundance ([M/H])}

We assume that from the point of view of model atmosphere analysis,
the metallicity [M/H] and iron abundance [Fe/H] are synonymous.
Throughout our abundance analysis we have iterated the metal abundance
of the models at each determination of the abundances to be the same
as those obtained on the previous run: Column 5 of
Table~\ref{tab:atpar} contains our final adopted model values of
[Fe/H].  The average uncertainty for the stars in the table is
0.14~dex. (See Section~\ref{sec:abund}, Table~\ref{tab:relabs_gil}).

\subsection{Abundance Determination}\label{sec:abund} 

Chemical abundances were determined by using the model atmosphere
techniques described in \citet[Section~3]{norris10},
\citet[Section~2]{yong13a}, and \citet[Section~3.1]{gilmore13}, to
which we refer the reader.  Here, we provide only some of the details
of our procedures.  As before, we adopted the ATLAS9 models of
\citet{castelli03}\footnote{http://wwwuser.oat.ts.astro.it/castelli/grids.html}
(plane-parallel, one-dimensional (1D), local thermodynamic equilibrium
(LTE)), with $\alpha$-enhancement, {\alphafe} = +0.4, and
microturbulent velocity {\vxi} = 2~{\kms}. The analysis of the
equivalent widths presented in Section~\ref{sec:ew} used the LTE
stellar-line-analysis program MOOG \citep{sneden73}, as modified by
\citet{sobeck11}, to include an improved treatment of continuum
scattering.  We took into account hyperfine splitting (HFS) for lines
of Sc, V, Mn, Co, Cu, and Ba, using data from \citet{kurucz95}; and
for La, we adopted HFS data from \citet{ivans06}.  For Ba, we assumed
the \citet{mcwilliam98} r-process isotopic composition, while for Cu,
we assumed solar isotope ratios. Once the atmospheric parameters
\teff, \logg, and chemical abundance [M/H] are given, one needs to
determine the microturbulent velocity, {\vxi}.  We achieved this by
requiring that the abundance determined from the Fe I lines should be
independent of their equivalent widths.  During this procedure, we
excluded Fe I lines that fell more than either 3$\sigma$ or 0.5~dex
from the mean value.  Values of {\vxi} are presented in Column 6 of
Table~\ref{tab:atpar}.  Other authors have noted that {\vxi} is a
function of {\logg}: we find {\vxi} = 2.392 -- 0.313$\times${\logg},
with an RMS scatter of 0.003~{\kms}, in excellent agreement with the
relationship of \citet{worley13}, who reported {\vxi} = 2.386 --
0.313$\times${\logg}. In our error analysis below, we shall adopt an
error $\sigma_{\xi_{t}}$ = 0.2~{\kms}.

Detailed results of the abundance analysis are presented in
Table~\ref{tab:abund_gil}.  In the 32 blocks of this table (one per
star) Columns 1 -- 4 contain the atomic species, the mean absolute
abundance log\,$\epsilon$(X) ( = log(N$_{\rm~X}$/N$_{\rm~H}$) + 12.0),
its standard error of the mean, s.e.$_{\log\epsilon}$, and the number
of lines analyzed, respectively.  Column 5 presents [Fe/H] or [X/Fe],
obtained by using the data in Column 2 and the solar abundances of
\citet{asplund09}.  In the calculation of [X/Fe], we
  adopted the value of [Fe/H] determined using the neutral species.
  Given the much larger number of Fe I lines, this leads to
  considerably higher precision in our relative abundances.  Had we
  adopted [Fe II/H], the precision of our [X/Fe] values would be
  considerably poorer given the small number (2 -- 12) of available Fe
  II lines.  The question that then remains is: how accurate are the
  [X/Fe] results?  In particular, what is the role of NLTE, which
  affects Fe~I more than Fe~II?  According to \citet[see their Figure
    2]{lind12}, in our parameter range, the NLTE corrections are
  $\le$+0.1 dex for Fe~I, and 0.0 for Fe~II.  We shall discuss the
  role of these small effects in Section~\ref{sec:relabund}.

\subsection{Abundance Uncertainties}\label{sec:errors}

The random internal error (s.e.$_{\log\epsilon}$) of the abundances in
Column 3 of Table~\ref{tab:abund_gil} is the standard error of the
mean of the abundances from lines analyzed for each element.  The
abundances are also subject to systematic errors resulting from the
uncertainties in our derived model parameters -- \teff, \logg, [M/H],
and \vxi.  The uncertainties we adopt for these quantities are
$\sigma_{T_{\rm eff}}$ = 80\,K (Section~\ref{sec:teff}), $\sigma_{{\rm
    log}g}$ = 0.07~dex (Section~\ref{sec:logg}), $\sigma_{\rm [M/H]}$
= 0.15 dex (see below in this section), and $\sigma_{\xi{_t}}$ =
0.2~{\kms} (Section~\ref{sec:abund}).  For each of our program stars
we varied \teff, \logg, [M/H], and \vxi, one at a time, by these
uncertainties to determine the corresponding abundance differences.
Since we shall be interested mainly in relative abundances, [X/Fe], we
have determined the corresponding uncertainties in those quantities,
while for iron we estimated the uncertainty in [Fe~I/H].  To
determine the total systematic error we co-added the four error
contributions as follows: first, given that the error in {\logg} is
determined primarily by uncertainty in {\teff}, we added the errors in
{\teff} and {\logg} linearly; and second, we then quadratically added
this error to those associated with [M/H] and {\vxi}.  For heuristic
purposes, Table~\ref{tab:errors} presents the average systematic
abundance errors for the 32 program stars, where Columns 1 -- 6
contain the species, errors in {\teff}, {\logg}, [M/H], and $\xi{_t}$,
and in the total systematic error, respectively.

To determine total error estimates, we adopted the following procedure
(see \citealt{norris10}).  The random errors in Column 3 of the
Table~\ref{tab:abund_gil} are based on the dispersion in what is often
a small number of lines, and hence is itself uncertain. We replace
this estimated random error, s.e.$_{\log\epsilon}$, from $N$ lines, by
max(s.e.$_{\log\epsilon}$, s.e.$_{\log\epsilon(\rm Fe~I)}$ ${\times}$
$\sqrt{N_{\rm Fe I}/N}$).  The second term is what one might expect
from a set of $N$ lines having the dispersion we obtained from our
more numerous ($N_{\rm Fe I}$) Fe I lines.  We then quadratically
combine this updated random error and that associated with uncertainty
in the atmospheric parameters from Column 6 of
Table~\ref{tab:abund_gil} to obtain the total error, $\sigma$[X/Fe],
which we present in Column 7 of Table~\ref{tab:abund_gil}.  The mean
of the total error $\sigma$[Fe/H] for the 32 stars in
Table~\ref{tab:abund_gil} is 0.14 dex, which agrees well with the
uncertainty $\sigma_{\rm [M/H]}$ = 0.15 dex we have adopted previously
in this section.

\subsection{Summary of Abundances, and Comparison of Our Atmospheric Parameters with Those of Others}\label{sec:finabs}

We summarize our essential abundance results -- [Fe/H] and relative
abundances [X/Fe] -- for the 32 red giants, in
Table~\ref{tab:relabs_gil}, where the column structure of the table
will be self-evident. For each star there are three rows: the first
presents abundances, the second the corresponding total abundance
errors, and the third the number of lines involved.

In Figure~\ref{fig:fe} we compare our [Fe/H] values with those
presented by \citet{shetrone03}, \citet{koch06}, \citet{venn12},
\citet{lemasle12}, and \citet{fabrizio12}.  The agreement in the
figure is quite satisfactory, except that our values differ
systematically from those of \citet{fabrizio12}, for which there is
considerable dispersion.  At least part of difference will result from
the fact (see Section~\ref{sec:ew}, Figure~\ref{fig:gof}) that the
equivalent widths of \citet{fabrizio12} are on average smaller that
those of the present work.

We are now in a position to compare our atmospheric parameters with
those of \citet{shetrone03}, \citet{koch08a}, \citet{venn12},
\citet{lemasle12}, and \citet{fabrizio12}.  In what follows we present
the mean differences in {\teff}, {\logg}, and [Fe/H] (in the sense
present work -- other work), together with N, the number of stars
involved.  For \citet{shetrone03} the differences are
$\langle$$\Delta${\teff}$\rangle$ = 42 $\pm$ 37\,K,
$\langle$$\Delta${\logg}$\rangle$ = 0.35 $\pm$ 0.07\,dex,
$\langle$$\Delta$[Fe/H]$\rangle$ = --0.01 $\pm$ 0.08, and N = 3; for
\citet{koch08a} they are are $\langle$$\Delta${\teff}$\rangle$ = --95
$\pm$ 59\,K, $\langle$$\Delta${\logg}$\rangle$ = --0.49 $\pm$
0.14\,dex, $\langle$$\Delta$[Fe/H]$\rangle$ = 0.01 $\pm$ 0.05, and 6;
for \citet{venn12} the corresponding values are 113 $\pm$ 26\,K, 0.08
$\pm$ 0.01\,dex, --0.07 $\pm$ 0.08, and 6; for \citet{lemasle12} they
are 104 $\pm$ 11\,K, 0.12 $\pm$ 0.03\,dex, 0.08 $\pm$ 0.04, and 9; and
for \citet{fabrizio12} they are 7 $\pm$ 8\,K, 0.10 $\pm$ 0.01\,dex,
0.33 $\pm$ 0.04, and 31.  The most significant difference in this
comparison is the abundance difference $\langle$[Fe/H]$\rangle$ = 0.33
$\pm$ 0.04 between \citet{fabrizio12} and the present work, given the
good agreement between their {\teff} and {\logg} values with ours.  We
note for completeness that \citet{fabrizio15} reported that their
[Fe/H] values are 0.37~dex smaller than those of \citet{venn12}.

In Figure~\ref{fig:xfe}, we compare our [O/Fe], [Na/Fe], [Mg/Fe], and
[Ca/Fe] relative abundances (one set in each of the vertical panels)
with those of other works for which data are available --
\citet{koch08a} in the leftmost panel, \citet{shetrone03} and
\citet{venn12} together in the next panel, then \citet{lemasle12}, and
finally \citet{fabrizio15} in the rightmost panel.  Clearly, the
agreement is less than ideal.  That said, with typical $S/N$ vales of 10
-- 30 and the small number of lines available for the elements in the
figure (O (1), Na (1), Mg (1 --2), and Ca ($\sim$10 -- 15), the
results are not unexpected.  Examination of the data for
Ca, with $\sim$10 -- 15 lines, where the scatter is smaller, is
supportive of this view.

In Figure~\ref{fig:xfe} one also sees that there is larger scatter in
the comparison with the abundances of \citet{fabrizio15}, with a
tendency to more supersolar values in the latter.  This is a somewhat
puzzling result, given that we would have expected the
\citet{fabrizio12, fabrizio15} values of both [Fe/H] and [X/H] (for
element X) to be smaller that ours by similar amounts, and to some
extent cancel out in the relative abundance, [X/Fe].  We also comment
that if one considers abundances relative to hydrogen, [X/H], rather
that relative abundances [X/Fe], the disagreement seen in
Figure~\ref{fig:xfe} is significantly reduced.  That is, if we define
$\Delta$[X/Fe] = [X/Fe]$_{\rm This~work}$ -- [X/Fe]$_{\rm Fabrizio}$
and $\Delta$[X/H] = [X/H]$_{\rm This~work}$ -- [X/H]$_{\rm Fabrizio}$,
for O, Na, Mg, and Ca, we find average values
$\langle$$\Delta$[O/Fe]$\rangle$ = 0.23 $\pm$ 0.04 ,
$\langle$$\Delta$[Na/Fe]$\rangle$ = 0.49 $\pm$ 0.07,
$\langle$$\Delta$[Mg/Fe]$\rangle$ = 0.19 $\pm$ 0.05, and
$\langle$$\Delta$[Ca/Fe]$\rangle$ = 0.22 $\pm$ 0.03, while in
comparison we have $\langle$$\Delta$[O/H]$\rangle$ = --0.09 $\pm$
0.05, $\langle$$\Delta$[Na/H]$\rangle$ = 0.15 $\pm$ 0.06,
$\langle$$\Delta$[Mg/H]$\rangle$ = --0.13 $\pm$ 0.03,
$\langle$$\Delta$[Ca/H]$\rangle$ = --0.10 $\pm$ 0.04.  For all four
elements, $\langle$$\Delta$[X/H]$\rangle$ is closer to zero than is
$\langle$$\Delta$[X/Fe]$\rangle$ .

Further work at higher $S/N$ is clearly needed if one is to
fully understand the chemical enrichment of Carina.

\section{ENLARGING THE HIGH-RESOLUTION SAMPLE FROM THE LITERATURE}\label{sec:litsam}

We have enlarged the sample in Table~\ref{tab:obsdat} by re-analyzing
the equivalent widths of high-resolution spectroscopic observations of
Carina red giants available in the literature.  Our aim is to
determine abundances for these stars using, as far as possible,
exactly the same techniques as described above, in order to create an
enlarged Carina sample with abundances on the same homogeneous system,
to provide clearer insight into the detailed abundance profile of the
system.

To this end we collated equivalent widths from \citet{shetrone03},
\citet{venn12}, and \citet{lemasle12} in the wavelength range 5300 --
6780\,{\AA}.  These works contain data for 5, 9, and 35 Carina giants,
respectively.  We note that for objects in common with our 32 stars,
their equivalent widths are in good agreement with ours (see
Section~\ref{sec:ew}).  In this compilation of 49 datasets there are
18 stars that are also present in our Table~\ref{tab:obsdat}.  In
total, there are 63 individual stars in the union of the present work
and the literature sample.  We refer the reader to our Appendix for
the cross-identification of names employed by the various authors for
the objects analyzed in the present work.

Basic data for the literature sample are presented in
Table~\ref{tab:litdat}. Columns 1 -- 3 present identification,
literature source, and coordinates, respectively; Columns 4 -- 9
contain $V$, $B-V$, $I$, $J$, $H$, and $K$ photometry from the sources
discussed in Section~\ref{sec:photometry}; and in Columns 10 -- 14 we
show our adopted model atmosphere parameters, \teff, its error
$\sigma${\teff}, surface gravity \logg, [M/H], and {\vxi}, determined
following the methods described in Section~\ref{sec:analysis} and in
Section~\ref{sec:litab} below. 

\subsection{Literature Abundance Re-analysis}\label{sec:litab}

Abundances of the literature stars were determined by using the
procedures outlined in Section~\ref{sec:analysis}, for lines in common
between our linelist and those of the literature sample.  There was
only one significant difference in the analysis of the literature
sample.  This was driven by concern about the accuracy of the
determination of microturbulence ({\vxi}) resulting from the
relatively small number of Fe~I lines ($\sim$15 -- 30) available for
some stars.  In Section~\ref{sec:abund}, we determined {\vxi} by
requiring that there should be no dependence of abundance on line
strength. It became clear in our analysis of the literature data that
having only a small number of Fe~I lines could lead to significant
uncertainty in {\vxi}.  As noted above in Section~\ref{sec:abund}, an
alternative estimate for {\vxi}, used by some workers, is to adopt a
generic value of {\vxi} as a function of {\logg} (e.g.,
\citealt{worley13}).  Insofar as the latter method is independent of
the number of Fe~I lines and {\vxi} is an artifact of one-dimensional
(as opposed to three-dimensional) model atmospheres \citep{asplund00},
we investigated this as follows.  In addition to determining {\vxi} as
described in Section~\ref{sec:abund}, we repeated the analysis of the
stars in Tables~\ref{tab:obsdat} and \ref{tab:litdat}, with {\vxi}
defined by {\vxi} = 2.39 -- 0.31$\times${\logg}, following
\citet{worley13}.  Ideally, the abundances derived by the two methods
should give the same results. To investigate how this might depend on
the number of Fe~I lines available, in Figure~\ref{fig:dfe_micro} we
plot $\Delta$[Fe/H], the difference between abundances obtained with
the two different prescriptions for {\vxi}, as a function of the
number of Fe I lines.  It is clear that the scatter increases
significantly when the number of Fe~I lines is smaller than $\sim$20.
To overcome this problem, in what follows we accepted the abundances
determined by using the formulaic value of {\vxi} for the six most
discrepant points in Figure~\ref{fig:dfe_micro} (Car-1087, MKV0556,
MKV0577, MKV0733, MKV0743, MKV0812 in Table~\ref{tab:litdat});
otherwise we adopted values obtained using {\vxi} as described in
Section~\ref{sec:abund}.

Our abundances for the literature samples are presented in
Table~\ref{tab:relabs_kvs} for the works of \citet{venn12} and
\citet{shetrone03} and in Table~\ref{tab:relabs_lem} for that of
\citet{lemasle12}.  The formatting in these tables is the same as in
Table~\ref{tab:relabs_gil}.

\section{THE ([FE/H]) METALLICITY DISTRIBUTION FUNCTION\label{sec:mdf}}

An essential observational ingredient which constrains the population
characteristics of Carina is its metallicity distribution function
(MDF), for which we take the distribution function of [Fe/H] as proxy.
We present the MDF for our 63 RGB star sample in panel (a) of
Figure~\ref{fig:mdf}.  The mean values and dispersions of the
distribution are --1.59 and 0.33~dex (where for the latter we have
applied a small correction to take into account the error of 0.15 dex
involved in the measurement of [Fe/H]), respectively, while the range
in [Fe/H] covers --2.68 to --0.64. For future reference we note that
the 63 stars discussed here represent a somewhat incomplete sample of
Carina's RGB due to two effects. First, the four investigations
leading to the sample were chosen for high-resolution spectroscopic
observations, leading to a bias towards brighter stars (see our
Figure~\ref{fig:cmd} and Figure 5 of Paper I).  Second, as discussed
in Paper I (Section~2.1.1), the sample is probably slightly incomplete
at the 5 -- 10\% level, insofar as the RGB {\bio} distributions of
\citet{koch06} and \citet{walker09} are slightly broader than that
observed for the present sample.  

Panel (b) of Figure~\ref{fig:mdf} presents the MDFs obtained from the
results of other high-resolution spectroscopic investigations, where
the thin line is based on the collective results of
\citet{shetrone03}, \citet{venn12}, and \citet{lemasle12}, while the
thick line comes from the work of \citet{fabrizio12} for the 31 stars
in common with the present investigation -- resulting from independent
analysis of the same spectra analyzed in this work (obtained in ESO
Proposal 180.B-0806(B) (PI Gilmore)).  We refer the reader to
Section~\ref{sec:finabs} for a discussion of the differences between
the abundances of \citet{fabrizio12}, on the one hand, and those of
the present work and of the collective results of \citet{shetrone03},
\citet{venn12}, and \citet{lemasle12}, on the other.  Panels (c) and
(d) contain the results of \citet{koch06} and \citet{starkenburg10},
respectively, based on their analysis of measurements of Ca~II IR
triplet data.  Here one sees that the MDFs derived from the Ca~II data
reach somewhat lower abundances than does our analysis.  It is unclear
if this is due to selection biases or errors in the application of the
CaT calibrations (e.g., errors in the assumed stellar ages).  We refer
the reader to Paper I (Sections 2.3 and 4.) where we have discussed
potential shortcomings of the calibrations of the triplet adopted in
these publications as applied to Carina.  Finally, in panel (e) of
Figure~\ref{fig:mdf}, we present the MDF of \citet{kordopatis16}, from
their analysis of medium resolution near-infrared spectra.  For these
data the median error of the metallicity is 0.29 dex.  We defer
further consideration of our MDF to Section~\ref{sec:distributions}.

\section{RELATIVE ELEMENT ABUNDANCES}\label{sec:relabund}

In Figures~\ref{fig:relab1} -- \ref{fig:relab6}, we present relative
abundances, [X/Fe], derived in the present work for 18 atomic species
in the range O -- Eu\footnote{We do not plot results for Al since this
  element was detected in only one star for which we have spectra.},
as a function of [Fe/H]. In each column, the top three panels contain
results from analysis of our equivalent widths, from those of
\citet{shetrone03} and \citet{venn12}, and from those of
\citet{lemasle12}, respectively.  In the bottom two panels, the
results for these three sub-samples are presented as average values
for the 63 independent stars in Tables~\ref{tab:relabs_gil},
\ref{tab:relabs_kvs}, and \ref{tab:relabs_lem}, as available, weighted
by the inverse squares of their total errors.  (These average values
and their errors may be found in Table~\ref{tab:relabs_aver} of the
Appendix.)  In these panels, the final uncertainties in the relative
abundances are shown in two ways: in the second row from the bottom,
the usual error bars are presented, while in the bottom row, the size
of the circles decreases linearly as the error increases.  The full
red line in each panel represents the mean values for Galactic halo
stars, based on the data presented by \citet{venn12}, which we shall
adopt in what follows in our discussion of the abundances of our
Carina sample.

In Figure~\ref{fig:contours} an alternative summary of the averaged
abundances is presented following \citet{yong13a}, in which the
results are represented as summed double generalized histogram contour
plots having Gaussian kernels equal to the individual total observational
errors.

In this section we present a brief overview of these results, where in
the comparison of Carina's average relative abundances with those of
Galactic halo values, the averages will be taken over the range --2.0
$\leq$ [Fe/H] $\leq$ --1.0.

\subsection{The Light Elements O and Na}

Oxygen and sodium abundances have been determined from only one line
each in this analysis (Na~I 5688.2\,{\AA}, [O~I] 6300.3\,{\AA}; see
Table~\ref{tab:ew1}).  We have selected to restrict our analysis to
these lines, which have equivalent width measurements throughout our
sample, are known to yield abundances in close accord with other
features when these are available, and are in good agreement between
various analyses (e.g., see \citealp{venn12}, Sections 4.4.2 and
4.4.3).  Our abundances do not appear to deviate wildly from Galactic
halo values, although some stars do show abundances lower than those
of the halo by 0.2 -- 0.3 dex, particularly for sodium.  As discussed
by \citet{venn12}, the production of sodium follows that of the
$\alpha$-elements until AGB stars contribute, and those latter
contributions can be metallicity dependent.  Thus, reasonable
differences in the mean metallicity of the AGB stars in the Galactic
halo versus those in Carina could account for the small differences in
some stars. 

The mean values, using weights equal to the inverse square of the
errors, are $\langle$[O I/Fe]$\rangle$ = 0.46 $\pm$ 0.04 and
$\langle$[Na~I/Fe]$\rangle$ = --0.28 $\pm$ 0.04, which may be compared
with the Galactiic halo values plotted in Figures~\ref{fig:relab1} of
0.60 and 0.00, respectively.  As pointed out to us by a referee, if
one has low $\alpha$ relative abundances for Mg -- Ti in Carina, as
seen in our Figures 9 -- 11, one might also expect a significant
reduction in oxygen as well.  We refer the reader to the work of
\citet{ramirez12}, which confirms the reduction of [O/Fe] in dwarfs of
the Galactic halo having [Fe/H] $<$--0.8 and low relative abundances
of the $\alpha$-elements, Mg -- Ti.  One might claim that the lower
values of [O/Fe] in some of the stars in Figure 9 might offer support
for this requirement.  That said, we are of the view that further work
of considerably higher $S/N$ than available here is needed to resolve
this issue.

The most significant result from our data, however, is that there is
no strong anti-correlation between the abundances of these two
elements, in clear contrast to what is seen in all of the Galactic
globular clusters.  We shall return to this matter in Section 11.

\subsection{The $\alpha$-elements: Mg, Si, Ca, and Ti}

The $\alpha$-elements are built through He-capture during various
stages in the evolution of massive stars, and dispersed through SNe II
events.  Thus, the {\alphafe} ratio is a key tracer of the relative
contributions of SN II to SN Ia products in a star forming region.  As
discussed in \citet{venn12}, Ti is not a true $\alpha$-element. It
behaves, however, like one since the dominant isotope $^{48}$Ti forms
through explosive Si-burning and the $\alpha$-rich freeze out during a
core collapse supernova event \citep{woosley95}. In our analysis, we
examine two lines of Mg I, three of Si I, 22 of Ca I, 11 of Ti I, and
5 of Ti II.  Our Ti I and Ti II abundances are in good agreement when
both are measured in the same star, and therefore we average the
abundances of the two species.  The Mg II lines near 5200\,{\AA} are
too strong for the present analysis.

The mean values, using weights equal to the inverse square of the
errors, are $\langle$[Mg/Fe]$\rangle$ = 0.12 $\pm$ 0.02,
$\langle$[Si/Fe]$\rangle$ = 0.28 $\pm$ 0.04, $\langle$[Ca/Fe]$\rangle$
= 0.04 $\pm$ 0.01, and $\langle$[Ti/Fe]$\rangle$ = 0.18 $\pm$ 0.01.
With the exception of Si, which is the least accurately measured, all
of these fall below halo values (0.30~dex), which is a well-known
signature of dSph systems, well documented for Carina by
\citet{shetrone03} and \citet{venn12}.  We shall discuss this group of
elements at some length in Section~\ref{sec:alphas}.  An interesting
effect which we shall emphasize there is that the quality of the data
varies strongly between the four elements.  Note the clearer tightness
of the [Ca/Fe] vs. [Fe/H] relationship in Figure~\ref{fig:relab2}, in
some contrast to the looser ones for the other $\alpha$-elements --
driven as we shall discuss in Section~\ref{sec:alphas} by the
considerably larger number of lines available for Ca in the wavelength
range observed in the present investigation.

\subsection{The odd-elements (Sc, V, Mn), Iron-peak Elements (Co, Cr, Ni), and Cu}

Massive stars (M $>$ 8~M$_\odot$) provide the first contributions to
the chemical elements due to their short life times, producing metals
both during their evolution and during core-collapse SNe.  It is
currently thought that the iron-peak and odd-Z elements are
synthesized in these core-collapse SNe through a variety of processes
(explosive burning ranging from He to Si, $\alpha$-rich freeze out, a
range in yields depending on neutron densities, explosion energies,
mass of the progenitor, metallicity and initial composition, and the
mass cut for the fraction expelled; see e.g., \citealp{woosley95}, \citealp{nomoto13}, and \citealp{pignatari16}).  Abundances of these elements tend to show
a strong odd-even effect in the predicted yields, where the odd-Z
nuclei have lower abundances than those with even-Z.  Only at later
times will lower mass stars contribute to these elements through SNe
Ia events, possibly dominating the total-iron group inventory.  The
diversity of SNe~Ia have been modelled by \citet{kobayashi15}, ranging
from single and double degenerate channels, to rare types from hydrid
white dwarfs, and their impact in metal-poor systems with stochastic
star formation, such as Carina.

In general, our odd-Z elements (Sc, V, Mn) are measured from a small
number of spectral lines (1 -- 10), and include hyperfine structure
corrections (see Section~\ref{sec:abund}).  The ranges seen in these
abundance ratios are quite large, with $\Delta$[Sc/Fe] and
$\Delta$[V/Fe] $\sim \pm0.5$.  The mean values for Sc and V, using
weights equal to the inverse square of the errors, are
$\langle$[Sc/Fe]$\rangle$ = 0.00 $\pm$ 0.02 and
$\langle$[V/Fe]$\rangle$ = --0.05 $\pm$ 0.03, while for Mn the value
is $\langle$[Mn/Fe]$\rangle$ = --0.37 $\pm$ 0.01.  In comparison the
corresponding Galactic halo values are 0.10, 0.10. and --0.40.
Only $\langle$[Mn I/Fe]$\rangle$ tends to be above the Galactic halo
data.  \citet{kobayashi15} have predicted that in inhomogeneously
mixed systems, an enhanced [Mn/Fe] value could indicate contributions
from SNe Iax models (a low metallicity, hybrid white dwarf, in a close
binary system).

Our iron-peak elements Cr and Ni are measured from a small number of
lines (4 -- 11) in most of the stars in our sample, and both are in
excellent agreement with the Galactic halo trends. The mean values,
using weights equal to the inverse square of the errors, are
$\langle$[Cr/Fe]$\rangle$ = --0.07 $\pm$ 0.02 and
$\langle$[Ni/Fe]$\rangle$ = --0.10 $\pm$ 0.01, which compare well with
Galactic halo values of 0.00 and 0.00 respectively.  We note that
larger dispersions and errors are reported for these elements from the
Lemasle et al. dataset.  Also, a small number of Co and Cu abundances
are available for $\le$10 stars from one spectral line each (Co I
5647.2\,{\AA} and Cu I 5700.2\,{\AA}).  Both are limited to the more
metal-rich stars in our sample.  The [Co/Fe] ratios are in good
agreement with the Galactic trend, while the [Cu/Fe] ratios may be
slightly below the Galactic halo values.  The nucleosynthesis of Cu is
more complicated than most of the iron-group and odd-Z elements; for
example, Cu can also have a significant contribution from the weak
s-process in massive stars (e.g., \citealp{pignatari16}).

\subsection{[Fe~II/Fe]}\label{sec:fe2}

[FeII/Fe] provides a necessary and fundamental requirement of any
chemical abundance analysis in which it can be determined. That is, a
sound analysis should yield [Fe II/Fe] = 0. Inspection of the middle
column of Figure 10 shows this expectation is not met, with the bulk
of [Fe II/Fe] values lying in the range of 0.0 to +0.5 dex.  If one
considers the results for the 31 stars plotted in the upper panel (for
which we presented new observational material and determined
abundances in Sections 2 and 3) the mean value, determined using
weights equal to the inverse square of the errors, is
$\langle$[Fe~II/Fe]$\rangle$ = 0.25 $\pm$ 0.02.

Inspection of Table~\ref{tab:errors} suggests that this is about two
times larger than the uncertainties in the analysis, which are
dominated for both elements by the uncertainties in {\teff}.  In the
table, one sees that an average error of $\Delta${\teff} = --80~K will
lead to an error of $\Delta$[Fe II/Fe] = +0.15.

A second consideration is the overionization of Fe I by the radiation
field in the atmosphere of a metal-poor red giant, which is neglected
under the assumption of LTE.  For the typical stellar parameters of
our sample ({\teff} = 4500 K, {\logg} = 1.0, [Fe/H] = --2.0 to --1.0), this
effect is estimated to be $\le$+0.1 dex \citep{lind12}.  On the other
hand, \citet{mashonkina16} find that for Fe I lines with equivalent
widths in the range 100 -- 200~m{\AA}, the effect can be as large as
+0.2 dex in the same stellar parameter regime.

Given the magnitude and sign of the NLTE corrections, the LTE FeI and
FeII abundances are not anomalous.

\subsection{The Heavy Neutron-capture Elements: Ba, La, Nd, Eu}\label{sec:ba}

We have determined element abundances for four heavy neutron-capture
elements. These are formed in massive stars and core collapse
supernovae through rapid neutron capture reactions, and those other
than Eu also form via slow neutron captures during the thermal pulsing
AGB stages in intermediate-mass stars.  The specific details of these
nucleosynthetic processes are a dynamic field of current research.
For the core collapse SNe, new models and calculations of their yields
include details of the SN explosion energies, explosion symmetries,
early rotation rates, and metallicity distributions (e.g., see
\citealp{kratz14}, \citealp{nishimura15}, \citealp{tsujimoto15}, and
\citealp{pignatari16}), as well as exploration of contributions from
compact binary mergers as a (or as the most) significant site for the
r-process (e.g., see \citealp{fryer12}, \citealp{korobkin12},
\citealp{perego14}, and \citealp{cote16}).  Similarly, predictive
yields from AGB stars by mass, age, metallicity distributions, and
details of convective-reactive mixing are also an active field of
research (e.g., see \citealp{lugaro14}, \citealp{cristallo15}, and
\citealp{pignatari16}).

Our abundances for Ba, La, Nd, and Eu are from only a few lines; for
example, Eu II 6645.1\,{\AA} and Nd II 5319.8\,{\AA} are from only the
one line each.  Barium and lanthanum are from 3 and 4 spectral lines
each, respectively, but the Ba II lines tend to be strong whereas the
La II lines tend to be weak (or absent).  There is a tendency in the
relative abundances of these four elements to be similar to or higher
than those for stars in the Galactic halo at metallicities [Fe/H] $\ge
-2.0$, where the highest abundances tend to occur for [La/Fe] and
[Ba/Fe] -- in all samples (\citealp{shetrone03}, \citealp{lemasle12},
\citealp{venn12}, and this paper).  We do not interpret this as due to
uncertainties in our abundances but rather to a real astrophysical
effect; for example, the bulk of our sample has a full range of
[Ba/Fe] $\sim$ 0.0 to +0.5, with $<$[Ba/Fe]$>$ = 0.17 $\pm$ 0.02
(weighted by the inverse square of the errors, and where we somewhat
arbitrarily exclude the two objects with [Ba/Fe] $<$ --0.50).  In
comparisom the Galactic halo value is [Ba/Fe] = --0.10.  The one sigma
offset due to $\Delta$Teff $= +80$ K, (see Table\ref{tab:errors})
leads to an error of $\Delta$[Ba/Fe] = $-0.08$, which lessens but does
not eliminate the offset from the bulk of the Galactic stars.  These
offsets are similar for La, Nd, and Eu.  This abundance signature is
typically seen in dwarf galaxies (see \citealp{tolstoy09}) and
interpreted in terms of higher yields of Ba (second-peak s-process
element, and thus also La and Nd) from metal-poor AGB stars.  What is
more surprising are the low [Ba/Fe] values (and even two stars with
low [Nd/Fe]) at high metallicities.  \citet{venn12} interpreted this
signature in terms of inhomogeneous mixing in the Carina interstellar
medium with pockets of iron-rich SNe Ia gas.  We shall return to this
topic in Section~\ref{sec:sne}.

\section{THE $\alpha$-ELEMENTS}\label{sec:alphas}

The $\alpha$-elements are of importance for a complete understanding
of Carina, given the role they play in the interpretation of its
abundance patterns and CMD.  Here we address their dependence on
metallicity, their relative abundances compared with solar values,
their distribution function, and what they have to tell us about
the manner in which chemical evolution within the system may have occurred
during their ejection from SNe and mixing with the ambient medium to
form subsequent generations.

\subsection{The $\alpha$-elements as a Function of [Fe/H]}\label{sec:alphas_vs_fe}

The $\alpha$-elements comprise O, Ne, Mg, Si, S, Ar, Ca, Ti (e.g.,
\citealp{vandenberg06}), while in practice in the present work the
available elements are O, Mg, Si, Ca, and Ti -- with O and Si
determined in only a minority of the sample.  This leads to a number
of complicating factors concerning the comparison between observation
and theory, of which we mention two.  The first is that given the
results of stellar evolution and nucleosynthesis calculations for
massive stars (e.g., \citealp{woosley95}) the possibility exists that
in the metal-poor regime the $\alpha$-elements might have a non-solar
relative distribution.  This leads to the difficulty that stellar
evolution isochrone grids that assume scaled-solar $\alpha$-ratios
will be inaccurate at some unknown level.  How does one best compare
observation with theory?  \citet{vandenberg15} have argued that since
Mg has a greater relative abundance than the other {$\alpha$}-elements
it should therefore be a better probe than the other elements (in
particular the more readily and accurately observed Ca) given it has a
greater effect on a star's evolution.  The second point is that, in
our opinion, only the abundance of Ca is well-determined in currently
available RGB samples of Carina.  This situation is driven by the
small numbers of lines available for analysis (on average 1.8, 0.8,
14.6, and 2.5 lines for Mg I, Si I, Ca I, and Ti II, respectively) and
the relatively low $S/N$ (7 -- 23) per 0.03 \AA\, pixel for the
majority of stars having full wavelength coverage in the present
analysis.  Against this background, we seek to address two questions.
First, is there evidence for differences in the relative abundances of
the $\alpha$-elements, one from another -- in particular, does [Mg/Ca]
differ from the solar value of zero?  Second, do the
$\alpha$-abundances change as a function of [Fe/H], both in size and
dispersion.

As a first approach, we present in Figure~\ref{fig:alphas} two
comparisons of light-$\alpha$ and heavy-$\alpha$ elements versus
[Fe/H], together with their difference as a function of [Fe/H], based
on the data presented in Figures~\ref{fig:relab1} -- \ref{fig:relab6}.
The upper panels of each column contain light-$\alpha$ elements, while
the middle panels present heavy ones.  The bottom panels show the
difference between the light and heavy species.  One sees that, below
[Fe/H] = --1.2, on average the difference between the two species is
not large, of order 0.1 dex, with a hint of a dependence on [Fe/H].
For the complete Carina sample in the right bottom panel,
$\langle$[Mg/Ca]$\rangle$ = 0.12 $\pm$ 0.03.  It will be interesting
to see if data of higher quality support these results.

A fundamental result of abundance studies of the dSph galaxies is that
the $\alpha$-elements exhibit a very different behavior with respect
to [Fe/H], when compared with that found in Galactic halo stars.  That
is, while the latter exhibit relatively constant, and enhanced, values
of {\alphafe} $\sim0.3 - 0.5$ for [Fe/H] $<$ --1.0, which decrease to
$\sim0.0$ as [Fe/H] decreases to 0.0, for dSph systems the decrease of
the $\alpha$-elements to solar values begins at much lower values of
[Fe/H]~$\sim-1.5$ to --2.0 (see e.g., \citealp{tolstoy09}).  This
behavior is understood in terms of different rates of star formation
between the two types of systems, in which the dwarf galaxies have
lower star formation rates and Type~Ia SNe play a larger role at
lower [Fe/H] in reducing {\alphafe} than occurred in the Galactic
halo (see e.g., \citealp{gilmore91}).  In Carina, the present data
suggest that on average the abundance of the {$\alpha$}-elements, in
particular that of the most accurately measured element, Ca, is lower
than Galactic halo values by $\sim$0.2 dex at all values of [Fe/H],
below $\sim -1.0$.  We shall return to this in Section~\ref{sec:incommix}.

\subsection{The {\alphafe} Distribution Function}\label{sec:disfun}

A second fundamental relationship involving {\alphafe} is its
distribution function; and of considerable importance is the
comparison of the Carina distribution with that of the Galactic halo.
In Figure~\ref{fig:alpha_v3} we present {\alphafe} values for Mg, Si,
Ca, and Ti determined for the Galactic halo, Carina, and the globular
cluster {\wcen}. (We include {\wcen} insofar as it may be of interest
given it is a Galactic globular cluster of baryonic mass 4.0
$\times10^{6} $M$_{\odot}$ (\citealp{dsousa13}) similar to that of
Carina, and which has chemically enriched itself not only in the light
elements, but also in iron.)  See the figure caption for sample
details.  We caution that while individual studies might be internally
consistent, there are likely to be small zero-point offsets between
the various works.  The format of Figure~\ref{fig:alpha_v3} is
similar to that of Figure~\ref{fig:alphas}, and the top four panels in
each column present results for Mg, Si, Ca, and Ti, taken from sources
listed in the figure caption, while in the bottom panel the value of
{\alphafe} is determined by averaging [Mg I/Fe], [Ca I/Fe], and [Ti
  II/Fe] when all three quantities are available.  In the interest of
obtaining the most reliable result for Carina we have required that a
value only be accepted when a star has no fewer than two lines
available for each element, and we have weighted the individual
abundances by the inverse square of their errors. (We made exceptions
for three stars (Car-1087, Car-5070, and Car-7002) which have only one
line of Mg and/or Ti, but the highest $S/N$ $\sim$ 30 in the sample.)
The data for the 30 Carina giants that meet these criteria are
presented in Table~\ref{tab:alphas_v2}, and were used to determine the
{\alphafe} distribution function presented in the lower panel of
Figure~\ref{fig:obs_distrib}, where the thick red line is based on the
means of Mg, Ca, and Ti determined by using weights inversely
proportional to the inverse square errors in abundance, while the thin
black line presents results when equal weights are assumed. In the
upper panel the distribution function is presented for the Galactic
halo (over the range --3.0 $<$ [Fe/H] $<$ --1.0).

We make the following points concerning Figures~\ref{fig:alpha_v3} and
~\ref{fig:obs_distrib}: (i) as noted above, for Carina, our
spectroscopic {\alphafe} distribution is determined in large part by
that of Ca, given the higher precision of the [Ca/Fe] values; but that
said the lower accuracy of Mg and Ti causes the two distributions in
Figures~\ref{fig:obs_distrib} to be offset by only
$\Delta${\alphafe}$\sim$~0.05 dex; (ii) the weighted {\alphafe}
distribution for Carina is smaller by $\sim 0.25$ dex than that of the
Galactic halo.  As noted above, the Galactic halo data come from a
number of studies and could be affected by zero-point offsets.
Inspection of the four halo datasets (\citealp{fulbright00},
\citealp{barklem05}, \citealp{preston00}, and \citealp{yong13a})
plotted in the left panel of Figure~\ref{fig:alpha_v3}, however,
suggests that this is not the case insofar as each of them has
$\langle${\alphafe}$\rangle$ $\sim 0.3$; (iii) in Carina, the observed
spread in {\alphafe} is small, with dispersion 0.13 $\pm$~0.02 dex;
(iv) while the number of stars is not large at lowest abundance, for
[Fe/H] $<$ --2.0 one finds $\langle${\alphafe}$\rangle$ = 0.04
$\pm$~0.08, which is small compared with the higher values of
{\alphafe} that one associates with the Galactic halo (i.e. {\alphafe}
$\sim~0.3 - 0.4$); and (v) while more data are required, there appear
to be stars with low values of {\alphafe} $\sim -0.2$ over the entire
range --2.0 $<$ [Fe/H] $<$ --1.0 (where some 80\% of the sample lie).

The final two points are of considerable importance. If point (iv)
obtains, the absence of Galactic-halo-like {\alphafe} values
({\alphafe}~$\sim~0.3 - 0.4$) on the RGB would suggest that all of the
Carina stars currently observed to date at high-resolution formed from
material enriched by ejecta from either Type Ia SNe
(e.g. \citealp{tolstoy09} and references therein) or from Type II SNe
in a population having a mass function skewed towards lowest masses
(e.g. \citealp{kobayashi06} and \citealp{kobayashi11}), both of which lead to lower
{\alphafe} values in subsequent generations.  We shall return to point
(v) in the next sub-section.

The improved data presented in this section may also be used to
re-address the question of the dependence of {\alphafe} on atomic
species discussed in Section~\ref{sec:alphas_vs_fe}.
Figure~\ref{fig:mgca30_v2} presents the dependence of [Mg/Ca] on
[Fe/H] for the Galactic halo and Carina, for the samples discussed in
the present section and defined in Figure~\ref{fig:alpha_v3}, for
stars for which both Mg and Ca abundances are available.  At noted
above, this Carina sample contains only 30 stars, with somewhat
conservative selection criteria, and as such might be expected to be
of higher quality that the results discussed in
Section~\ref{sec:alphas_vs_fe}.  The result in the right panel of
Figure~\ref{fig:mgca30_v2} leads essentially to the same conclusion as
reported in Section~\ref{sec:alphas_vs_fe}: the difference between the
[Mg/Fe] and [Ca/Fe] distributions of Carina are essentially the same,
with a mean value of $\langle$[Mg/Ca]$\rangle$ = 0.13 $\pm$ 0.03.  For
the halo sample in Figure~\ref{fig:mgca30_v2} the mean value is
$\langle$[Mg/Ca]$\rangle$ = --0.025 $\pm$ 0.008.

\subsection{Incomplete Mixing}\label{sec:incommix}

We complete this section with a comment on the asymmetry of the
{\alphafe} distribution function seen in Figure~\ref{fig:obs_distrib}
and the distribution of stars in the ([Ca/Fe], Fe/H]) -- plane, where
we use [Ca/Fe] as proxy for {\alphafe}, recalling that Ca has the most
accurately determined abundances among the $\alpha$-elements.  The
main point is that the asymmetry of the {\alphafe} distribution
function to smaller values of {\alphafe} in
Figure~\ref{fig:obs_distrib} is clear.  To investigate this further,
in Figure~\ref{fig:cafe}, in the upper panel, we reproduce the [Ca/Fe]
results from Figure~\ref{fig:relab2} (second panel from the bottom),
and in the lower panel present only the higher quality data in the
sample that have errors $\sigma$[Ca/Fe] $<$ 0.15.  As noted in the
previous section (point (v)), and seen more clearly in the lower
panel, there are stars with low values of [Ca/Fe] (i.e. below $\sim
0.0$ dex) at all values of [Fe/H], most significantly in the range
--2.0 $<$ [Fe/H] $<$ --1.0, where the density of points is highest.
While more and better data are needed, the existence of low [Ca/Fe]
values at all [Fe/H] drives the asymmetry of the {\alphafe}
distribution function at all values of [Fe/H] and is very likely
indicative of inhomogeneous mixing of the ejecta of supernovae within
each of Carina's generations, between the relatively
$\alpha$-element-poor ejecta from Type Ia SNe and the ambient medium
during the several formation epochs of Carina's populations.

\section{THE TYPE IA SUPERNOVAE ENRICHMENT OF CARINA}\label{sec:sne}

\citet{venn12} first demonstrated quite generally that Carina exhibits
enrichment by Type Ia SNe.  In particular, their star Car-612, with
[Fe/H] = --1.3, exhibits element deficiencies of $\Delta$[X$_{\rm
    i}$/Fe] $\sim0.7$ for all elements other than those in the
Fe-peak, relative to Galactic halo stars having the same [Fe/H].  They
interpret this result in terms of Car-612 having formed from material
that was substantially enriched by the ejecta of Type Ia SNe which
contained principally Fe-rich material.  We have adapted the formalism
of their Section 5.4 to search for similar stars in the present
sample.  For each star we determined $\Delta$[X$_{\rm i}$/Fe]$_{\rm
  Carina}$ = [X$_{\rm i}$/Fe]$_{\rm Carina}$ -- [X$_{\rm i}$/Fe]$_{\rm
  Halo}$, where [X$_{\rm i}$/Fe]$_{\rm Halo}$ is the value determined
at the observed [Fe/H] and obtained using the halo averages of
\citet{venn12} (and plotted in our Figures~\ref{fig:relab1} --
\ref{fig:contours}).  We then compared these with the Type Ia and Type
II SNe models of \citet{iwamoto99} as follows.  For each of the seven
\citet{iwamoto99} Type Ia models we created a set of admixtures of
Type Ia and Type II models, similar to that which one might envisage
having formed in Carina as the ejecta of Type Ia SNe appeared in the
system.  Then for each admixture we computed the abundance difference
$\Delta$[X$_{\rm i}$/Fe]$_{\rm Model}$ = [X$_{\rm i}$/Fe]$_{\rm
  Admixture}$ -- [X$_{\rm i}$/Fe]$_{\rm {Type~II}}$.  For each Iwamoto
et al.\ model admixture we then determined the summation S =
$\Sigma(\Delta$[X$_{\rm i}$/Fe]$_{\rm Carina}$ -- $\Delta$[X$_{\rm
    i}$/Fe]$_{\rm Model})^{2}$, for the set of elements i = O, Na, Mg,
Si, Ca, Ti, V, Cr, Mn, Co, and Ni, and accepted the admixture having
the smallest S as representing the best model for the star in
question.

Figure~\ref{fig:iwam6} shows an example of the process for the
\citet{venn12} star Car-612 (CC07452 in our Table~\ref{tab:obsdat}).
Here each cell represents a comparison of the observations with the
model admixture described in the legend of that cell, where the
abscissa is atomic number, and the ordinate represents the difference
between the abundance of the admixture and that of Type II material,
for both observation and theory. Each of the seven columns pertains to one
of the \citet{iwamoto99} models, and for a given column the fraction
of Type Ia material increases from zero Type Ia material in the bottom
cell to an admixture of 0.6 Type Ia plus 0.4 Type II material at the
top.  In the case of Car-612 the best fit occurs for the
\citet{iwamoto99} W70 Type Ia for an admixture of 0.4 Type Ia + 0.6
Type II material, which is presented in blue in the second column from
the left and the third row from the top.

The results of this exercise for the 63 red giants are presented in
Table~\ref{tab:iwamoto}, where columns (1) -- (3) contain the
starname, the \citet{iwamoto99} Type Ia model admixed in the
best-fitting model, and the fraction of Type Ia material involved.
For seven stars, which we identify in the table, abundances are
available for only four of the 11 elements that we include in the
above analysis, which leads to less accurate, and inconclusive
results, and which we do not consider further here.

The present formalism differs from that of \citet{venn12} insofar as
the analysis of \citet{iwamoto99} includes no species past the
iron-peak, while \citet{venn12} also include heavy-neutron-capture
elements.  To investigate the possible significance of this we extend
the analysis to include barium, and compute the mean difference,
$\Delta$[X$_{\rm i}$/Fe], between the relative abundance [X$_{\rm
    i}$/Fe]$_{\rm Carina}$ and [X$_{\rm i}$/Fe]$_{\rm Halo}$ (weighed
by the inverse square of the errors) for the non-iron-peak elements O,
Na, Mg, Si, Ca, Ti, and Ba.  The resulting $\langle\Delta$[X$_{\rm
    i}$/Fe]$\rangle$ values and $\Delta$[Ba/Fe] are presented in columns (4)
and (5) of Table~\ref{tab:iwamoto}.

Figure~\ref{fig:histograms} shows generalized histograms of the
logarithm of the Type Ia fractions (Fraction$_{\rm Type~Ia}$) and
$\langle\Delta$[X$_{\rm i}$/Fe]$\rangle$. In the left and right panels
one sees that the bulk of the stars fall within the ranges 0.0 -- 0.2
and --0.3 -- 0.0, respectively, and in both panels there are a few
objects above and below these ranges, respectively.  These latter
objects are the stars that show the characteristics of Type Ia
enrichment\footnote{The reader may ask what causes there to be a few
  stars having values above 0.0 in the right panel of
  Figure~\ref{fig:histograms} but none below 0.0 in the left.  This
  results from the definitions: put most simply,
  $\langle\Delta$[X$_{\rm i}$/Fe]$\rangle$ admits positive values,
  while Fraction$_{\rm Type~{Ia}}$ does not permit negative ones.}.

In order to proceed with the discussion, we identify Type Ia enriched
objects as those in which (i) the average of Fraction$_{\rm Type~Ia}$
and --$\langle\Delta$[X$_{\rm i}$/Fe]$\rangle$ is greater than 0.20,
and (ii) $\Delta$[Ba/Fe] = [Ba/Fe]$_{\rm Carina}$ -- [Ba/Fe]$_{\rm
  Halo}$ $<$ 0.0 (if data are available).  There are seven objects
that meet these requirements, for six of which we plot in
Figure~\ref{fig:iwam_venn} (left panels) the abundances relative to
those of the Galactic halo, $\Delta$[X$_{\rm i}$/Fe] (= [X$_{\rm
    i}$/Fe]$_{\rm Carina}$ -- [X$_{\rm i}$/Fe]$_{\rm Halo}$), as a
function of atomic number.  We also plot as a line the best fitting
admixed Iwamato model described above.  We note for completeness that
CC07452 and CC96633 were first described by \citet{venn12} as having
been formed from material that had been enriched by ejecta from Type
Ia SNe.  MKV0740 and MKV0743 were first analyzed by \citet{lemasle12}.
It will be interesting to see if future observations support our
identification of them as Type Ia enriched.  Our seventh candidate,
Car-1087, is shown in the top, right panel.  Given the high value of
$\Delta$[Mg/Fe] in this star, we are reluctant to identify it as Type
Ia enriched.  For completeness, the remainder of the right panel in
the figure contains stars that show little if any evidence for Type Ia
enrichment.

If the six stars in the left panels of Figure~\ref{fig:iwam_venn} are
indeed all Type Ia enriched, this would suggest that some 10\% of the
56 Carina RGB sample presented in Table~\ref{tab:iwamoto} have this
property.  While the statistics available for the Galactic halo are
small, the Carina fraction appears large compared with the 1 -- 2\%
reported as Galactic halo ``Fe-enhanced'' stars by \citet{yong13a}
(who used the same analysis techniques as adopted here).

\section{AGE ESTIMATES}\label{sec:age}

The location of the RGB in the color-magnitude diagram (or
alternatively in the luminosity, effective temperature plane) is
principally determined by [Fe/H], age, and to a lesser extent by
{\alphafe} and helium.  For representative Carina giants in the
present sample, with [Fe/H]/{\alphafe}/Age = --1.50/0.0/7.0 and
absolute magnitude {\mv} = --1.5, for example, uncertainties in
{\teff} and [Fe/H] of 50\,K and 0.1 dex each lead to age differences
of $\sim 3$ Gyr.

There are also, as we shall see, intriguing age differences of the
same order between different stellar evolution modelling formalisms.
It is thus well appreciated that this is not the place to anticipate
accurate age determinations.  This caveat notwithstanding, in this
section we shall investigate what may be learned from determining ages
from our current {\teff}, [Fe/H], and {\alphafe}, interpreted using
currently available isochrones.  A further uncertainty in these
assumptions lies in our choice of {\alphafe}.  Following the
discussion in Section~\ref{sec:disfun}, we recall that our values of
{\alphafe} are determined principally by [Ca/Fe], given the greater
relative accuracy with which the determination could be made.  As
noted there, this value differs by only $\sim$0.08~dex from that
obtained for [Mg/Fe].

With our M$_{\rm V}$, {\teff}, [Fe/H], and {\alphafe} (using [Ca/Fe]) we
have interpolated in the Yale-Yonsei
\citep{demarque04}\footnote{http://www.astro.yale.edu/demarque/yyiso.html},
Victoria-Regina
\citep{vandenberg06}\footnote{http://www.cadc-ccda.hia-iha.nrc-cnrc.gc.ca/community/VictoriaReginaModels/},
and Dartmouth
\citep{dotter08}\footnote{http://stellar.dartmouth.edu/models/isolf\_new.html}
isochrones, assuming that all of our stars lie on the RGB, and have
ages in the range 2 -- 14 Gyr.  In the cases where extrapolation in age
is required we have adopted values of 2 or 14 Gyr, as appropriate.

The comparison between the ages for the three sets of isochrones is
shown in Figure~\ref{fig:plages}, where the three panels show the
various combinations.  The strong sensitivity of age to the modelling
assumptions is clear; and lest one might conclude that the better
agreement between Yale-Yonsei and Dartmouth is indicative of
superiority, we recall that these two formalisms originate from a
common source.  Another estimate of the age uncertainties may be
obtained by comparing the present values with those of
\citet{lemasle12}, who have provided ages of Carina red giants based
on the isochrones of \citet{pietrinferni04, pietrinferni06}.  Here, by
way of example, we compare our ages obtained using the Dartmouth
isochrones with the results of \citet{lemasle12}, and investigate the
mean differences in age, $\Delta$Age (in the sense Age$_{\rm
  This~work}$ -- Age$_{\rm Lemasle}$), for stars in common between the
two investigations with neither involving extrapolation in the age
determination.  We find a mean difference
$\langle$$\Delta$Age$\rangle$ = --0.5 $\pm$ 0.6 and an age dispersion
of $\sigma$($\Delta$Age) = 3.3 $\pm$ 0.4 Gyr.  This dispersion is of
the same order as might have been expected from the above discussion.

In Figure~\ref{fig:agehist}, the left panel presents the generalized
histograms of age for the three isochrone sets (obtained using a
Gaussian kernel with $\sigma$ = 3.0~Gyr). The best that may be said of
this diagram is that two main groupings are evident: one which is old,
and the other young and in the range $\sim2 - 7$ Gyr\footnote{We refer
  the reader to a new and independent approach to the ages of the
  Carina red giants suggested by \citet{monelli14} and
  \citet{fabrizio15}, based on the $C_{U,B,I}$ index to separate
  giants from the old and intermediate-age sub-populations discussed
  here.}.  For comparison, in the right panel of the figure, we
present a generalized histogram (obtained with the same Gaussian
kernel) based on the variable {\alphafe} population model in Paper I,
where we used synthetic CMD techniques that led to four components for
Carina, described in the legend in the figure in terms of the
[Fe/H]/{\alphafe}/Age parameters of the populations.  The four
components are distinguished in the diagram.  One would have to say
that the agreement between the two panels is at best indicative,
driven (as noted above) by the large errors associated with the
determination of ages based on the positions of red giants in the CMD.

Finally, Figure~\ref{fig:amr} presents the age-metallicity
relationship (AMR) for Carina.  On the left, Dartmouth
ages\footnote{We note that while we have chosen to use the Dartmouth
  ages in the presentation, the same essential behavior is seen when
  we use the Victoria-Regina and Yale-Yonsei isochrones.} are plotted
as a function of [Fe/H], where the size of the symbols increase as
their [Ca/Fe] values decrease.  The first thing to note about this
panel is that for abundances [Fe/H] $<$ --1.8, all stars appear to
have ages of 14 Gyr.  This is of no physical significance, since, as
stated above, when upward extrapolation in the age was required, we
set an upper limit of 14 Gyr.  That said, in this panel one sees
evidence for Type Ia enrichment (decreasing [Ca/Fe]) not only at all
[Fe/H] as discussed in Section~\ref{sec:sne} but also at all ages. The
right panel presents the relationship for the four populations
determined in our synthetic CMD analysis for the variable {\alphafe}
case presented in Paper I, where the bars represent the ranges in age
and metallicity of the populations.  The form of Carina's AMR is of
considerable interest.  Both determinations in Figure~\ref{fig:amr}
show that the oldest population, including Carina's metal-poor
pre-enrichment tail, brought Carina's metallicity up to [Fe/H] $\sim$
--1.5 dex, and that gas must have been retained for the following star
formation, given that the younger populations join smoothly in
metallicity (and, importantly, element ratios) to the older ones.  It
follows that the ensuing two thirds of star formation managed to
regulate itself against gas loss in such a way that very little net
extra chemical enrichment occurred. We conjecture that star formation
was self-limiting against gas winds over almost the whole of Carina's
star forming duration.

\section{COMPARISON BETWEEN OBSERVED AND SYNTHETIC-CMD METALLICITY AND [$\alpha$/FE] DISTRIBUTION FUNCTIONS} \label{sec:distributions}

In Paper I and the present work we have determined Carina's MDF and
{\alphafe} distribution function using two very different and
basically independent methods.  Paper I presented an analysis of the
CMD in terms of synthetic CMDs based on the fitting of isochrones to
the observations, while in the present work this has been achieved by
the analysis of high-resolution, moderate $S/N$ spectra of its red
giants.  A comparison of the results is presented in
Figure~\ref{fig:cardist}, where the thick (red) and thin (black) lines
refer to the present spectroscopic results and the CMD analysis of
Paper I, respectively.  The MDFs appear on the left and the {\alphafe}
distribution function on the right, while the upper and lower panels
present results pertaining to the synthetic CMD analysis when
{\alphafe} is allowed to vary (top panels) or held constant (bottom).
Finally, the legends contain the defining physical parameters
[Fe/H]/{\alphafe}/Age of the four populations that were determined in
the CMD analysis of Paper I.  We recall here that in Paper I we
referred to these four groups in terms of increasing age, as the
``first'', ``second'', ``third'', and ``fourth'' populations.  We
shall adopt this nomenclature in what follows.  In each panel the
areas under the two distributions have been normalized to be equal.

While the fits are not outstanding, they show basic agreement between
the two methods. Consider first the {\alphafe} distribution
function. As might have been expected, given the observed variations
in the abundances of the $\alpha$-elements, the variable-$\alpha$ case
(upper right panel) provides a better and reasonable fit, except for
the asymmetry of the observations to lower values of {\alphafe}, which
is not included in the models.  As discussed in
Section~\ref{sec:incommix} we regard this as evidence of incomplete
mixing of the ejecta from Type Ia SNe with the ambient medium during
the formation of Carina's populations.

The fits of the MDFs are found wanting in three areas: the
observations show a small offset to lower abundance, an asymmetry
favoring lower abundances, and a somewhat broader distribution.  Given
the possibility that the RGB sample in incomplete, and the
likelihood that there are small differences between the abundance
scales of the stellar atmosphere and stellar evolution formalisms,
among other possibilities, the agreement could have been worse.

\subsection{The Need for a More Complicated ``First'' Population}

In Section 5 of Paper I, we discussed the shortcoming of our first
population in that while we described it as monomodal with mean
metallicity [Fe/H] = --1.85 and with very little spread ($\Delta$[Fe/H]
$\sim$~0.1), there are in the literature high resolution spectroscopic
abundance analyses of some five Carina red giants with abundances in
the range --2.9 $<$ [Fe/H] $\leq$ --2.5 The present work essentially
confirms this result: while in our Table~\ref{tab:relabs_aver} there
is only one star with [Fe/H] $<$ --2.5, there are eight with [Fe/H]
$<$ --2.0.  Of these, three are also present in the literature sample
with --2.9 $<$ [Fe/H] $\leq$ --2.5.

In Paper I we determined a mean age and metallicity for each of the
four populations, with which we associated spreads in age and
metallicity.  As discussed there, however, given available
information, these spreads were not well constrained.  There are a
number of possible explanations for the absence of stars with [Fe/H]
$<$ --2.0 in our model.  One is that our first population is
multimodal, and that in Paper I we essentially identified only a major
sub-component of the oldest population, while there exists a more
metal-poor, presumably older and minor sub-component, as well
\footnote{A referee made the following related comment on the large
  spread in [Fe/H] in our Figure 26. ``Data plotted in the left panel
  indicate that the old stellar subpopulation experienced, at fixed
  age, a metal enrichment of the order of one dex.  Is this
  metal-enrichment supported by the spread in magnitude of the
  subgiant branch of the old population?''.  In Carina's ({\mv},
  {\bio}) CMD, the SGB of this ``first'' population has a magnitude
  spread at a given color of $\Delta${\mv}~$\sim$~0.40 mag.  We first
  compare this with the isochrones of Dotter et al. (2008) for two
  models having $\Delta$[Fe/H] = 0.8, and age difference $\Delta$Age =
  0.0 -- [Fe/H]/[{\alphafe}/Age = --2.4/+0.2/13.0 and --1.6/+0.2/13.0.
    In the region of the observed SGB these isochrones are parallel
    curves separated by $\Delta${Mv} $\sim$~0.40 mag that fit within
    the observations.  More to the conjecture that we make here, if
    one admits an age difference $\Delta$Age = --2.0 Gyr, (i.e. the
    more metal-poor stars are older) and consider isochrones
    [Fe/H]/{\alphafe}/Age = --2.4/+0.2/14.0 and --1.6/+0.2/12.0, the
    corresponding sequences, while falling well within the region of
    the observed SGB, are closer together, with
    $\Delta$$M_{V}$~$\sim$~0.1 -- 0.3 mag (diverging with increasing
    color).}.  While this is outside the scope of the present work and
  we shall not consider it further, it is tempting to conjecture that
  this minor component contains the first stars to form in Carina.
  Other, more detailed possibilities include simple inhomogeneous
  mixing of the early star forming gas (e.g., \citealp{revaz12}),
  and/or active accretion of lower mass systems in the earliest stages
  of Carina's formation, such as described by the models of
  \citet{wise12}.

\section{A SEARCH FOR THE [NA/FE] VS. [O/FE] ANTI-CORRELATION}\label{sec:nao}

It has become clear in the past three decades that all Galactic
globular clusters exhibit strong and anti-correlated abundance
variations within the light elements C, N, O, Na, Mg, and Al
(\citealp{gratton04}, and references therein).  Indeed, the effect is
so ubiquitous and strong that \citet{carretta10a} have suggested,
based on their study of the Na-O -- anti-correlation in some 17
clusters, ``a new definition of {\it{bona fide}} globular clusters [,] as
the stellar aggregates showing the Na-O anti-correlation''.  No
consensus, however, exists as to the cause of these anti-correlations.
The principal explanations, mostly with an emphasis on two
intra-cluster populations of roughly commensurate size, include (i)
chemical enrichment of the later generation by the ejecta of AGB stars
of an earlier one (e.g., \citealp[and references therein]{dercole10}),
(ii) enrichment of a later generation by the ejecta of earlier
rotating massive stars (\citealp[and references therein]{decressin07}),
(iii) a scenario in which ``low-mass pre-main-sequence stars accrete
enriched material released from interacting massive binary and rapidly
rotating stars on to their circumstellar discs'' \citep{bastian13},
and (iv) supermassive black holes early in the cluster lifetime
\citep{denissenkov14}.  In stark contrast, only 3\% of the Galactic
halo stellar population exhibits these anti-correlated abundance
patterns \citep{martell11}.

What then do we know about the existence of these abundance
anti-correlations in the dSph systems? To the best of our knowledge,
the most comprehensive studies relevant to this are those of
\citet{shetrone03}, \citet{mcwilliam13}, and \citet{fabrizio15}.
\citet{shetrone03} found only subsolar values of [Na/Fe] in Sculptor
(5/5 stars), Fornax (4/5 stars), and Carina (5/5 stars), more like
that of the Galactic halo than its globular clusters; for the
Sagittarius dSph \citet{mcwilliam13} reported ``no detectable
signature of pollution by proton-burning products''; and
\citet{fabrizio15} presented results for [O/Fe] in nine stars in
Carina, all but one of which have [O/Fe] $>$ +0.4 (albeit all of these
having [Na/Fe] in the range --0.1 to +0.7). We confirm their
results\footnote{We are unable to comment on the existence or
  otherwise of an Al-O anti-correlation in Carina as is found in
  globular clusters, since Al was detected in only one star for which
  we have spectra.}.  Figure~\ref{fig:onaal} presents [Na/Fe]
vs. [O/Fe] for the stars in Carina for which we have spectra
(Table~\ref{tab:obsdat}), together with results for the Galaxy's
globular clusters M3, M4, M5, M13 from \citet{kraft_m3_92,
  kraft_m13_97} and \citet{ivans_m4_99, ivans_m5_01})\footnote{These
  clusters have [Fe/H] in the range --1.5 to --1.2, within which all
  but one of the Carina stars in Figure~\ref{fig:onaal} having both Na
  and O abundances lie (allowing for an error of measurement 0.15 dex
  for the [Fe/H] values).}. In the figure, one sees that the Carina
stars are confined to high [O/Fe] and low [Na/Fe] similar to what is
expected from the results of \citet {martell11} for Galactic halo
stars, while those in the clusters occupy not only that region, but
also stretch to low [O/Fe], high [Na/Fe] values, driven by the
operation of nuclear (p,$\gamma$)-reactions (a solution first
described by \citet{denisenkov90} to explain the Na-O
anti-correlation), peculiar to the clusters.

What drives this basic difference between the globular clusters and
Carina? The obvious physical candidates are that the dSph is dark
matter dominated, is of considerably lower central density
($\sim$0.3~M$_{\odot}$/pc$^{3}$ \citep{walker09b}, compared with
$\sim$10000M$_{\odot}$/pc$^{3}$ for a typical cluster
\citep{pryor93}), and had star formation episodes lasting considerably
longer (say 2 -- 3 Gyr) than occurred in the globular clusters ($<$1
Gyr).  There are obvious differences in formation histories -- Carina
has some 3.4~$\times10^{6}$ M$_{\odot}$ of dark matter within its
half-light radius \citep{walker09b}, while the clusters are currently
dark matter free.  Finally, Carina formed well away from other stellar
systems, while the Galactic globular clusters initially evolved in an
environment more challenging for the survival of their gaseous
components.  It will be interesting to see which, if any, of these
comparisons lies at the bottom of the abundance pattern differences
between the dSph and globular clusters.

\section{THE EXTREMELY LITHIUM ENHANCED STAR CC11560}\label{sec:li}

Among the stars in Table~\ref{tab:obsdat}, for which we have
high-resolution spectra, CC11560 exhibits an exceptionally strong
6707\AA\ Li line. For all other stars in the table, this feature
cannot be detected.  Canonical stellar evolution models predict that
lithium is readily destroyed as stars ascend the red giant branch
\citep{iben67}.  There are, however, a small number of lithium-rich
red giant stars in globular clusters (\citealp{wallerstein82},
\citealp{brown89}, and \citealp{kraft99}) and dwarf spheroidal galaxies
\citep{kirby12}.  We refer the reader to Kirby et al. for a
comprehensive discussion of the origin of this effect.  Suffice it
here to say that the most likely source of the excess Li is the
so-called Cameron--Fowler mechanism \citep{cameron71} which was
proposed to occur during helium shell flashes and/or ``cool bottom
processing'' in which it is produced in the hydrogen burning shell
\citep{sackmann99}.  This Li is then destroyed during deep convective
mixing as the star moves further up the giant branch.

We measured the lithium abundance in CC11560 via spectrum synthesis of
the 6707\AA\ resonance line as well as from the 6103\AA\ subordinate
line (see Figure~\ref{fig:lithium}). We computed line profiles
adopting {\loggf} = 0.17 and 0.58 for the 6707\AA\ and 6103\AA\ lines
\citep{lindgard77}, respectively, assuming LTE, and applied NLTE
corrections following \citet{lind09} and obtain an average abundance
of A(Li)$_{\rm NLTE}$ = +3.36. To our knowledge, CC11560 is the first
Li-rich red giant found in the Carina dwarf galaxy.

\citet{kirby12} studied 15 Li-rich red giants in dwarf spheroidal
galaxies covering a range in magnitudes brighter than the RGB bump,
and reported that they represent $\sim$1\% of the RGB population in
that region of the CMD.  This is of the same order as the ratio of
1/32 for the RGB stars of which we have spectra.  In their sample, the
abundances ranged over +1.76 $<$ A(Li)$_{\rm NLTE}$ $<$ +3.85.
CC11560 lies within this range.  Finally, \citet{kirby12} reported
that there was no correlation between lithium enhancement and any
other parameter and concluded that these stars are likely experiencing
a brief and possibly universal phase of stellar evolution.  In
CC11560, also, no element other than Li is unusual with respect to the
values found for other program stars.  It thus appears that CC11560
belongs to the population of lithium-rich stars discussed by
\citet{kirby12}.  We refer the reader to their work for a more
detailed discussion of the phenomenon.

\section{SUMMARY}\label{sec:summary}

\begin{itemize}

\item

We have presented high-resolution, moderate $S/N$ observations of 32
RGB stars in the Carina dSph galaxy, which have been analyzed using
1D/LTE model atmosphere techniques to produce chemical abundances for
$\sim 20$ elements.  The analysis has been extended to Carina red
giants for which observational data of a similar quality are available
in the literature.  Abundances are available for a total sample of 63
independent stars within this system.

\item

The total sample, which originates from four sources, is incomplete in
two ways. There are relatively fewer fainter stars than in a complete
sample, and a small dearth of Carina members blueward of its
well-defined RGB (evident from comparison with complete
radial-velocity samples, as pointed out in Paper I).

\item 

Bearing in mind the above caveats, the sample has
$\langle$[Fe/H]$\rangle$ = --1.59, with dispersion $\sigma$[Fe/H] =
0.33 dex, and an {\alphafe} distribution with mean value
$\langle${\alphafe}$\rangle$ = 0.07 and dispersion $\sigma${\alphafe}
= 0.13~dex.

\item

Consideration of the available $\alpha$-elements shows little evidence
for significant differences in abundances of the individual
$\alpha$-elements relative to that of iron.  For example,
$\langle$[Mg/Fe]$\rangle$ and $\langle$[Ca/Fe]$\rangle$ differ by less
that 0.1 dex.  The mean value of {\alphafe} (the average of Mg, Ca,
and Ti) for Carina is smaller by $\sim 0.25$ dex than that of the
Galactic halo.

\item 

Calcium, which has the most accurately determined abundance of the
$\alpha$-elements, shows an asymmetric distribution in [Ca/Fe] towards
smaller values at all [Fe/H], most significantly in the range --2.0 $<$
[Fe/H] $<$ --1.0, where the density of points is highest -- suggestive
of incomplete mixing of the ejecta of Type Ia SNe and the ambient
medium of each of Carina's generations.

\item

Comparison of the abundance distributions of the individual stars with
the abundance predictions of Type Ia SNe models of \citet{iwamoto99},
together with the formalism of \citet{venn12}, show evidence for
large Type Ia enrichment within the material that formed some 10\% of
the stars in our sample, confirming the results of \citet{venn12}.

\item

Very approximate ages, internally accurate to only 3 -- 4 Gyr, have been
determined by comparing the positions of the stars in the CMD with the
isochrones of \citet{demarque04}, \citet{vandenberg06}, and
\citet{dotter08}.  A comparison of the results suggests that the
external errors from the use of different isochrones is also
considerable.

\item

That said, the age-metallicity relationships from the present work and
from the CMD analysis presented in Paper I are in general agreement,
with both presenting a monotonic progression from (Age, [Fe/H]) $\sim$
(13 Gyr. --2.0) to $\sim$ (1 -- 2 Gyr, --1.0).  There is, however, an
observed excess of stars below [Fe/H] $\sim -2.0$ in the
high-resolution spectroscopic data.  (See also Paper I (Section 5).)
It will be important to understand whether this is due to a selection
effect favoring metal-poor stars, to spectroscopic measurement error,
or perhaps to the existence of stars with [Fe/H] $\sim$ --2.0 that
were formed before the first population identified in the synthetic
CMD analysis, which has $\langle$[Fe/H]$\rangle$ = --1.85 $\pm$ 0.05.
We conjecture in the present work that our ``first'' (and oldest)
population in Paper I is multimodal, and that we essentially
identified only a major sub-component of the oldest population, while
there exists a more metal-poor, presumably older and minor
sub-component as well, perhaps resulting from simple inhomogeneous
mixing of the early star forming gas and/or active accretion of lower
mass systems in the earliest stages of Carina's formation.

\item

We demonstrate that the Na-O anti-correlation found in all of the
Galactic globular clusters is not present in Carina, confirming the
results of \citet{shetrone03}, who first demonstrated that the
abundance of Na is sub-solar in Carina (and in three other dSph
systems), \citet{mcwilliam13}, and \citet{fabrizio15}.

\item

We report the serendipitous discovery of an extremely lithium
enhanced red giant, CC11560, which has A(Li)$_{\rm NLTE}$ = +3.36, a
characteristic exhibited by only $\sim$ 1\% of the red giants in the
Galactic globular clusters and its dwarf spheroidal satellites.

\end{itemize}

  \appendix{APPENDIX: CROSS IDENTIFICATION AND AVERAGED ABUNDANCES\\ FOR THE 63 INDEPENDENT CARINA RED GIANTS}\label{app:crossid}

Table~\ref{tab:appendix} presents cross-identifications and
coordinates of the 63 independent stars analyzed in the present work.
The names in Columns 1 -- 4 are those used here and in the works of
\citet{venn12}, \citet{shetrone03}, and \citet{lemasle12},
respectively.

Finally, Table~\ref{tab:relabs_aver} presents the average relative
abundances, together with their errors, for these 63 stars in
Tables~\ref{tab:relabs_gil}, \ref{tab:relabs_kvs}, and
\ref{tab:relabs_lem}, which were introduced in
Section~\ref{sec:relabund}.  In cases where a star appears in more
than one table, the abundances were averaged with weighting by the
inverse squares of their errors in the relevant tables.  In
Table~\ref{tab:relabs_aver} there are 63 two-line blocks, in which the
first and second lines contain the average abundances and their
errors, respectively.  Stellar identifications are given in the first
column, and in cases where multiple observations are involved, the two
starnames are given in the first column of the two lines of the block.

\acknowledgements

We are extremely grateful to P.~B. Stetson, M.~J. Irwin, and
M. Gullieuszik for their generosity in providing photometry for the
objects investigated in the present work.  Studies at RSAA, ANU, of
the Galaxy's most metal-poor stars and its dwarf galaxy satellite
systems by J,E.N. and D.Y. are supported by Australian Research
Council grants DP0663562, DP0984924, DP120100475, DP150100862, and
FT140100554.  K.A.V. acknowledges support from the Canadian NSERC
Discovery Grants program.  This work was partly supported by the
European Union FP7 program through ERC grant number 320360.

\noindent{\it Facilities:} {VLT:Kueyen(UVES)}

\clearpage

\clearpage

\clearpage

\clearpage

\clearpage

\begin{figure}[htbp]
\begin{center}

\includegraphics[width=10.0cm,angle=0]{fig5.eps}
  \caption{\label{fig:fe} \scriptsize Comparison of the abundances,
    [Fe/H], of the present work with those of (a) \citet{shetrone03},
    (b) \citet{koch06} (using the \citet{carretta97} calibration), (c)
    \citet{venn12}, (d) \citet{lemasle12}, and (e) \citet{fabrizio12}.
    The black line is the 1-1 relationship.}
\end{center}
\end{figure}

\clearpage

\begin{figure}[htbp]
\begin{center}

\includegraphics[width=10.0cm,angle=0]{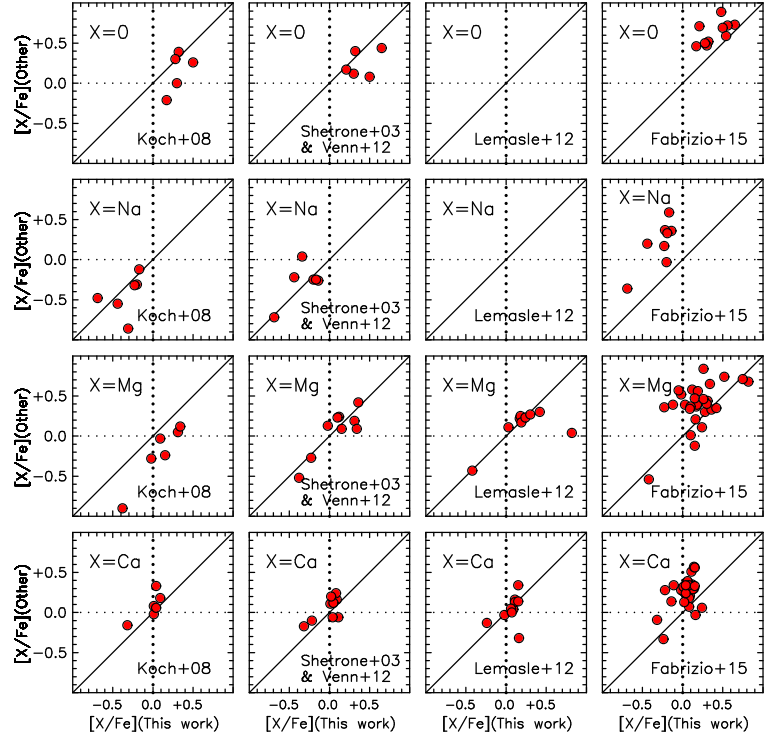}
  \caption{\label{fig:xfe} \scriptsize {Comparison of
        the relative abundances (from top to bottom) [O/Fe], [Na/Fe],
        [Mg/Fe], and [Ca/Fe] of the present work with those (from left
        to right) of \citet{koch08a}, of \citet{shetrone03} and
        \citet{venn12}, of \citet{lemasle12}, and of
        \citet{fabrizio15}.  The line is the 1-1 relationship.  See
        text for discussion.}}
\end{center}
\end{figure}

\clearpage

\begin{figure}[htbp]
\begin{center}

\includegraphics[width=10.0cm,angle=0]{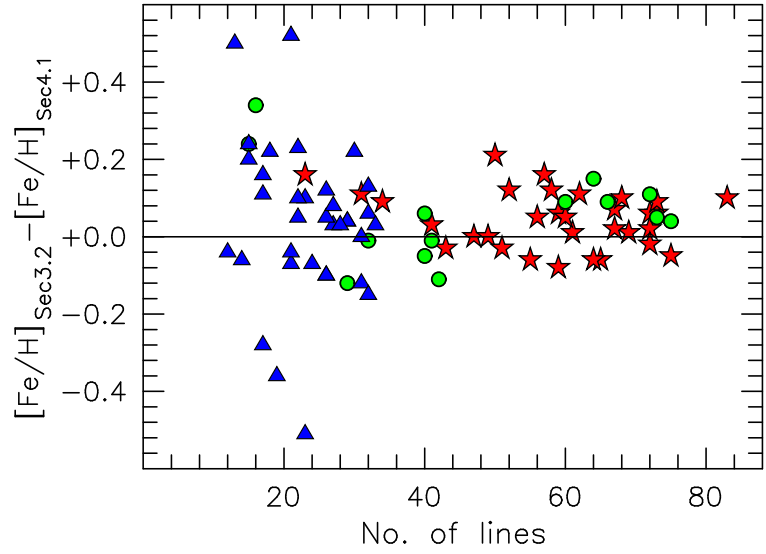}
  \caption{\label{fig:dfe_micro} \scriptsize The difference between
    [Fe/H] obtained using microturbulence as described in
    Sections~\ref{sec:abund} and ~\ref{sec:litab}, as a function of
    the number of Fe I lines, using data from the present work (star
    symbols); from \citet{shetrone03} and \citet{venn12} (circles);
    and from \citet{lemasle12} (triangles).}
\end{center}
\end{figure}

\clearpage

\begin{figure}[htbp!]
\begin{center}

\includegraphics[width=5.0cm,angle=0]{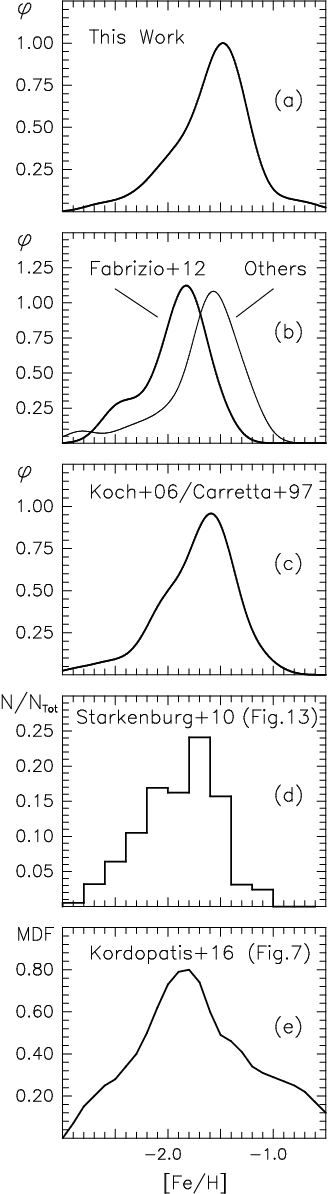}

  \caption{\label{fig:mdf} \scriptsize {The generalized
      [Fe/H] metallicity distribution function of Carina based on
      high-resolution spectroscopy for (a) the 63 RGB sample on the
      present work and (b) from \citet{fabrizio12} (thick line) and
      the collective data of \citet{shetrone03}, \citet{venn12}, and
      \citet{lemasle12} (thin line); (c) contains results for the RGB
      sample of \citet{koch06} obtained from their analysis of Ca~II
      IR triplet data, adopting the [Fe/H] calibration of
      \citet{carretta97}.  For these three panels a Gaussian kernel of
      0.15 dex, appropriate to the accuracy of the abundance analyses,
      has been adopted. Panel (d) contains the histogram of the Carina
      MDF from \citet[their Figure 13]{starkenburg10} obtained from
      analysis of the Ca~II IR triplet data of \citet{koch06}; and
      panel (e) presents the MDF of \citet[their Figure
        7]{kordopatis16}.}}

\end{center}
\end{figure}

\clearpage

\begin{figure}[htbp!]
\begin{center}

\includegraphics[width=11.0cm,angle=-90]{fig9.eps}
  \caption{\label{fig:relab1} \scriptsize [O/Fe], [Na/Fe] and [Mg/Fe]
    vs. [Fe/H] (left to right). From top to bottom the first three
    rows contain results for the present sample; from our analysis of
    data from \citet{shetrone03} and \citet{venn12}; and from our
    analysis of the data of \citet{lemasle12}, respectively.  The
    bottom two rows contain the final weighted average values of the
    data in the top three rows; in the penultimate row total error
    bars are included; and in the bottom row the size of the symbols
    decreases linearly as the error increases.  The red lines present
    average values for Milky Way halo stars following \citet{venn12}.}

\end{center}
\end{figure}

\clearpage
  
\begin{figure}[htbp!]
\begin{center}

\includegraphics[width=11.0cm,angle=-90]{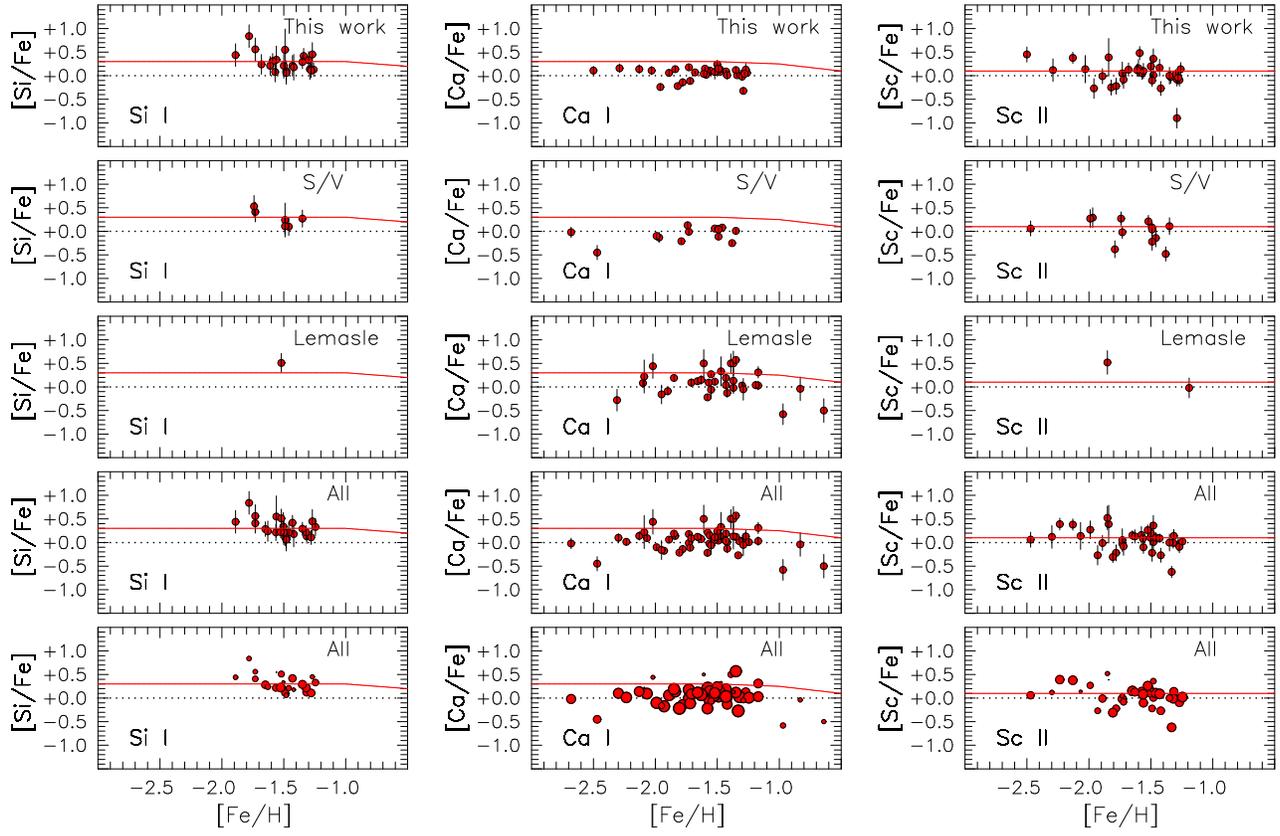}
  \caption{\label{fig:relab2}\scriptsize [Si/Fe], [Ca/Fe] and [Sc/Fe]
    vs. [Fe/H].  The format is as described in
    Figure~\ref{fig:relab1}.}
\end{center}
\end{figure}

\clearpage

\begin{figure}[htbp!]
\begin{center}

\includegraphics[width=11.0cm,angle=-90]{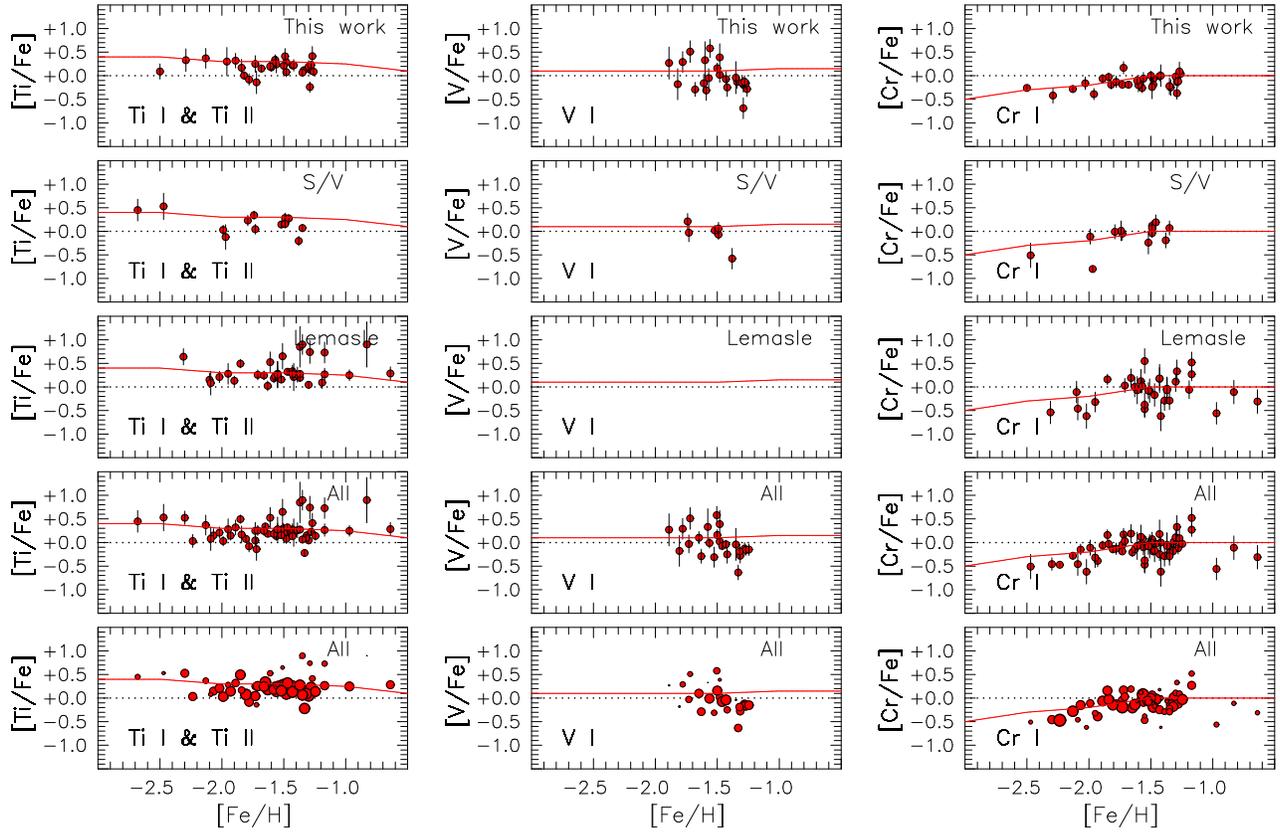}
  \caption{\label{fig:relab3}\scriptsize [Ti/Fe], [V/Fe] and [Cr/Fe]
    vs. [Fe/H].  The format is as described in
    Figure~\ref{fig:relab1}. }
\end{center}
\end{figure}

\clearpage

\begin{figure}[htbp!]
\begin{center}

\includegraphics[width=11.0cm,angle=-90]{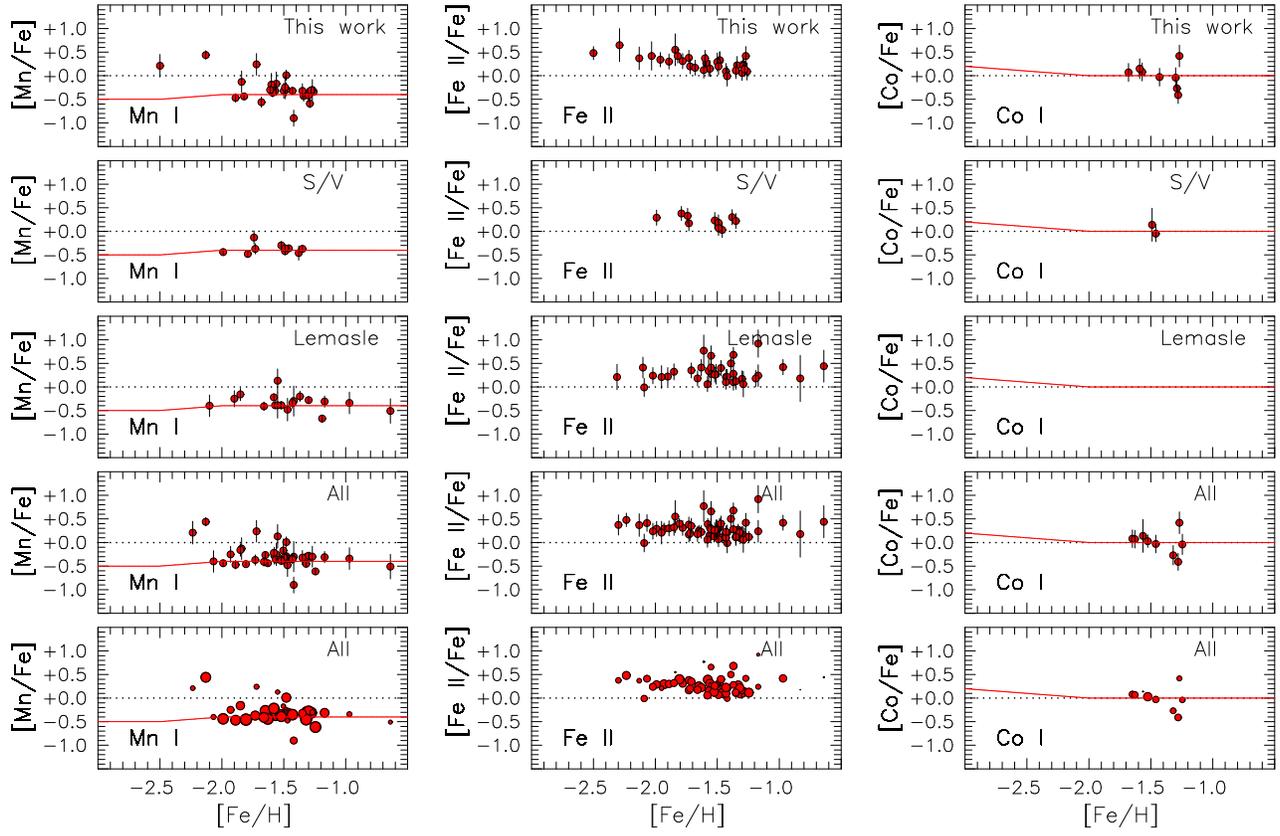}
  \caption{\label{fig:relab4}\scriptsize [Mn/Fe], [Fe~II/Fe] and
    [Co/Fe] vs. [Fe/H].  The format is as described in
    Figure~\ref{fig:relab1}.}
\end{center}
\end{figure}

\clearpage
\begin{figure}[htbp!]
\begin{center}

\includegraphics[width=10.0cm,angle=-90]{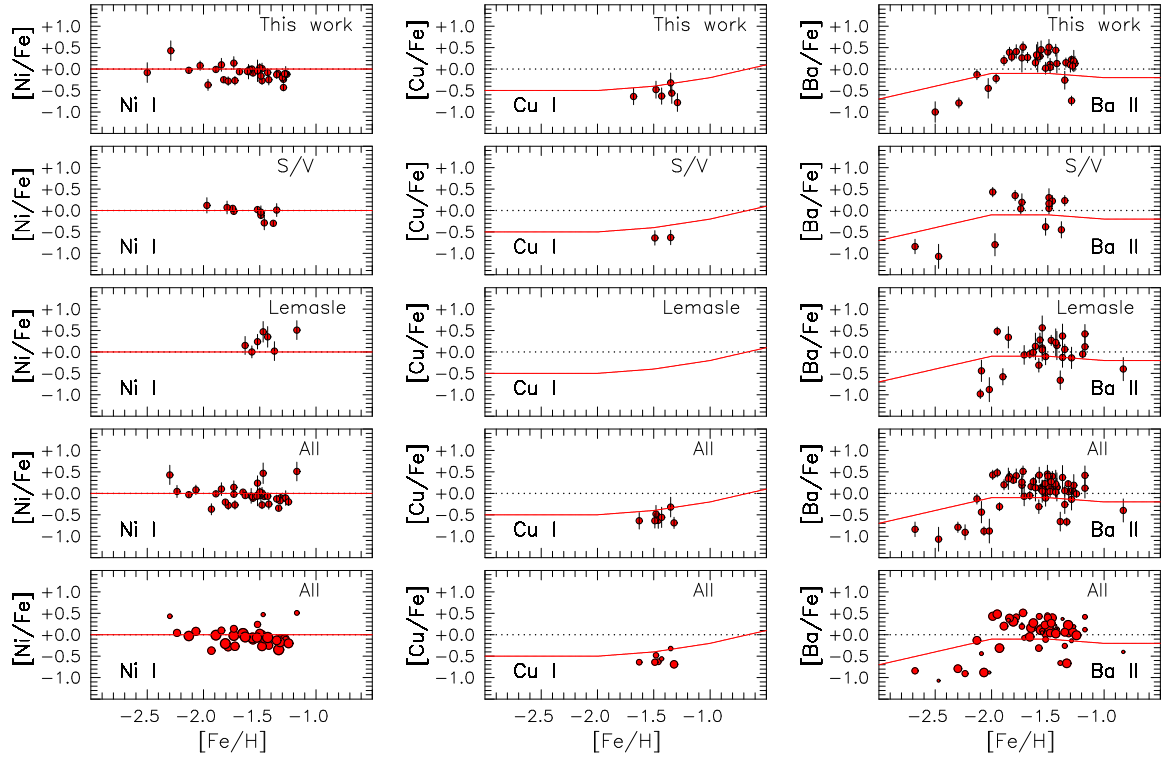}
  \caption{\label{fig:relab5}\scriptsize [Ni/Fe], [Cu/Fe] and [Ba/Fe] vs. [Fe/H].
    The format is as described in Figure~\ref{fig:relab1}. }
\end{center}
\end{figure}

\clearpage
\begin{figure}[htbp!]
\begin{center}

\includegraphics[width=10.0cm,angle=-90]{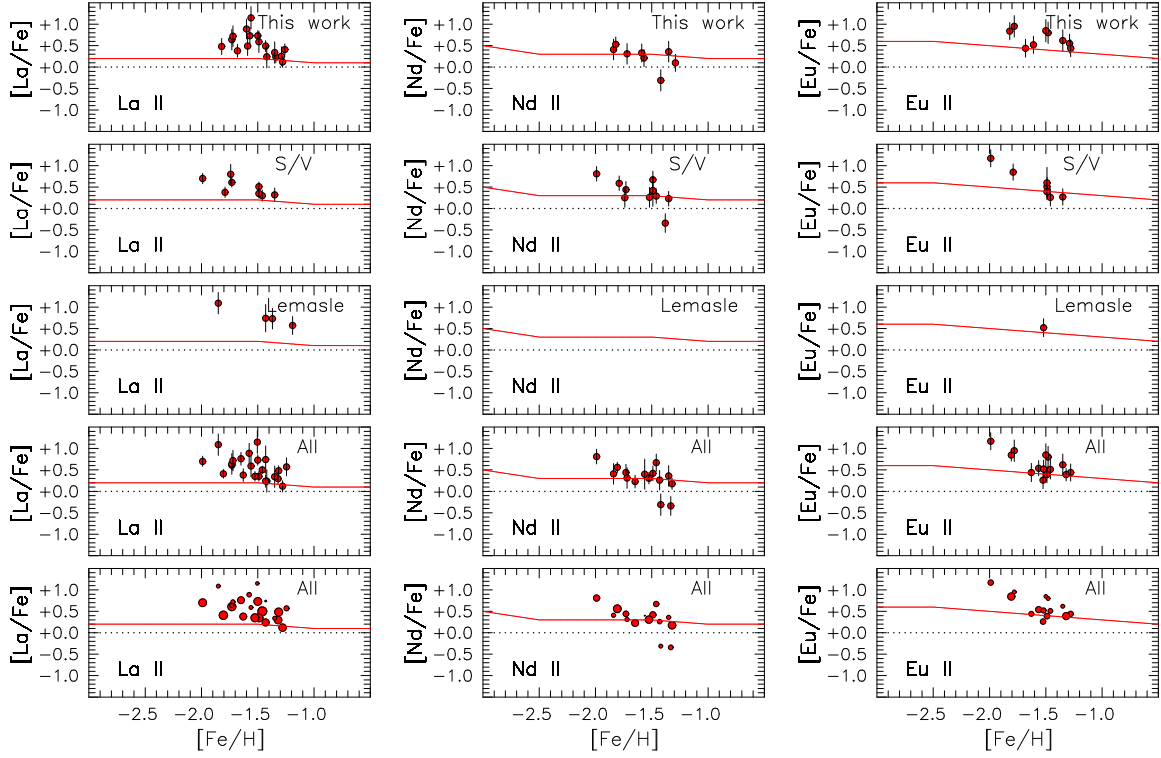}
  \caption{\label{fig:relab6}\scriptsize [La/Fe], [Nd/Fe] and [Eu/Fe] vs. [Fe/H].
    The format is as described in Figure~\ref{fig:relab1}.}
\end{center}
\end{figure}

\clearpage
\begin{figure}[htbp!]
\begin{center}

\includegraphics[width=10.0cm,angle=0]{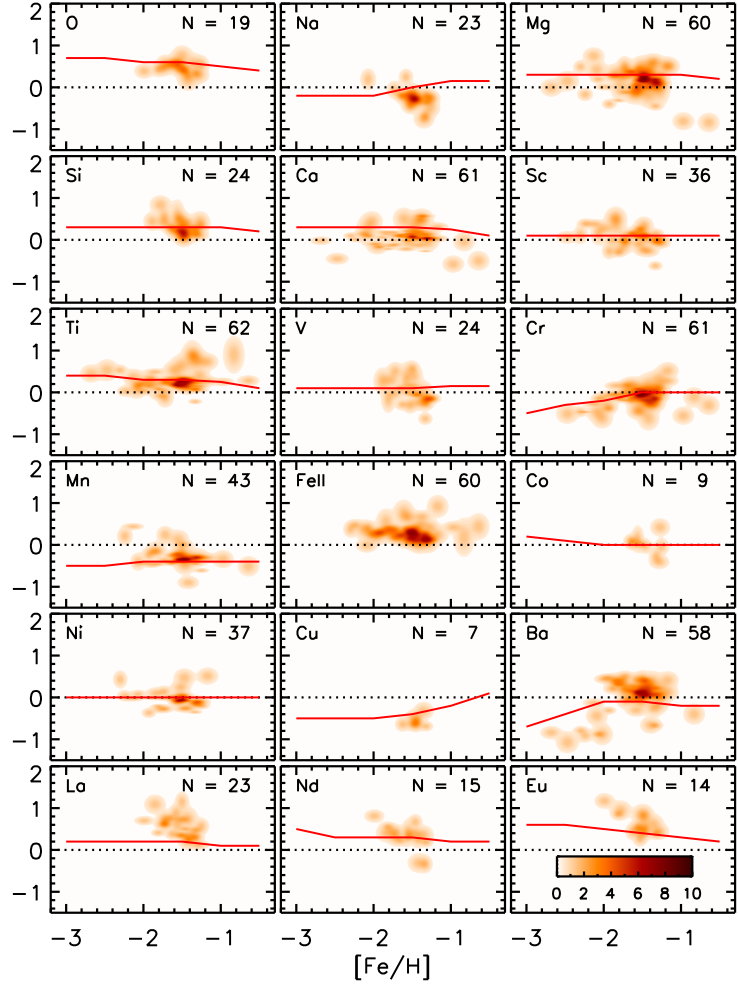}
\end{center}
  \caption{\label{fig:contours}\scriptsize Contours of relative
    abundance -- [X/Fe] vs. [Fe/H].  Here each star is represented by
    a double Gaussian having kernels equal to the total observational
    errors. The red lines present average values for Milky Way halo
    stars following \citet{venn12}.}

\end{figure}

\clearpage

\begin{figure}[htbp]
\begin{center}

\includegraphics[width=9.0cm,angle=-90]{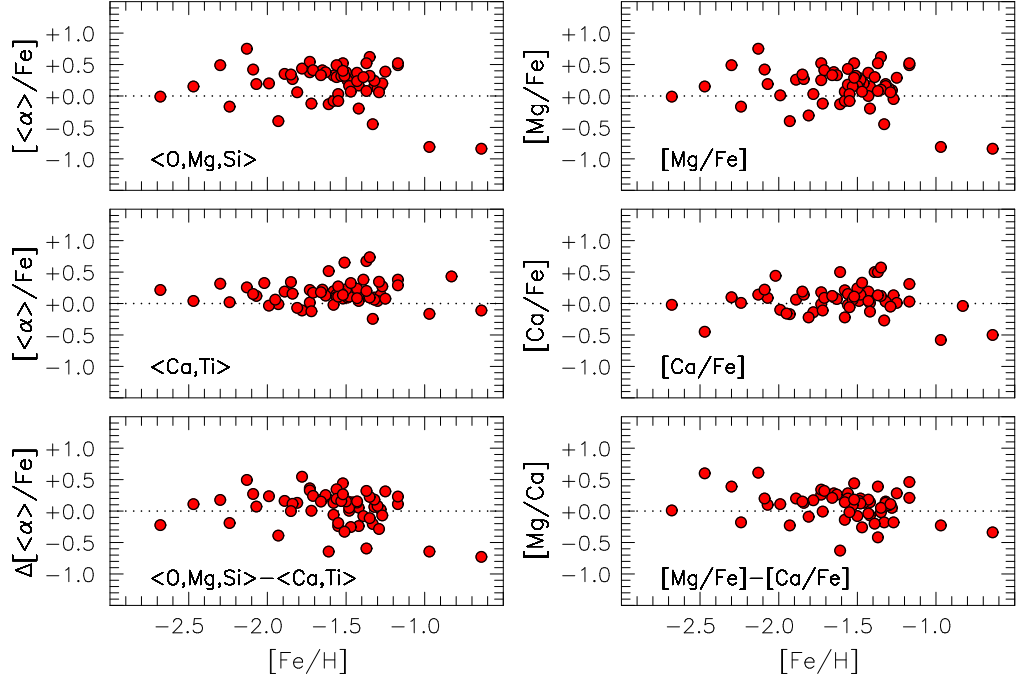}

  \caption{\label{fig:alphas}\scriptsize [$\alpha$-related indices/Fe]
    vs. [Fe/H]. Left: In the upper panel
    $\langle$[$\alpha$/Fe]$\rangle$ is the average of the light
    $\alpha$s O, Mg, and Si, in the middle it is the average of the
    heavy $\alpha$s Ca and Ti, and the bottom panel presents the
    difference between the mean light and heavy $\alpha$s presented in
    the upper and middle panels.  Right: Same format as in the left
    panels, for the light $\alpha$-element, Mg, and the heavy
    $\alpha$-element, Ca.  See text for discussion.}

\end{center}
\end{figure}

\clearpage

\begin{figure}[htbp!]
\begin{center}

\includegraphics[width=15.5cm,angle=0]{fig17.eps}
 
 \caption{\label{fig:alpha_v3}\small A comparison of {\alphafe}
   abundances for (left panels) the Galactic halo (data from
   \citealp{fulbright00}, \citealp{barklem05}, \citealp{preston00},
   and \citealp{yong13a}); (middle panels) Carina (this work), where
   the size of the symbols decreases linearly as the error increases;
   and (right panels) {\wcen} \citep{norris95}.}

\end{center}
\end{figure}

\clearpage
\begin{figure}[htbp!]
\begin{center}

\includegraphics[width=9.0cm,angle=0]{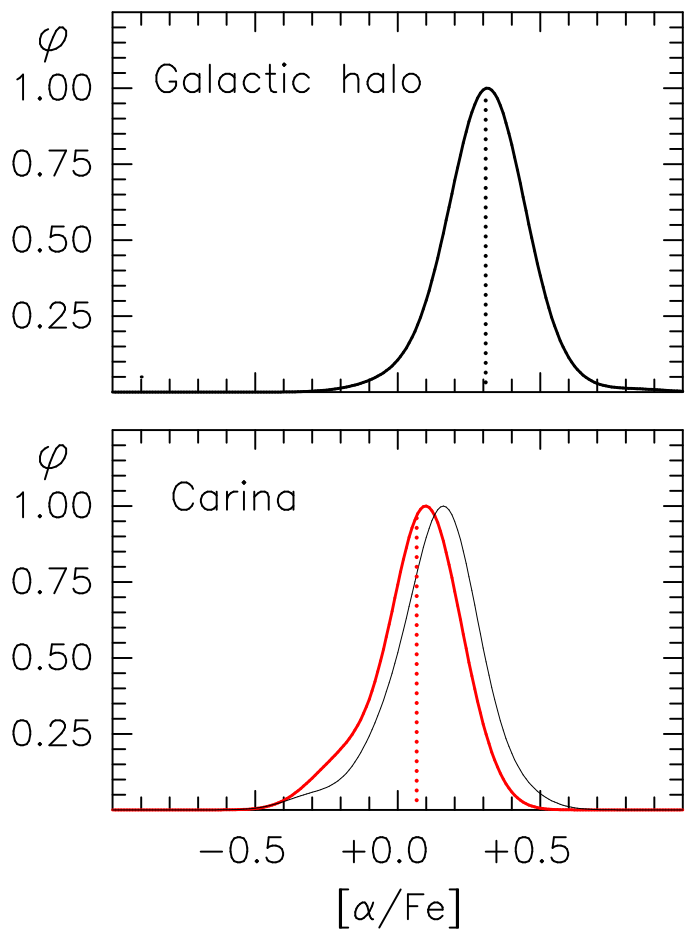}

  \caption{\label{fig:obs_distrib} \small The Carina {\alphafe}
    distribution function, based on stars for which abundances are
    available for all of Mg, Ca, and Ti.  The thick red line
    represents average abundances that are the weighted means using
    weights inversely proportional to the inverse square errors, while
    the thin line results when equal weights are assumed.  In the
    upper panel the result is presented for the Galactic halo. Here
    the data for have been taken from the samples listed in the
    caption to Figure~\ref{fig:alpha_v3}.  A Gaussian kernel of 0.10
    dex has been adopted in preparing the diagram.  Note that the
    results for Carina's {\alphafe} is sensitive to the weighting
    procedure at only the $\sim$ 0.1 dex level, and also that there is
    an {\alphafe} $\sim$ 0.25 dex offset between the Carina (when
    error dependent weights are adopted) and Galactic halo
    samples.}

\end{center}
\end{figure}

\clearpage

\begin{figure}[htbp!]
\begin{center}
\includegraphics[width=15.0cm,angle=0]{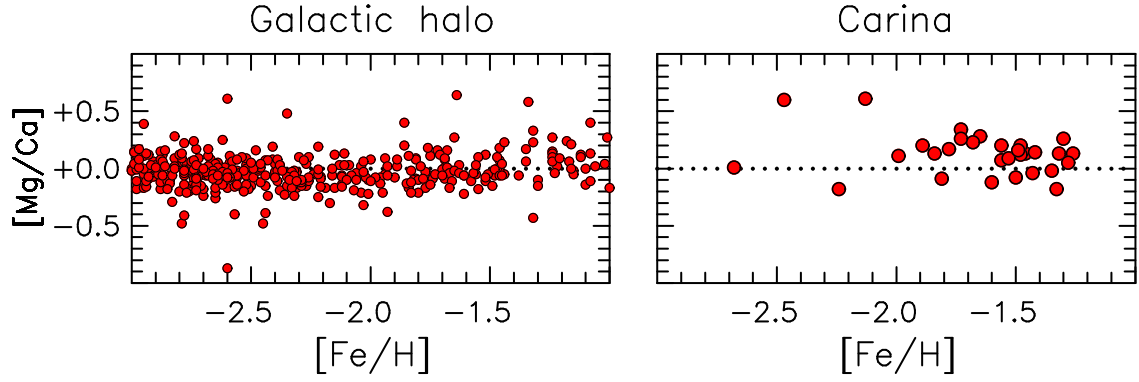}
 
 \caption{\label{fig:mgca30_v2}\small A comparison of the relative
   magnesium to calcium ratio, [Mg/Ca], as a function of [Fe/H] for
   (left panel) the Galactic halo (data from \citealp{fulbright00,
     barklem05, preston00, yong13a}) and (right panel) Carina.}

\end{center}
\end{figure}

\clearpage

\begin{figure}[htbp!]
\begin{center}

\includegraphics[width=10.0cm,angle=0]{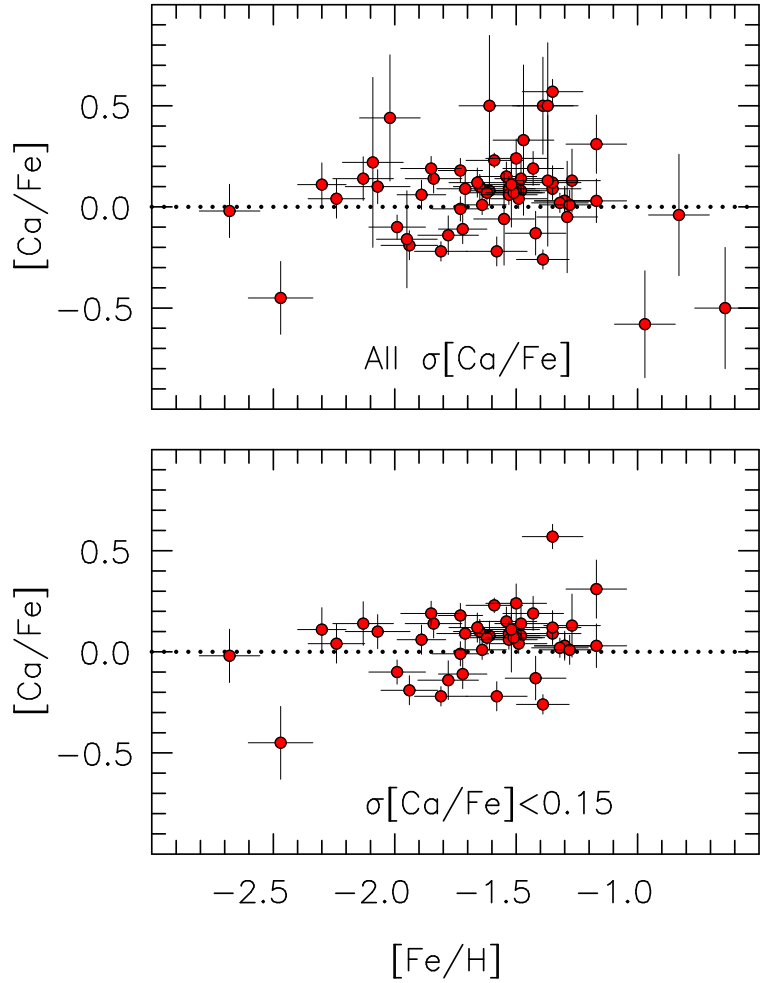}
 
 \caption{\label{fig:cafe}\small [Ca/Fe] vs [Fe/H] for the Carina RGB
   stars of the present investigation. In the upper panel all stars
   are plotted, while in the lower panel only those with
   $\sigma$[Ca/Fe] $<$ 0.15 are presented, permitting clearer insight
   into the [Ca/Fe] distribution.  See text for discussion.}

\end{center}
\end{figure}

\clearpage

\begin{figure}[htbp!]
\begin{center}

\includegraphics[width=10.0cm,angle=-90]{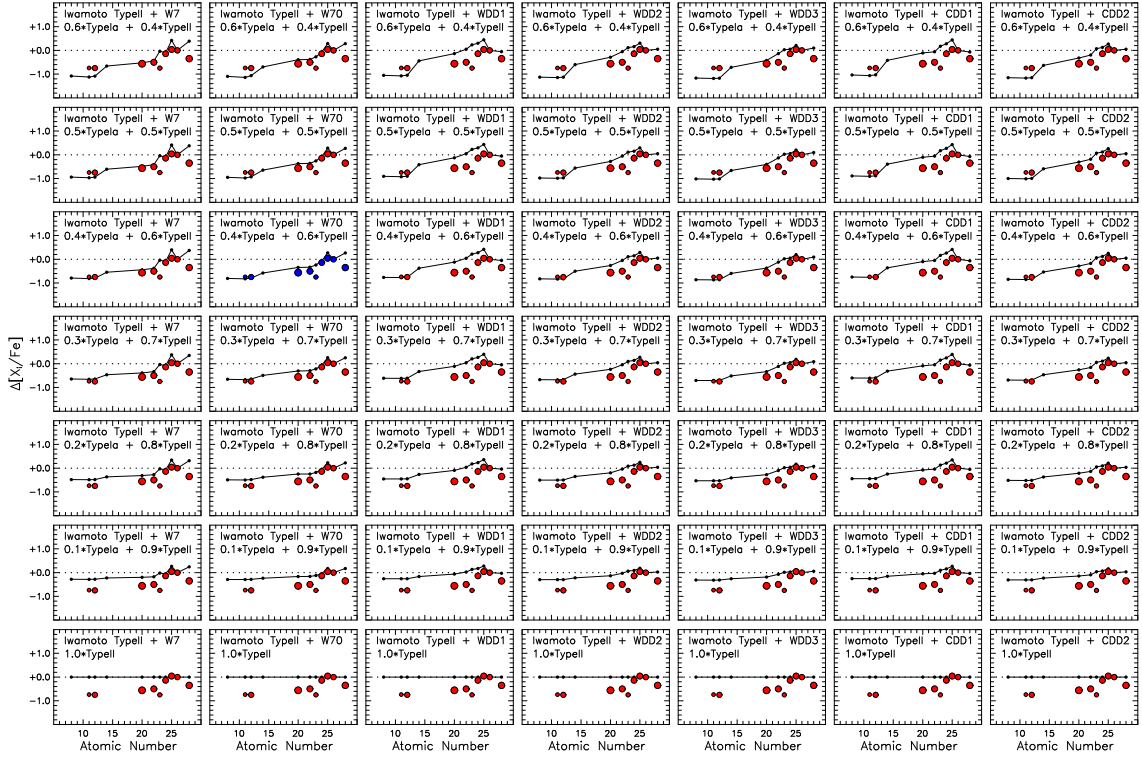}
  \caption{\label{fig:iwam6} \scriptsize The comparison of model
    admixtures of Type Ia and Type II SNe (small black dots) and
    observed abundances (red circles) as a function of atomic number
    for Car-612 = CC07452. For the model, the ordinate is
    $\Delta$[X$_{\rm i}$/Fe] = [X$_{\rm i}$/Fe]$_{\rm
      Admixture}$ -- [X$_{\rm i}$/Fe]$_{\rm {Type~II}}$, while for the
    observations it is $\Delta$[X$_{\rm i}$/Fe] = [X$_{\rm
        i}$/Fe]$_{\rm Carina}$ -- [X$_{\rm i}$/Fe]$_{\rm Halo}$. Blue
    symbols are used for the best fit case.  The size of the
    observational symbols decreases linearly as the error increases.
    See text for more details.}

\end{center}
\end{figure}

\clearpage

\begin{figure}[htbp!]
\begin{center}

\includegraphics[width=12.0cm,angle=0]{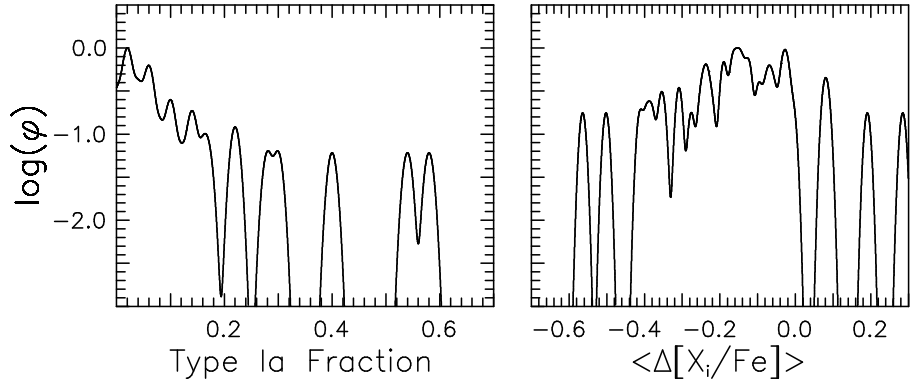}

  \caption{\label{fig:histograms} \scriptsize Histograms of the
    logarithm of Fraction$_{\rm Type~{Ia}}$(left) and
    $\langle\Delta$[X$_{\rm i}$/Fe]$\rangle$ (right) of the Carina red giants
    analyzed in this work, where a Gaussian kernel of 0.008 has been
    adopted in both cases, to facilitate identification of individual stars.  }

\end{center}
\end{figure}

\clearpage

\begin{figure}[htbp!]
\begin{center}

\includegraphics[width=9.0cm,angle=0]{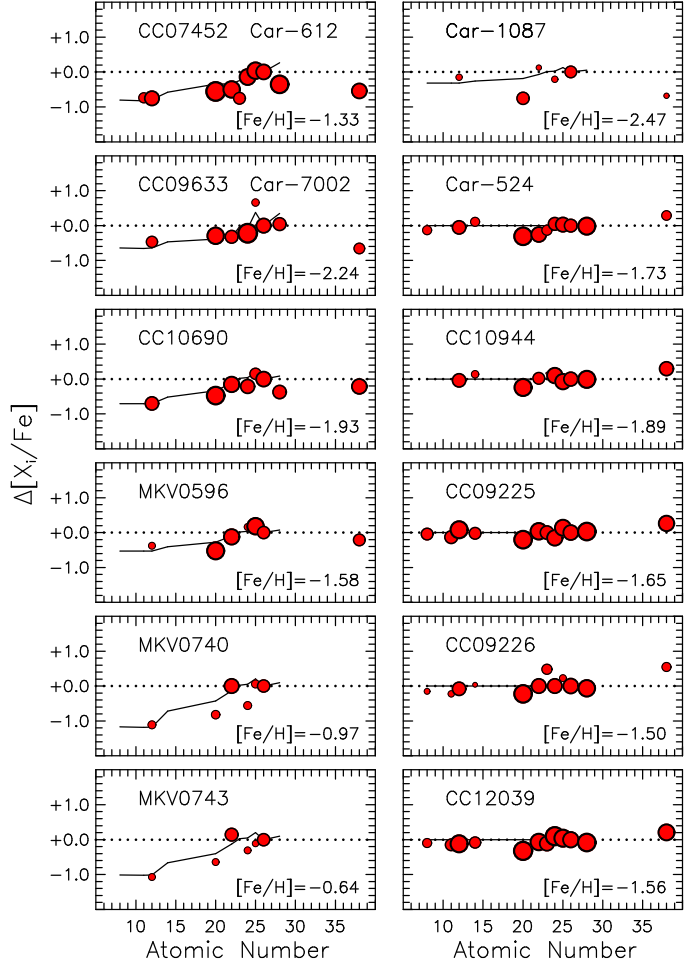}
  \caption{\label{fig:iwam_venn} \scriptsize $\Delta$[X$_{\rm i}$/Fe]
    as a function of atomic number. (Left panels): The comparison
    between the best fit Iwamoto et al Type II and Type Ia admixture
    models (black line) with observations (red circles; size of the
    symbols decreases linearly as the error increases) of the six
    Carina giants that show the strongest evidence for Type Ia
    enhancement. For the model, the ordinate is $\Delta$[X$_{\rm
        i}$/Fe] = [X$_{\rm i}$/Fe]$_{\rm Admixture}$ -- [X$_{\rm
        i}$/Fe]$_{\rm {Type~II}}$, while for the observations it is
    $\Delta$[X$_{\rm i}$/Fe] = [X$_{\rm i}$/Fe]$_{\rm Carina}$ --
    [X$_{\rm i}$/Fe]$_{\rm Halo}$. (Right panels): The top panel shows
    a seventh Type Ia enhanced candidate, which we reject as
    inconclusive.  The other five stars present abundances typical of
    Carina red giants with no Type Ia enhancement.}

\end{center}
\end{figure}

\clearpage

\begin{figure}[htbp]
\begin{center}

\includegraphics[width=5.0cm,angle=0]{fig24.eps}
  \caption{\label{fig:plages}\scriptsize The comparison of the ages of
    the red giants of Carina determined using the isochrones of
    \citet{demarque04} (Yale-Yonsei), \citet{vandenberg06}
    (Victoria-Regina), and \citet{dotter08} (Dartmouth).  The diagram
    demonstrates the difficulty in age determination for red giants
    from analysis of their position in the CMD, resulting from the
    relatively small sensitivity of age to temperature and abundance
    and small differences between the different stellar isochrones.}
\end{center}
\end{figure}

\clearpage

\begin{figure}[htbp]
\begin{center}

\includegraphics[width=10.0cm,angle=0]{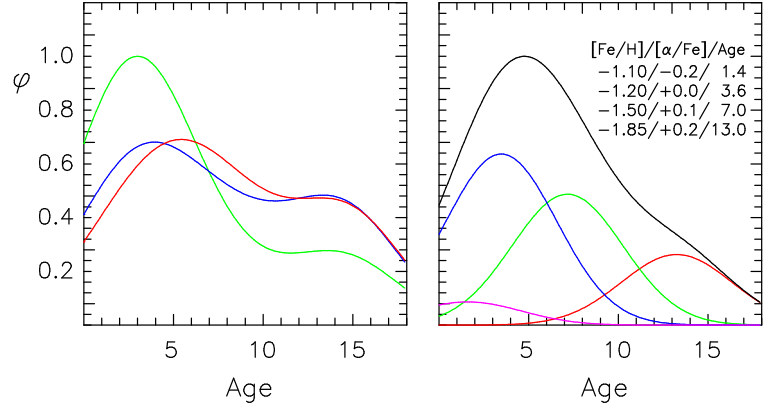}

  \caption{\label{fig:agehist}\scriptsize Left: Generalized histograms
    of the age distributions of the Yale-Yonsei (blue),
    Victoria-Regina (green) and Dartmouth (red) isochrones. The three
    distributions have been normalized to have the same area. Right:
    Histogram of the age distribution of a four component model of
    Carina based on Paper I, for which the population parameters are
    given in the legend.  Red, green, blue, and magenta refer to first
    through fourth populations, respectively, while their summation is
    presented in black.  A Gaussian kernel having $\sigma$ = 3.0~Gyr
    has been adopted in both panels.}

\end{center}
\end{figure}

\clearpage

\begin{figure}[htbp]
\begin{center}

\includegraphics[width=10.0cm,angle=0]{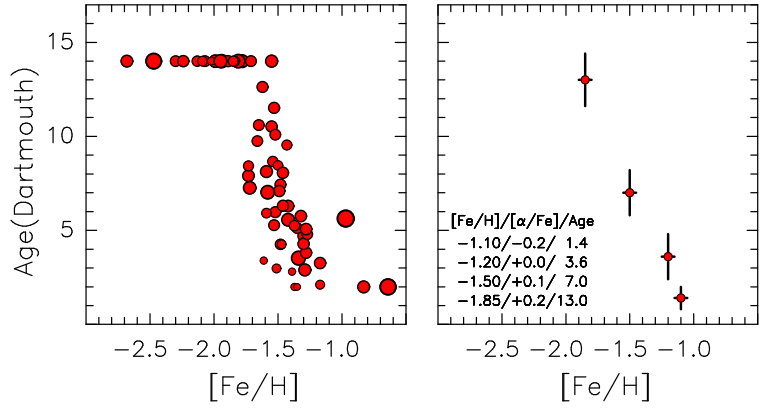}

  \caption{\label{fig:amr}\scriptsize {The age-metallicity relationship
    for Carina.  Left: Age vs [Fe/H] for 63 RGB members, where the
    size of the symbols increase as their [Ca/Fe] values
    decrease. Right: Data for the four CMD populations (variable
    {\alphafe} case) discussed in Paper I, where the bars represent
    the [Fe/H] and age spreads determined in that work. The population
    parameters are given in the legend.}}

\end{center}
\end{figure}

\clearpage

\begin{figure}[htbp!]
\begin{center}

\includegraphics[width=10.0cm,angle=-90]{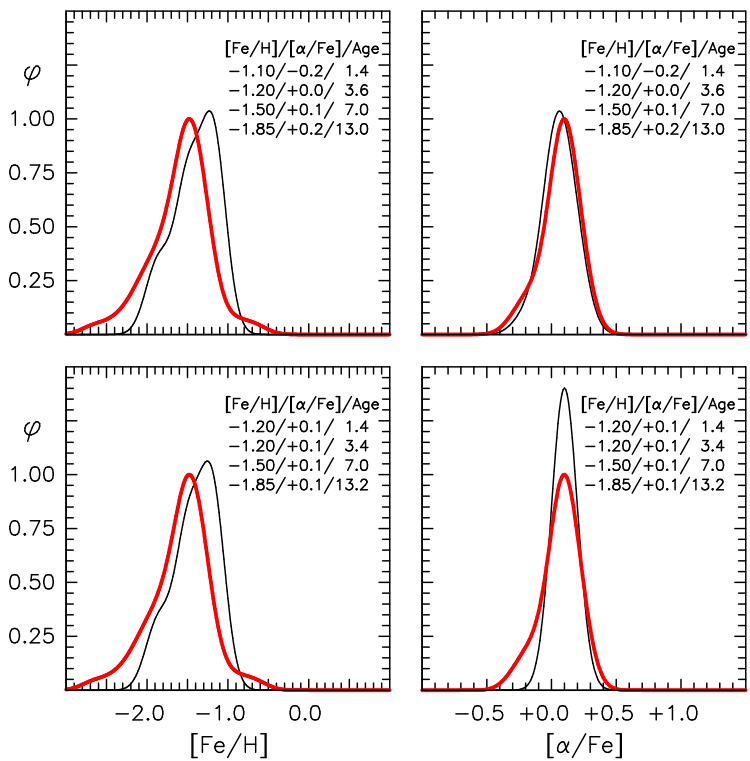}

  \caption{\label{fig:cardist}\small Comparison between the observed
    thick (red) and synthetic ([Fe/H]) MDF and {\alphafe} distribution
    functions (thin black) of the Carina upper RGB, for variable
    {\alphafe} (upper panels) and fixed {\alphafe} (lower
    panels). Gaussian kernels having $\sigma$ values of 0.15~dex and
    0.10~dex were used for the [Fe/H] and {\alphafe} distributions,
    respectively. The synthetic CMD population parameters are given in
    the legends.}

\end{center}
\end{figure}
 
\clearpage
\begin{figure}[htbp]
\begin{center}

\includegraphics[width=12.0cm,angle=0]{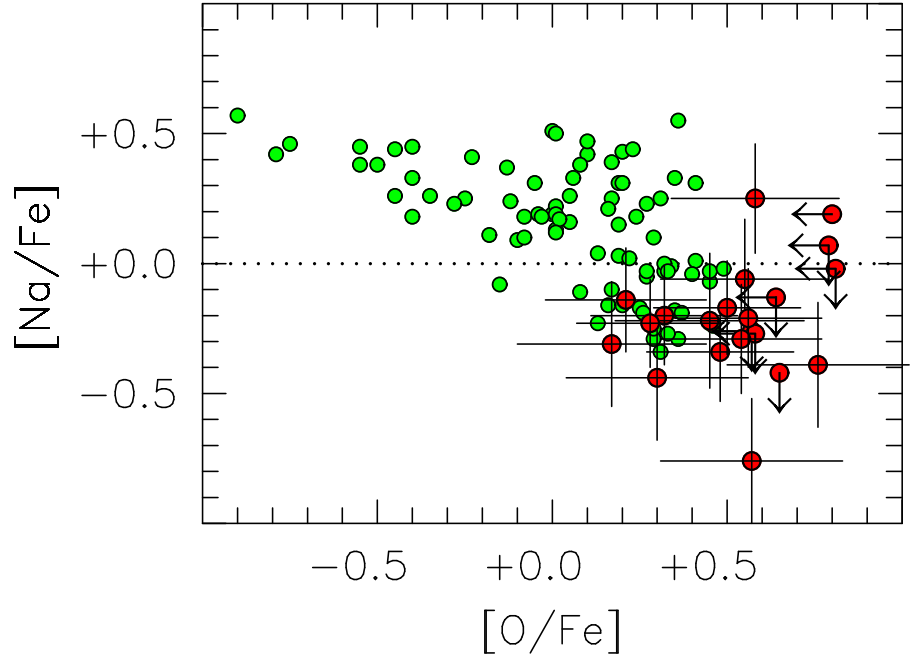}
  \caption{\label{fig:onaal}\scriptsize [Na/Fe] vs. [O/Fe] for red
    giants in Carina (large red circles, present work) and in the
    globular clusters M3, M4, M5, M13 (small green circles).  We note
    that all estimates assume LTE. Arrows are used for Carina to
    indicate when only an upper limit is available for the [O/Fe]
    and/or [Na/Fe] values.  See text for discussion of the difference
    between the two distributions.}

\end{center}
\end{figure}

\clearpage
\begin{figure}[htbp]
\begin{center}

\includegraphics[width=9.0cm,angle=0]{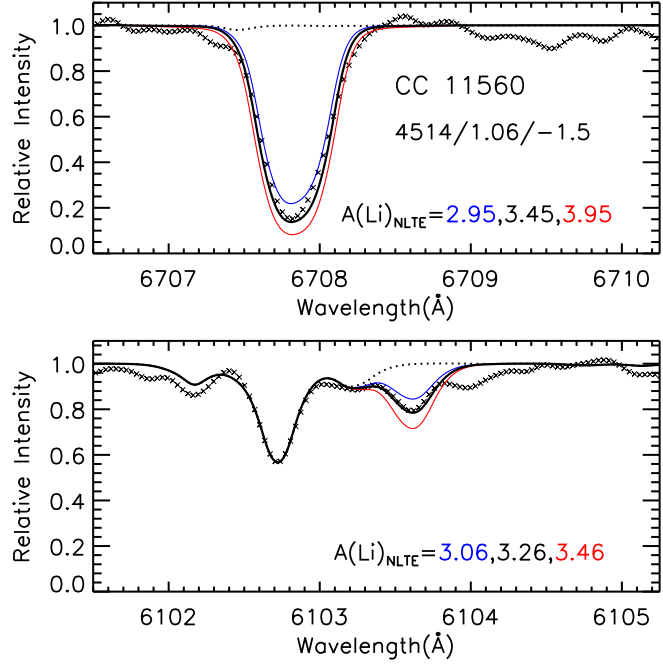}
  \caption{\label{fig:lithium}\scriptsize Comparison between the
    observed spectrum of the lithium enhanced CC11560 (crosses) and
    LTE model atmospheric synthetic spectra.  The full lines represent
    the adopted fit together with two lines that bracket the best fit
    by $\pm$ 0.5 dex (upper panel) and $\pm$ 0.2 dex (lower panel),
    while the dotted line in each panel is the case of zero lithium.
    The adopted lithium abundances, on the NLTE scale, are given in the
    legends.  The {\teff}/{\logg}/[Fe/H] values for CC11560 are also
    presented. }
\end{center}
\end{figure}

\clearpage 
 
\begin{deluxetable}{cccccccccc}                                                                                                                   
\tablecolumns{10}                                                                                                                            
\tablewidth{0pt}

\tabletypesize{\scriptsize}
\tablecaption{\label{tab:obsdat} Coordinates, Photometry, and Radial Velocities for 32 Carina Red Giants}                  
\tablehead{                                                                                                                              
  \colhead{Object} & {RA (2000)  Dec (2000) } & {$V$}       & {$B-V$}        &    {$I$} &    {$J$}  &     {$H$} &     {$K$} &  {Vrad} &  {$\sigma$(Vrad)} \\  
            {(1)}   & {(2)}              & {(3)}       & {(4)}          &    {(5)} &    {(6)}  &    {(7)}  &     {(8)} &   {(9)} &  {(10)}  
  }                                                                                                                                       
\startdata                                                                                                                               
CC06122 &   06 41 39.64 $-$50 49 58.8 &   17.60 &    1.43 &   16.18 &   15.23 &   14.48 &   14.34 & 230.31 & 0.08  \\
CC06486 &   06 41 46.00 $-$50 51 00.6 &   18.17 &    1.24 &   16.89 &   16.01 &   15.41 &   15.27 & 226.23 & 0.19  \\
CC06975 &   06 42 10.86 $-$50 52 20.6 &   18.38 &    1.11 &   17.13 &   16.23 &   15.65 &   15.55 & 229.57 & 0.18  \\
CC07452 &   06 40 58.46 $-$50 53 35.5 &   17.80 &    1.29 &   16.41 &   15.52 &   14.88 &   14.74 & 225.23 & 0.09  \\
CC07889 &   06 41 40.83 $-$50 54 45.8 &   18.55 &    1.06 &   17.34 &   16.53 &   15.97 &   15.83 & 236.29 & 0.27  \\
CC08447 &   06 41 10.54 $-$50 55 53.0 &   17.99 &    1.25 &   16.65 &   15.72 &   15.09 &   14.94 & 232.52 & 0.10  \\
CC08469 &   06 42 17.43 $-$50 55 55.3 &   17.81 &    1.29 &   16.43 &   15.47 &   14.80 &   14.64 & 223.63 & 0.11  \\
CC08615 &   06 41 49.58 $-$50 56 11.5 &   18.50 &    1.05 &   17.32 &   16.45 &   15.93 &   15.78 & 223.91 & 0.21  \\
\enddata                                                                                                      
\end{deluxetable}      

\begin{center}                                                                                                          
\begin{deluxetable}{lccrrrrrrrrrrrr}                                                                                    
\tabletypesize{\tiny}                                                                                                   
\tablecolumns{15}                                                                                                       
\tablewidth{0pt}                                                                                                        
\rotate                                                                                                                 
\tablecaption{\label{tab:ew1} Atomic Data and Equivalent Widths (m{\AA}) for 11 Carina Red Giants\tablenotemark{b}}                          
\tablehead{                                                                                                             
\colhead{Species} & {Wavelength} &  {$\chi$} & {log$gf$} & {CC06122} & {CC06486} & {CC06975} & {CC07452} & {CC07889} & {CC08447} & {CC08469} & {CC08615} & {CC08695} & {CC08788} & {CC09000} \\
{(1)} & {(2)} & {(3)} & {(4)} & {(5)} & {(6)} & {(7)} & {(8)} & {(9)} & {(10)} & {(11)} & {(12)} & {(13)} & {(14)} & {(15)}
}                                                                                                                       
\startdata                                                                                                              
$S/N$\tablenotemark{a}  &  &  &  &      17 &       10 &        8 &       18 &        9 &       16 &       11 &       11 &       14 &       10 &       23 \\
W(min)\,(m{\AA})  &  &  &  &      26 &       28 &       31 &       14 &       34 &       24 &       24 &       20 &       23 &       23 &       20 \\
Fe I  &  5302.30 &   3.28 & $-$0.880 &  153.5 &    ... &  126.3 &  144.5 &   66.5 &  122.7 &  147.7 &  115.0 &  143.3 &  121.7 &  140.7 \\
Fe I  &  5307.37 &   1.61 & $-$2.912 &  183.0 &   90.5 &  113.7 &  162.0 &   57.0 &  136.3 &  166.7 &  117.3 &  138.3 &  109.0 &  145.0 \\
Fe I  &  5324.19 &   3.21 & $-$0.100 &  186.5 &    ... &    ... &    ... &  118.5 &  186.5 &  182.0 &    ... &  181.0 &    ... &  193.0 \\
Fe I  &  5332.90 &   1.56 & $-$2.780 &    ... &    ... &  103.0 &  192.5 &   68.5 &    ... &    ... &    ... &  168.5 &    ... &    ... \\
Fe I  &  5339.93 &   3.27 & $-$0.720 &  139.5 &  102.0 &    ... &  160.5 &   29.5 &  144.3 &  147.3 &  110.0 &  157.3 &  143.5 &  147.3 \\
\enddata                                                                                                                
\tablenotetext{a}{$S/N$ per $\sim$0.28\,{\AA}~pixel at 5300 -- 5700\,{\AA}.}                                                      
\tablenotetext{b}{This table is available in its entirety in a machine-readable form in the online journal. A portion is shown here for guidance regarding its form and content.}
\end{deluxetable}                                                                                                       
\end{center}

\begin{center}                                                                                                          
\begin{deluxetable}{lccrrrrrrrrrrrr}                                                                                    
\tabletypesize{\tiny}                                                                                                   
\tablecolumns{15}                                                                                                       
\tablewidth{0pt}                                                                                                        
\rotate                                                                                                                 
\tablecaption{\label{tab:ew2} Atomic Data and Equivalent Widths (m{\AA}) for 11 Carina Red Giants\tablenotemark{b}}                          
\tablehead{                                                                                                             
\colhead{Species} & {Wavelength} &  {$\chi$} & {log$gf$} & {CC09179} & {CC09225} & {CC09226} & {CC09430} & {CC09507} & {CC09633} & {CC09869} & {CC09929} & {CC10038} & {CC10194} & {CC10318} \\
{(1)} & {(2)} & {(3)} & {(4)} & {(5)} & {(6)} & {(7)} & {(8)} & {(9)} & {(10)} & {(11)} & {(12)} & {(13)} & {(14)} & {(15)}
}                                                                                                                       
\startdata                                                                                                              
$S/N$\tablenotemark{a}  &  &  &  &      13 &       18 &       14 &       12 &       13 &       11 &       12 &        8 &       12 &        9 &       10 \\
W(min)\,(m{\AA})  &  &  &  &      20 &       20 &       20 &       25 &       22 &       29 &       25 &       30 &       22 &       26 &       25 \\
Fe I  &  5302.30 &   3.28 & $-$0.880 &   83.7 &  129.0 &   98.0 &   73.0 &   97.0 &   54.0 &  127.7 &  118.0 &  133.0 &  124.0 &  129.0 \\
Fe I  &  5307.37 &   1.61 & $-$2.912 &   90.3 &  132.3 &  116.5 &   89.5 &  101.0 &   62.0 &  162.0 &   72.0 &  131.7 &  179.5 &    ... \\
Fe I  &  5324.19 &   3.21 & $-$0.100 &    ... &    ... &  147.5 &    ... &    ... &   94.5 &  186.0 &    ... &    ... &    ... &    ... \\
Fe I  &  5332.90 &   1.56 & $-$2.780 &    ... &    ... &    ... &    ... &  119.0 &    ... &  158.0 &    ... &    ... &    ... &  145.5 \\
Fe I  &  5339.93 &   3.27 & $-$0.720 &  130.0 &  148.7 &  130.7 &  102.0 &  116.0 &   46.0 &  136.0 &    ... &  121.7 &  140.5 &  153.7 \\
\enddata                                                                                                                
\tablenotetext{a}{$S/N$ per $\sim$0.28\,{\AA}~pixel at 5300 -- 5700\,{\AA}.}                                                      
\tablenotetext{b}{This table is available in its entirety in a machine-readable form in the online journal. A portion is shown here for guidance regarding its form and content.}
\end{deluxetable}                                                                                                       
\end{center}                                                                    

\begin{center}                                                                                                          
\begin{deluxetable}{lccrrrrrrrrrrr}                                                                                     
\tabletypesize{\tiny}                                                                                                   
\tablecolumns{14}                                                                                                       
\tablewidth{0pt}                                                                                                        
\rotate                                                                                                                 
\tablecaption{\label{tab:ew3} Atomic Data and Equivalent Widths (m{\AA}) for 10 Carina Red Giants\tablenotemark{b}}                   
\tablehead{                                                                                                             
\colhead{Species} & {Wavelength} & {$\chi$} & {log$gf$} & {CC10414} & {CC10686} & {CC10690} & {CC10802} & {CC10944} & {CC11217} & {CC11388} & {CC11560} & {CC12038} & {CC12039} \\
{(1)} & {(2)} & {(3)} & {(4)} & {(5)} & {(6)} & {(7)} & {(8)} & {(9)} & {(10)} & {(11)} & {(12)} & {(13)} & {(14)}
}                                                                                                                       
\startdata                                                                                                              
$S/N$\tablenotemark{a}  &  &  &  &       9 &       20 &       14 &        7 &       16 &       16 &       10 &       11 &        8 &       18 \\
W(min)\,(m{\AA})  &  &  &  &      26 &       15 &       25 &       33 &       20 &       22 &       26 &       23 &       24 &       22 \\
Fe I  &  5302.30 &   3.28 & $-$0.880 &  118.5 &  126.3 &   97.0 &    ... &  101.0 &  155.0 &    ... &  106.5 &  132.5 &  141.3 \\
Fe I  &  5307.37 &   1.61 & $-$2.912 &  166.7 &  134.0 &  115.5 &  120.5 &  121.0 &  166.3 &  145.5 &  109.0 &   98.3 &  139.7 \\
Fe I  &  5324.19 &   3.21 & $-$0.100 &    ... &    ... &  130.0 &    ... &    ... &    ... &    ... &    ... &    ... &  178.0 \\
Fe I  &  5332.90 &   1.56 & $-$2.780 &    ... &    ... &    ... &    ... &    ... &    ... &    ... &  155.0 &    ... &    ... \\
Fe I  &  5339.93 &   3.27 & $-$0.720 &  162.5 &  126.3 &  109.0 &  133.7 &  106.7 &  164.3 &    ... &  109.5 &  132.5 &  143.3 \\
\enddata                                                                                                                
\tablenotetext{a}{$S/N$ per $\sim$0.28\,{\AA}~pixel at 5300 -- 5700\,{\AA}.}                                                      
\tablenotetext{b}{This table is available in its entirety in a machine-readable form in the online journal. A portion is shown here for guidance regarding its form and content.}
\end{deluxetable}                                                                                                       
\end{center}                                                                                                            

\begin{deluxetable}{ccccccc}                                                                                                                  
\tablecolumns{6}                                                                                                                            
\tablewidth{0pt}
\tabletypesize{\scriptsize}
\tablecaption{\label{tab:atpar} Model Atmosphere Parameters for 32 Carina Red Giants}                  
\tablehead{                                                                                                                              
  \colhead{Object}  & {\teff} & {$\sigma${\teff}} & {\logg} & {[M/H]} & {\vxi}     \\  
             {}     & {(K)}   & {(K)}             & {(cgs)} & {(dex)}  & {(\kms)}    \\        
           {(1)}    & {(2)}   & {(3)}             & {(4)}   & {(5)}    & {(6)}   
  }                                                                                                                                       

\startdata                                                                                                                               
CC06122 & 4284 &  60 & 0.68 &  $-$1.5 &  2.30  \\
CC06486 & 4524 &  62 & 1.07 &  $-$2.0 &  1.95  \\
CC06975 & 4563 &  84 & 1.18 &  $-$1.8 &  2.30  \\
CC07452 & 4437 &  65 & 0.87 &  $-$1.3 &  2.30  \\
CC07889 & 4675 &  55 & 1.31 &  $-$2.3 &  2.30  \\
CC08447 & 4418 &  72 & 0.93 &  $-$1.5 &  2.15  \\
CC08469 & 4337 &  68 & 0.81 &  $-$1.5 &  2.25  \\
CC08615 & 4651 &  84 & 1.28 &  $-$1.7 &  2.40  \\
\enddata                                                                       
\end{deluxetable}      


\begin{deluxetable}{lrrrrrr}                                                                                                                  
\tablecolumns{7}                                                                                                                            
\tablewidth{0pt}
\tabletypesize{\tiny}
\tablecaption{\label{tab:abund_gil} Chemical Abundances for 32 Carina Red Giants\tablenotemark{b}}                  
\tablehead{                                                                                                                              
  \colhead{Species}  & {log\,$\epsilon$} & {s.e.$_{\log\epsilon}$} & {N}   & {[X/Fe]}  & {$\sigma_{\rm {Sys}}$[X/Fe]}  & {$\sigma_{\rm {Tot}}$[X/Fe]}    \\  
           {(1)}     & {(2)}             & {(3)}                  & {(4)} & {(5)}      & {(6)}                        & {(7)}    
  }                                                                                                                                       
\startdata                                                                                                                               
    CC06122 &          &            &            &            &            &            \\ 
   Fe I    &     6.07   &     0.02   &       74   &    $-$1.43\tablenotemark{a}   &     0.14   &     0.14   \\ 
   O I     &     7.58   &     0.17   &        1   &     0.32   &     0.13   &     0.21   \\ 
   Na I    &     4.61   &     0.17   &        1   &    $-$0.20   &     0.08   &     0.19   \\ 
   Mg I    &     6.48   &     0.18   &        2   &     0.31   &     0.03   &     0.18   \\ 
   Al I    &  $<$5.38   &     0.17   &        1   &  $<$0.36   &     0.10   &     0.20   \\ 
   Si I    &     6.29   &     0.17   &        1   &     0.21   &     0.14   &     0.22   \\ 
   Ca I    &     5.09   &     0.04   &       19   &     0.09   &     0.02   &     0.04   \\
\enddata 
\tablenotetext{a}{[Fe/H]}                                                                                                     
\tablenotetext{b}{This table is available in its entirety in a machine-readable form in the online journal. A portion is shown here for guidance regarding its form and content.}
\end{deluxetable}      


\begin{deluxetable}{lrrrrr}                                                                                                                  
\tablecolumns{6}                                                                                                                            
\tablewidth{0pt}                                                                                                                  
\tablecaption{\label{tab:errors}Average Abundance Uncertainties in [Fe/H] and [X/Fe]}                   
\tablehead{                                                                                                                              
  \colhead{Species} & {$\sigma_{T_{\rm eff}}$} & {$\sigma_{{\rm log}g}$} & {$\sigma_{\rm [M/H]}$} & {$\sigma_{\xi{_t}}$} & {$\sigma_{\rm [X/Fe]}$} \\   
             {}     & {(80~K)}             & {(0.07~dex)}           & {(0.15~dex)}            & {(0.2~{\kms})}     &                         \\
           {(1)}    & {(2)}                & {(3)}                  &{(4)}                   &{(5)}               & {(6)}        
}                                                                                                                                       
\startdata                                                                                                          
   Fe I\tablenotemark{a}   &    0.109  &   $-$0.001  &   $-$0.015  &   $-$0.082  &    0.136 \\
   O I    &   $-$0.099  &      0.029  &    0.097  &    0.078  &    0.128 \\
   Na I   &   $-$0.043  &   $-$0.004  &    0.000  &    0.070  &    0.084 \\
   Mg I   &   $-$0.035  &   $-$0.008  &    0.002  &    0.018  &    0.047 \\
   Al I   &   $-$0.048  &   $-$0.002  &    0.004  &    0.077  &    0.092 \\
   Si I   &   $-$0.099  &      0.007  &    0.030  &    0.078  &    0.123 \\
   Ca I   &   $-$0.015  &   $-$0.005  & $-$0.010  &    0.016  &    0.027 \\
   Sc II  &   $-$0.115  &      0.028  &    0.074  &    0.042  &    0.111 \\
\enddata                                                                                                      
\tablenotetext{a}{Errors pertain to uncertainties in [Fe/H]}
\end{deluxetable}      


\begin{deluxetable}{lrrrrrrrrrrrrrrrrrrrrr}                                                                                    
\tabletypesize{\tiny}                                                                                                   
\tablecolumns{22}                                                                                                       
\tablewidth{0pt}                                                                                                        
\rotate                                                                                                                 
\tablecaption{\label{tab:relabs_gil} Summary\tablenotemark{a} of [Fe/H], [X/Fe], and Total Errors $\sigma_{\rm {Tot}}$ for 32 Carina Red Giants\tablenotemark{b}}                          

\tablehead{                                                                                                                              
  \colhead{Object} & {[Fe/H]} & {O}  & {Na} & {Mg} & {Al} & {Si} & {Ca} & {Sc} & {Ti~I} & {Ti~II} & {V} & {Cr} & {Mn} & {Fe~II} & {Co} & {Ni} & {Cu} & {Ba} & {La} & {Nd} & {Eu}\\  
  }
\startdata    
    CC06122    &  $-$1.43  &   0.32 &  $-$0.20 &   0.31 & $<$ 0.36 &   0.21 &   0.09 &   0.17 &   0.15 &   0.34 &  $-$0.07 &  $-$0.01 &  $-$0.32 &   0.09 &  $-$0.03 &  $-$0.08 &  $-$0.63 &   0.44 &   0.49 &    ... &    ... \\
               &   0.14   &   0.21 &   0.19 &   0.18 &   0.20 &   0.22 &   0.04 &   0.18 &   0.10 &   0.13 &   0.16 &   0.10 &   0.07 &   0.17 &   0.19 &   0.09 &   0.19 &   0.14 &   0.13 &    ... &    ... \\
               &     74   &      1 &      1 &      2 &      1 &      1 &     19 &      5 &      8 &      2 &     10 &      4 &      5 &      8 &      1 &     10 &      1 &      3 &      3 &    ... &    ... \\
               &          &        &        &        &        &        &        &        &        &        &        &        &        &        &        &        &        &        &        &        &        \\
    CC06486    &  $-$2.03 & $<$ 0.80 &   0.19 &   0.17 & $<$ 1.28 &    ... &   0.11 &   0.14 &    ... &    ... &    ... &  $-$0.16 &    ... &   0.42 &    ... &   0.08 &    ... &  $-$0.45 &    ... &    ... &    ... \\
               &   0.16   &   0.29 &   0.28 &   0.12 &   0.28 &    ... &   0.09 &   0.29 &    ... &    ... &    ... &   0.13 &    ... &   0.30 &    ... &   0.11 &    ... &   0.23 &    ... &    ... &    ... \\
               &     42   &      1 &      1 &      2 &      1 &    ... &     13 &      1 &    ... &    ... &    ... &      3 &    ... &      1 &    ... &      5 &    ... &      3 &    ... &    ... &    ... \\
               &        &        &        &        &        &        &        &        &        &        &        &        &        &        &        &        &        &        &        &        &        \\
\enddata                                                                        
\tablenotetext{a}{For each object the three rows contain [Fe/H] and [X/Fe]; $\sigma_{\rm {Tot}}$; and the number of lines, respectively.}
\tablenotetext{b}{This table is available in its entirety in a machine-readable form in the online journal. A portion is shown here for guidance regarding its form and content.}
\end{deluxetable}      


\begin{deluxetable}{lcrrrrrrrrrrrr}                                                                                                                   
\tablecolumns{14}                                                                                                                            
\tablewidth{0pt}
\tabletypesize{\scriptsize}
\rotate
\tablecaption{\label{tab:litdat} Coordinates, Photometry, and Atmospheric Parameters for 49 Carina Red Giants from the Literature}                  
\tablehead{                                                                                                                              
  \colhead{Object} & {Source\tablenotemark{a}} & {RA   (2000)  Dec } & {$V$}       & {$B-V$}        &    {$I$}  &    {$J$}  &     {$H$} &     {$K$} &  {\teff} &  {  $\sigma$$T_{\rm eff}$}& {\logg} &  {[Fe/H]}& {\vxi}  \\  
            {(1)}  & {(2)}    &         {(3)}       & {(4)}       &    {(5)}       &    {(6)} &    {(7)}  &     {(8)} &     {(9)} &  {(10)} &        {(11)}    &  {(12)}  &   {(13)} & {(14)}   
  }                                                                                      
\startdata                                                                                                                               
Car-484    & 1 &   06 41 39.64 $-$50 49 58.8 &   17.60 &    1.43 &   16.18 &   15.23 &   14.48 &   14.34    & 4284 &  60 &  0.68 & $-$1.53   & 2.25 \\
Car-524    & 1 &   06 41 14.71 $-$50 51 10.1 &   17.62 &    1.31 &   16.29 &   15.37 &   14.75 &   14.59    & 4436 &  71 &  0.79 & $-$1.72   & 2.50 \\
Car-612    & 1 &   06 40 58.46 $-$50 53 35.5 &   17.80 &    1.29 &   16.41 &   15.52 &   14.88 &   14.74    & 4437 &  65 &  0.87 & $-$1.34   & 2.35 \\
Car-705    & 1 &   06 42 17.43 $-$50 55 55.3 &   17.81 &    1.29 &   16.43 &   15.47 &   14.80 &   14.64    & 4337 &  68 &  0.81 & $-$1.47   & 2.25 \\
Car-769    & 1 &   06 41 19.70 $-$50 57 25.6 &   18.01 &    1.21 &   16.71 &   15.74 &   15.14 &   15.02    & 4438 &  93 &  0.95 & $-$1.64   & 2.30 \\
Car-1013    & 1 &   06 41 21.98 $-$51 03 43.3 &   17.78 &    1.27 &   16.42 &   15.50 &   14.83 &   14.71    & 4412 &  65 &  0.84 & $-$1.62   & 2.35 \\
Car-1087    & 1 &   06 41 15.47 $-$51 01 16.0 &   18.03 &    1.14 &   16.74 &   15.84 &   15.28 &   15.14    & 4542 &  78 &  1.02 & $-$2.79   & 2.05 \\
Car-5070    & 1 &   06 41 53.85 $-$50 58 11.2 &   17.88 &    1.21 &   16.53 &   15.62 &   15.04 &   14.90    & 4487 &  80 &  0.93 & $-$2.24   & 1.80 \\
\enddata                                                                                                      
\tablenotetext{a} {Sources: 1 = \citet{venn12}; 2 = \citet{shetrone03}; and 3 = \citet{lemasle12}}
\end{deluxetable}      


\begin{deluxetable}{lrrrrrrrrrrrrrrrrrrrrr}                                                                                    
\tabletypesize{\tiny}                                                                                                   
\tablecolumns{22}                                                                                                       
\tablewidth{0pt}                                                                                                        
\rotate                                                                                                                 
\tablecaption{\label{tab:relabs_kvs} Summary\tablenotemark{a} of [Fe/H], [X/Fe], and Total Errors $\sigma_{\rm {Tot}}$ for 14 Literature (\citet{venn12} \& \citet{shetrone03}) Carina Red Giants\tablenotemark{b}}                          
\tablehead{                                                                                                                              
  \colhead{Object} & {[Fe/H]} & {O}  & {Na} & {Mg} & {Al} & {Si} & {Ca} & {Sc} & {Ti~I} & {Ti~II} & {V} & {Cr} & {Mn} & {Fe~II} & {Co} & {Ni} & {Cu} & {Ba} & {La} & {Nd} & {Eu}\\  
  }
\startdata    
     Car$-$484   &  $-$1.49 &   0.50 &  $-$0.34 &   0.16 &    ... &    ... &   0.05 &   0.08 &   0.14 &   0.24 &  $-$0.07 &  $-$0.04 &  $-$0.37 &   0.12 &    ... &  $-$0.07 &    ... &   0.30 &   0.51 &   0.67 &   0.51 \\
               &   0.14 &   0.22 &   0.20 &   0.10 &    ... &    ... &   0.04 &   0.12 &   0.09 &   0.12 &   0.10 &   0.09 &   0.06 &   0.15 &    ... &   0.07 &    ... &   0.21 &   0.11 &   0.20 &   0.22 \\
               &     75 &      1 &      1 &      2 &    ... &    ... &     20 &      8 &      7 &      4 &      8 &      3 &      6 &      7 &    ... &      9 &    ... &      1 &      3 &      1 &      1 \\
               &        &        &        &        &        &        &        &        &        &        &        &        &        &        &        &        &        &        &        &        &        \\
     Car$-$524   &  $-$1.73 &   0.47 &    ... &   0.25 &    ... &   0.41 &  $-$0.01 &  $-$0.02 &   0.07 &   0.03 &  $-$0.03 &  $-$0.05 &  $-$0.37 &   0.17 &    ... &  $-$0.02 &    ... &   0.19 &   0.61 &   0.44 &    ... \\
               &   0.14 &   0.21 &    ... &   0.13 &    ... &   0.21 &   0.05 &   0.13 &   0.17 &   0.13 &   0.19 &   0.14 &   0.11 &   0.16 &    ... &   0.06 &    ... &   0.20 &   0.11 &   0.19 &    ... \\
               &     64 &      1 &    ... &      2 &    ... &      1 &     19 &      7 &      2 &      4 &      1 &      3 &      3 &      8 &    ... &      6 &    ... &      1 &      2 &      1 &    ... \\
               &        &        &        &        &        &        &        &        &        &        &        &        &        &        &        &        &        &        &        &        &        \\
\enddata
\tablenotetext{a}{For each object the three rows contain [Fe/H] and [X/Fe]; $\sigma_{\rm {Tot}}$; and the number of lines, respectively.}
\tablenotetext{b} {This table is available in its entirety in a machine-readable form in the online journal. A portion is shown here for guidance regarding its form and content.}
\end{deluxetable}


\begin{deluxetable}{lrrrrrrrrrr}                                                                                    
\tabletypesize{\tiny}                                                                                                   
\tablecolumns{11}                                                                                                       
\tablewidth{0pt}                                                                                                        
\tablecaption{\label{tab:relabs_lem} Summary\tablenotemark{a} of [Fe/H], [X/Fe], and Total Errors $\sigma_{\rm {Tot}}$ for 35 \citet{lemasle12} Carina Red Giants\tablenotemark{b}}                          
\tablehead{                                                                                                                              
  \colhead{Object} & {[Fe/H]} & {Mg} & {Ca} & {Ti~I} & {Ti~II} & {Cr} & {Mn} & {Fe~II} & {Ni} & {Ba} \\  
  }
\startdata    
     MKV0397   &  $-$2.02 &    ... &   0.44 &    ... &   0.21 &  $-$0.62 &    ... &   0.24 &    ... &  $-$0.88 \\
               &   0.15 &    ... &   0.26 &    ... &   0.12 &   0.26 &    ... &   0.17 &    ... &   0.28 \\
               &     17 &    ... &      1 &    ... &      2 &      1 &    ... &      5 &    ... &      1 \\
               &        &        &        &        &        &        &        &        &        &        \\
     MKV0458   &  $-$1.47 &   0.07 &   0.33 &   0.31 &   0.33 &  $-$0.17 &  $-$0.48 &   0.40 &   0.47 &   0.27 \\
               &   0.15 &   0.24 &   0.31 &   0.10 &   0.10 &   0.24 &   0.24 &   0.16 &   0.24 &   0.12 \\
               &     26 &      1 &      2 &      3 &      2 &      1 &      1 &      5 &      1 &      2 \\
               &        &        &        &        &        &        &        &        &        &        \\
\enddata
\tablenotetext{a}{For each object the three rows contain [Fe/H] and [X/Fe]; $\sigma_{\rm {Tot}}$; and the number of lines, respectively.}
\tablenotetext{b}{This table is available in its entirety in a machine-readable form in the online journal. A portion is shown here for guidance regarding its form and content.}
\end{deluxetable}      


\begin{deluxetable}{rlrrrrrrrrrrrr}                                                                                                                   
\tablecolumns{14}                                                                                                                            
\tablewidth{0pt}
\tabletypesize{\scriptsize}
\tablecaption{\label{tab:alphas_v2} {\alphafe} Values and Errors\tablenotemark{a} for 30 Carina Red Giants}                  
\tablehead{                                                                                                                              
  \colhead{Name 1}   & {Name2}  & {[Fe/H]} & {$\sigma$Fe}   & {[Mg/Fe]} & {$\sigma$Mg}  & {[Si/Fe]} & {$\sigma$Si}  & {[Ca/Fe]} & {$\sigma$Ca}  & {[Ti/Fe]} & {$\sigma$Ti}  & {{\alphafe}} & {$\sigma${\alphafe}} \\  
            {(1)}    & {(2)}    & {(3)}    & {(4)}          & {(5)}     & {(6)}         & {(7)}     & {(8)}         & {(9)}     & {(10)}        & {(11)}    & {(12)}        & {(13)}       & {(14)}   
  }                                                                                                                                       
\startdata 
CC06122   &  Car-484  &   $-$1.46  &    0.10  &    0.20  &    0.09  &    0.21  &    0.22  &    0.07  &    0.03  &    0.29  &    0.09  &      0.099  &   0.026  \\
CC06975   &           &   $-$1.84  &    0.14  &    0.27  &    0.05  &     ...  &     ...  &    0.14  &    0.06  &    0.17  &    0.18  &      0.215  &   0.038  \\
CC07452   &  Car-612  &   $-$1.33  &    0.10  &   $-$0.45  &    0.11  &     ...  &     ...  &   $-$0.27  &    0.03  &   $-$0.16  &    0.09  &     $-$0.272  &   0.030  \\
CC08447   &           &   $-$1.48  &    0.13  &    0.28  &    0.23  &    0.10  &    0.22  &    0.08  &    0.04  &    0.55  &    0.13  &      0.125  &   0.038  \\
CC08469   &  Car-705  &   $-$1.43  &    0.10  &    0.00  &    0.07  &    0.42  &    0.16  &    0.04  &    0.03  &    0.16  &    0.10  &      0.037  &   0.025  \\
CC08615   &           &   $-$1.73  &    0.11  &    0.52  &    0.24  &    0.56  &    0.24  &    0.18  &    0.05  &    0.05  &    0.25  &      0.189  &   0.048  \\
 CC8695   &  M12      &   $-$1.32  &    0.10  &    0.16  &    0.10  &    0.14  &    0.22  &    0.03  &    0.04  &    0.26  &    0.10  &      0.069  &   0.034  \\
CC08788   &           &   $-$1.48  &    0.13  &    0.26  &    0.14  &    0.06  &    0.24  &    0.14  &    0.07  &    0.05  &    0.10  &      0.132  &   0.053  \\
\enddata                                                                        
\tablenotetext{a} {The errors appear in the columns following the abundances; and the column header (e.g.) $\sigma$Mg stands for $\sigma$[Mg/Fe].} 
\end{deluxetable}      


\begin{deluxetable}{lccrr}                                                                                    
\tabletypesize{\scriptsize}                                                                                                   
\tablecolumns{5}                                                                                                       
\tablewidth{0pt}                                                                                                        
\tablecaption{\label{tab:iwamoto}Fraction of Type Ia Material, $\langle\Delta$[X$_{\rm i}$/Fe]$\rangle$, and [Ba/Fe] in the Carina Red Giants\tablenotemark{c} }                          
\tablehead{                                                            
  \colhead   {Star} & {Iwamato Type\tablenotemark{a}}  & {Type Ia Fraction}  & {$\langle\Delta$[X$_{\rm i}$/Fe]$\rangle$}  & $\Delta$[Ba/Fe] \\  
  \colhead   {(1)}  & {(2)}                            & {(3)}                & {(4)}           & {(5)}      
  }
\startdata    

MKV0842     & Type~II   &   0.00   &  $-$0.07 &  $-$0.54 \\
CC10194     & WDD3     &   0.06   &  $-$0.19 &   0.25 \\
MKV0677     & Type~II   &   0.02   &  $-$0.10 &   0.05 \\
MKV0628     & Type~II   &   0.02   &  $-$0.12 &   0.03 \\
MKV0640     & WDD1     &   0.06   &  $-$0.02 &   0.44 \\
MKV0914     & Type~II   &   0.02   &  $-$0.12 &  $-$0.29 \\
MKV0925     & Type~II   &   0.00   &   0.28 &   0.19 \\
Car-7002\tablenotemark{b}    & ...      &    ...   &    ... &    ... \\
\enddata
\tablenotetext{a}{Admixed \citet{iwamoto99} Type Ia supernova. ``Type II'' implies zero Type Ia}
\tablenotetext{b}{Insufficient information: too few elements observable}
\tablenotetext{c}{This table is available in its entirety in a machine-readable form in the online journal. A portion is shown here for guidance regarding its form and content.}
\end{deluxetable}      


\begin{deluxetable}{cllcc}                                                                                    
\tabletypesize{\scriptsize}                                                                                                   
\tablecolumns{5}                                                                                                       
\tablewidth{0pt}                                                                                                        
\tablecaption{\label{tab:appendix} Cross-identification and Coordinates of Program Stars}                          
\tablehead{                                                                                                                              
  \colhead{This work\tablenotemark{a}} & {Venn et al.\tablenotemark{a}}  & {Shetrone et al.\tablenotemark{a}}  & {Lemasle et al.\tablenotemark{a}} & {RA (2000) Dec} \\   
  \colhead   {(1)}    & {(2)}          & {(3)}              & {(4)}            & {(5)}       
  }
\startdata    
          &           &           & MKV0842   & 06 40 22.51  $-$50 59 01.9 \\
CC10194   &           &           &           & 06 40 30.82  $-$50 59 15.0 \\
          &           &           & MKV0677   & 06 40 31.12  $-$50 55 24.5 \\
          &           &           & MKV0628   & 06 40 35.37  $-$50 54 07.6 \\
          &           &           & MKV0640   & 06 40 40.82  $-$50 54 29.2 \\
          &           &           & MKV0914   & 06 40 42.49  $-$51 00 42.7 \\
          &           &           & MKV0925   & 06 40 43.17  $-$51 01 06.6 \\
          & Car-7002  &           &           & 06 40 49.15  $-$51 00 32.9 \\
\enddata
\tablenotetext{a}{The headers in Columns 1 -- 4 refer to the present work, \citet{venn12}, \citet{shetrone03}, and \citet{lemasle12}, respectively.}
\end{deluxetable}      


\begin{deluxetable}{lrrrrrrrrrrrrrrrrrrr}                                                                                    
\tabletypesize{\tiny}                                                                                                   
\tablecolumns{20}                                                                                                       
\tablewidth{0pt}                                                                                                        
\rotate                                                                                                                 
\tablecaption{\label{tab:relabs_aver} Average [Fe/H], [X/Fe], and Errors\tablenotemark{a} for the 63 Independent Carina Red Giants\tablenotemark{c}}                          
\tablehead{                                                                                                                              
  \colhead{Object\tablenotemark{b}} & {[Fe/H]} & {O}  & {Na} & {Mg} & {Si} & {Ca} & {Sc} & {Ti} & {V} & {Cr} & {Mn} & {Fe~II} & {Co} & {Ni} & {Cu} & {Ba} & {La} & {Nd} & {Eu}\\  
  }
\startdata    
MKV0842  & $-$1.39 &   ... &   ... &  0.30 &   ... &  0.50 &   ... &  0.26 &   ... & $-$0.29 &   ... &  0.50 &   ... &   ... &   ... & $-$0.66 &   ... &   ... &   ... \\       
         &  0.15 &   ... &   ... &  0.21 &   ... &  0.20 &   ... &  0.16 &   ... &  0.21 &   ... &  0.19 &   ... &   ... &   ... &  0.22 &   ... &   ... &   ... \\       
 & & & & & & & & & & & & & & & & & & & \\                                                                                                                                 
CC10194  & $-$1.42 &  0.17 & $-$0.31 &  0.15 &  0.18 &  0.01 & $-$0.27 &  0.22 & $-$0.25 &  0.00 & $-$0.90 & $-$0.01 &   ... & $-$0.25 &   ... &  0.13 &  0.24 & $-$0.31 &   ... \\       
         &  0.15 &  0.27 &  0.24 &  0.14 &  0.27 &  0.08 &  0.15 &  0.12 &  0.19 &  0.23 &  0.17 &  0.21 &   ... &  0.12 &   ... &  0.22 &  0.26 &  0.25 &   ... \\       
 & & & & & & & & & & & & & & & & & & & \\
\enddata                                                                        
\tablenotetext{a}{For each object the two rows contain [Fe/H] and [X/Fe]; and $\sigma_{\rm {Tot}}$, respectively.}
\tablenotetext{b}{In each block, one or two names are given when one or two observations are available.}
\tablenotetext{c}{This table is available in its entirety in a machine-readable form in the online journal. A portion is shown here for guidance regarding its form and content.}
\end{deluxetable}      


\begin{thebibliography}{119}
\expandafter\ifx\csname natexlab\endcsname\relax\def\natexlab#1{#1}\fi

\bibitem[{{Alonso} {et~al.}(1999){Alonso}, {Arribas}, \&
  {Mart{\'{\i}}nez-Roger}}]{alonso99}
{Alonso}, A., {Arribas}, S., \& {Mart{\'{\i}}nez-Roger}, C. 1999, \aaps, 140,
  261

\bibitem[{{Argast} {et~al.}(2002){Argast}, {Samland}, {Thielemann}, \&
  {Gerhard}}]{argast02}
{Argast}, D., {Samland}, M., {Thielemann}, F.-K., \& {Gerhard}, O.~E. 2002,
  \aap, 388, 842

\bibitem[{{Asplund} {et~al.}(2009){Asplund}, {Grevesse}, {Sauval}, \&
  {Scott}}]{asplund09}
{Asplund}, M., {Grevesse}, N., {Sauval}, A.~J., \& {Scott}, P. 2009, \araa, 47,
  481

\bibitem[{{Asplund} {et~al.}(2000){Asplund}, {Nordlund}, {Trampedach}, {Allende
  Prieto}, \& {Stein}}]{asplund00}
{Asplund}, M., {Nordlund}, {\AA}., {Trampedach}, R., {Allende Prieto}, C., \&
  {Stein}, R.~F. 2000, \aap, 359, 729

\bibitem[{{Barklem} {et~al.}(2005){Barklem}, {Christlieb}, {Beers}, {Hill},
  {Bessell}, {Holmberg}, {Marsteller}, {Rossi}, {Zickgraf}, \&
  {Reimers}}]{barklem05}
{Barklem}, P.~S., {Christlieb}, N., {Beers}, T.~C.,  {et~al.} 2005, \aap, 439,
  129

\bibitem[{{Bastian} {et~al.}(2013){Bastian}, {Lamers}, {de Mink}, {Longmore},
  {Goodwin}, \& {Gieles}}]{bastian13}
{Bastian}, N., {Lamers}, H.~J.~G.~L.~M., {de Mink}, S.~E.,  {et~al.} 2013,
  \mnras, 436, 2398

\bibitem[{{Battaglia} {et~al.}(2008){Battaglia}, {Irwin}, {Tolstoy}, {Hill},
  {Helmi}, {Letarte}, \& {Jablonka}}]{battaglia08}
{Battaglia}, G., {Irwin}, M., {Tolstoy}, E.,  {et~al.} 2008, \mnras, 383, 183

\bibitem[{{Bono} {et~al.}(2010){Bono}, {Stetson}, {Walker}, {Monelli},
  {Fabrizio}, {Pietrinferni}, {Brocato}, {Buonanno}, {Caputo}, {Cassisi},
  {Castellani}, {Cignoni}, {Corsi}, {Dall'Ora}, {Degl'Innocenti}, {Fran{\c
  c}ois}, {Ferraro}, {Iannicola}, {Nonino}, {Moroni}, {Pulone}, {Smith}, \&
  {Thevenin}}]{bono10}
{Bono}, G., {Stetson}, P.~B., {Walker}, A.~R.,  {et~al.} 2010, \pasp, 122, 651

\bibitem[{{Brown} {et~al.}(1989){Brown}, {Sneden}, {Lambert}, \&
  {Dutchover}}]{brown89}
{Brown}, J.~A., {Sneden}, C., {Lambert}, D.~L., \& {Dutchover}, Jr., E. 1989,
  \apjs, 71, 293

\bibitem[{{Cameron} \& {Fowler}(1971)}]{cameron71}
{Cameron}, A.~G.~W., \& {Fowler}, W.~A. 1971, \apj, 164, 111

\bibitem[{{Carretta} {et~al.}(2010){Carretta}, {Bragaglia}, {Gratton},
  {Recio-Blanco}, {Lucatello}, {D'Orazi}, \& {Cassisi}}]{carretta10a}
{Carretta}, E., {Bragaglia}, A., {Gratton}, R.~G.,  {et~al.} 2010, \aap, 516,
  A55

\bibitem[{{Carretta} \& {Gratton}(1997)}]{carretta97}
{Carretta}, E., \& {Gratton}, R.~G. 1997, \aaps, 121, 95

\bibitem[{{Casagrande} {et~al.}(2006){Casagrande}, {Portinari}, \&
  {Flynn}}]{casa06}
{Casagrande}, L., {Portinari}, L., \& {Flynn}, C. 2006, \mnras, 373, 13

\bibitem[{{Castelli} \& {Kurucz}(2003)}]{castelli03}
{Castelli}, F., \& {Kurucz}, R.~L. 2003, in IAU Symp. 210, Modelling of Stellar
  Atmospheres, ed.\ N.\ Piskunov, W.\ W.\ Weiss, \& D.\ F.\ Gray (San
  Francisco, CA: ASP), A20

\bibitem[{{Cayrel}(1988)}]{cayrel88}
{Cayrel}, R. 1988, in IAU Symposium, Vol. 132, The Impact of Very High S/N
  Spectroscopy on Stellar Physics, ed. G.~{Cayrel de Strobel} \& M.~{Spite},
  345

\bibitem[{{C{\^o}t{\'e}} {et~al.}(2016){C{\^o}t{\'e}}, {Belczynski}, {Fryer},
  {Ritter}, {Paul}, {Wehmeyer}, \& {O'Shea}}]{cote16}
{C{\^o}t{\'e}}, B., {Belczynski}, K., {Fryer}, C.~L.,  {et~al.} 2016,
  arXiv:1610.02405

\bibitem[{{Cristallo} {et~al.}(2015){Cristallo}, {Straniero}, {Piersanti}, \&
  {Gobrecht}}]{cristallo15}
{Cristallo}, S., {Straniero}, O., {Piersanti}, L., \& {Gobrecht}, D. 2015,
  \apjs, 219, 40

\bibitem[{{de Boer} {et~al.}(2014){de Boer}, {Tolstoy}, {Lemasle}, {Saha},
  {Olszewski}, {Mateo}, {Irwin}, \& {Battaglia}}]{deboer14}
{de Boer}, T.~J.~L., {Tolstoy}, E., {Lemasle}, B.,  {et~al.} 2014, \aap, 572,
  A10

\bibitem[{{Decressin} {et~al.}(2007){Decressin}, {Meynet}, {Charbonnel},
  {Prantzos}, \& {Ekstr{\"o}m}}]{decressin07}
{Decressin}, T., {Meynet}, G., {Charbonnel}, C., {Prantzos}, N., \&
  {Ekstr{\"o}m}, S. 2007, \aap, 464, 1029

\bibitem[{{Demarque} {et~al.}(2004){Demarque}, {Woo}, {Kim}, \&
  {Yi}}]{demarque04}
{Demarque}, P., {Woo}, J., {Kim}, Y., \& {Yi}, S.~K. 2004, \apjs, 155, 667

\bibitem[{{Denisenkov} \& {Denisenkova}(1990)}]{denisenkov90}
{Denisenkov}, P.~A., \& {Denisenkova}, S.~N. 1990, Soviet Astronomy Letters,
  16, 275

\bibitem[{{Denissenkov} \& {Hartwick}(2014)}]{denissenkov14}
{Denissenkov}, P.~A., \& {Hartwick}, F.~D.~A. 2014, \mnras, 437, L21

\bibitem[{{D'Ercole} {et~al.}(2010){D'Ercole}, {D'Antona}, {Ventura},
  {Vesperini}, \& {McMillan}}]{dercole10}
{D'Ercole}, A., {D'Antona}, F., {Ventura}, P., {Vesperini}, E., \& {McMillan},
  S.~L.~W. 2010, \mnras, 407, 854

\bibitem[{{Dolphin}(2002)}]{dolphin02}
{Dolphin}, A.~E. 2002, \mnras, 332, 91

\bibitem[{{Dotter} {et~al.}(2008){Dotter}, {Chaboyer}, {Jevremovi{\'c}},
  {Kostov}, {Baron}, \& {Ferguson}}]{dotter08}
{Dotter}, A., {Chaboyer}, B., {Jevremovi{\'c}}, D.,  {et~al.} 2008, \apjs, 178,
  89

\bibitem[{{D'Souza} \& {Rix}(2013)}]{dsousa13}
{D'Souza}, R., \& {Rix}, H.-W. 2013, \mnras, 429, 1887

\bibitem[{{Fabrizio} {et~al.}(2012){Fabrizio}, {Merle}, {Th{\'e}venin},
  {Nonino}, {Bono}, {Stetson}, {Ferraro}, {Iannicola}, {Monelli}, {Walker},
  {Buonanno}, {Caputo}, {Corsi}, {Dall''Ora}, {Degl''Innocenti}, {Fran{\c
  c}ois}, {Gilmozzi}, {Marconi}, {Pietrinferni}, {Prada Moroni}, {Primas},
  {Pulone}, {Ripepi}, \& {Romaniello}}]{fabrizio12}
{Fabrizio}, M., {Merle}, T., {Th{\'e}venin}, F.,  {et~al.} 2012, \pasp, 124,
  519

\bibitem[{{Fabrizio} {et~al.}(2015){Fabrizio}, {Nonino}, {Bono}, {Primas},
  {Th{\'e}venin}, {Stetson}, {Cassisi}, {Buonanno}, {Coppola}, {da Silva},
  {Dall'Ora}, {Ferraro}, {Genovali}, {Gilmozzi}, {Iannicola}, {Marconi},
  {Monelli}, {Romaniello}, \& {Walker}}]{fabrizio15}
{Fabrizio}, M., {Nonino}, M., {Bono}, G.,  {et~al.} 2015, \aap, 580, A18

\bibitem[{{Frebel} \& {Norris}(2015)}]{frebel15}
{Frebel}, A., \& {Norris}, J.~E. 2015, \araa, 53, 631

\bibitem[{{Fryer} {et~al.}(2012){Fryer}, {Belczynski}, {Wiktorowicz},
  {Dominik}, {Kalogera}, \& {Holz}}]{fryer12}
{Fryer}, C.~L., {Belczynski}, K., {Wiktorowicz}, G.,  {et~al.} 2012, \apj, 749,
  91

\bibitem[{{Fulbright}(2000)}]{fulbright00}
{Fulbright}, J.~P. 2000, \aj, 120, 1841

\bibitem[{{Gilmore} {et~al.}(2013){Gilmore}, {Norris}, {Monaco}, {Yong},
  {Wyse}, \& {Geisler}}]{gilmore13}
{Gilmore}, G., {Norris}, J.~E., {Monaco}, L.,  {et~al.} 2013, \apj, 763, 61

\bibitem[{{Gilmore} \& {Wyse}(1991)}]{gilmore91}
{Gilmore}, G., \& {Wyse}, R.~F.~G. 1991, \apjl, 367, L55

\bibitem[{{Governato} {et~al.}(2010){Governato}, {Brook}, {Mayer}, {Brooks},
  {Rhee}, {Wadsley}, {Jonsson}, {Willman}, {Stinson}, {Quinn}, \&
  {Madau}}]{governato10}
{Governato}, F., {Brook}, C., {Mayer}, L.,  {et~al.} 2010, \nat, 463, 203

\bibitem[{{Gratton} {et~al.}(2004){Gratton}, {Sneden}, \&
  {Carretta}}]{gratton04}
{Gratton}, R., {Sneden}, C., \& {Carretta}, E. 2004, \araa, 42, 385

\bibitem[{{Hernandez} {et~al.}(2000){Hernandez}, {Gilmore}, \&
  {Valls-Gabaud}}]{hernandez00}
{Hernandez}, X., {Gilmore}, G., \& {Valls-Gabaud}, D. 2000, \mnras, 317, 831

\bibitem[{{Hinkle} {et~al.}(2000){Hinkle}, {Wallace}, {Valenti}, \&
  {Harmer}}]{hinkle00}
{Hinkle}, K., {Wallace}, L., {Valenti}, J., \& {Harmer}, D. 2000, {Visible and
  Near Infrared Atlas of the Arcturus Spectrum 3727-9300 A}

\bibitem[{{Hurley-Keller} {et~al.}(1998){Hurley-Keller}, {Mateo}, \&
  {Nemec}}]{hurley98}
{Hurley-Keller}, D., {Mateo}, M., \& {Nemec}, J. 1998, \aj, 115, 1840

\bibitem[{{Iben}(1967)}]{iben67}
{Iben}, Jr., I. 1967, \araa, 5, 571

\bibitem[{{Ivans} {et~al.}(2001){Ivans}, {Kraft}, {Sneden}, {Smith}, {Rich}, \&
  {Shetrone}}]{ivans_m5_01}
{Ivans}, I.~I., {Kraft}, R.~P., {Sneden}, C.,  {et~al.} 2001, \aj, 122, 1438

\bibitem[{{Ivans} {et~al.}(2006){Ivans}, {Simmerer}, {Sneden}, {Lawler},
  {Cowan}, {Gallino}, \& {Bisterzo}}]{ivans06}
{Ivans}, I.~I., {Simmerer}, J., {Sneden}, C.,  {et~al.} 2006, \apj, 645, 613

\bibitem[{{Ivans} {et~al.}(1999){Ivans}, {Sneden}, {Kraft}, {Suntzeff},
  {Smith}, {Langer}, \& {Fulbright}}]{ivans_m4_99}
{Ivans}, I.~I., {Sneden}, C., {Kraft}, R.~P.,  {et~al.} 1999, \aj, 118, 1273

\bibitem[{{Iwamoto} {et~al.}(1999){Iwamoto}, {Brachwitz}, {Nomoto},
  {Kishimoto}, {Umeda}, {Hix}, \& {Thielemann}}]{iwamoto99}
{Iwamoto}, K., {Brachwitz}, F., {Nomoto}, K.,  {et~al.} 1999, \apjs, 125, 439

\bibitem[{{Kirby} {et~al.}(2012){Kirby}, {Fu}, {Guhathakurta}, \&
  {Deng}}]{kirby12}
{Kirby}, E.~N., {Fu}, X., {Guhathakurta}, P., \& {Deng}, L. 2012, \apjl, 752,
  L16

\bibitem[{{Kobayashi} \& {Nakasato}(2011)}]{kobayashi11}
{Kobayashi}, C., \& {Nakasato}, N. 2011, \apj, 729, 16

\bibitem[{{Kobayashi} \& {Nomoto}(2009)}]{kobayashi09}
{Kobayashi}, C., \& {Nomoto}, K. 2009, \apj, 707, 1466

\bibitem[{{Kobayashi} {et~al.}(2015){Kobayashi}, {Nomoto}, \&
  {Hachisu}}]{kobayashi15}
{Kobayashi}, C., {Nomoto}, K., \& {Hachisu}, I. 2015, \apjl, 804, L24

\bibitem[{{Kobayashi} {et~al.}(1998){Kobayashi}, {Tsujimoto}, {Nomoto},
  {Hachisu}, \& {Kato}}]{kobayashi98}
{Kobayashi}, C., {Tsujimoto}, T., {Nomoto}, K., {Hachisu}, I., \& {Kato}, M.
  1998, \apjl, 503, L155

\bibitem[{{Kobayashi} {et~al.}(2006){Kobayashi}, {Umeda}, {Nomoto}, {Tominaga},
  \& {Ohkubo}}]{kobayashi06}
{Kobayashi}, C., {Umeda}, H., {Nomoto}, K., {Tominaga}, N., \& {Ohkubo}, T.
  2006, \apj, 653, 1145

\bibitem[{{Koch} {et~al.}(2008){Koch}, {Grebel}, {Gilmore}, {Wyse}, {Kleyna},
  {Harbeck}, {Wilkinson}, \& {Wyn Evans}}]{koch08a}
{Koch}, A., {Grebel}, E.~K., {Gilmore}, G.~F.,  {et~al.} 2008, \aj, 135, 1580

\bibitem[{{Koch} {et~al.}(2006){Koch}, {Grebel}, {Wyse}, {Kleyna}, {Wilkinson},
  {Harbeck}, {Gilmore}, \& {Evans}}]{koch06}
{Koch}, A., {Grebel}, E.~K., {Wyse}, R.~F.~G.,  {et~al.} 2006, \aj, 131, 895

\bibitem[{{Koposov} {et~al.}(2011){Koposov}, {Gilmore}, {Walker}, {Belokurov},
  {Wyn Evans}, {Fellhauer}, {Gieren}, {Geisler}, {Monaco}, {Norris}, {Okamoto},
  {Pe{\~n}arrubia}, {Wilkinson}, {Wyse}, \& {Zucker}}]{koposov11}
{Koposov}, S.~E., {Gilmore}, G., {Walker}, M.~G.,  {et~al.} 2011, \apj, 736,
  146

\bibitem[{{Kordopatis} {et~al.}(2016){Kordopatis}, {Amorisco}, {Evans},
  {Gilmore}, \& {Koposov}}]{kordopatis16}
{Kordopatis}, G., {Amorisco}, N.~C., {Evans}, N.~W., {Gilmore}, G., \&
  {Koposov}, S.~E. 2016, \mnras, 457, 1299

\bibitem[{{Korobkin} {et~al.}(2012){Korobkin}, {Rosswog}, {Arcones}, \&
  {Winteler}}]{korobkin12}
{Korobkin}, O., {Rosswog}, S., {Arcones}, A., \& {Winteler}, C. 2012, \mnras,
  426, 1940

\bibitem[{{Kraft} {et~al.}(1999){Kraft}, {Peterson}, {Guhathakurta}, {Sneden},
  {Fulbright}, \& {Langer}}]{kraft99}
{Kraft}, R.~P., {Peterson}, R.~C., {Guhathakurta}, P.,  {et~al.} 1999, \apjl,
  518, L53

\bibitem[{{Kraft} {et~al.}(1992){Kraft}, {Sneden}, {Langer}, \&
  {Prosser}}]{kraft_m3_92}
{Kraft}, R.~P., {Sneden}, C., {Langer}, G.~E., \& {Prosser}, C.~F. 1992, \aj,
  104, 645

\bibitem[{{Kraft} {et~al.}(1997){Kraft}, {Sneden}, {Smith}, {Shetrone},
  {Langer}, \& {Pilachowski}}]{kraft_m13_97}
{Kraft}, R.~P., {Sneden}, C., {Smith}, G.~H.,  {et~al.} 1997, \aj, 113, 279

\bibitem[{{Kratz} {et~al.}(2014){Kratz}, {Farouqi}, \& {M{\"o}ller}}]{kratz14}
{Kratz}, K.-L., {Farouqi}, K., \& {M{\"o}ller}, P. 2014, \apj, 792, 6

\bibitem[{{Kurucz} \& {Bell}(1995)}]{kurucz95}
{Kurucz}, R., \& {Bell}, B. 1995, Atomic Line Data (R.L.~Kurucz and B.~Bell)
  Kurucz CD-ROM No.~23.~Cambridge, Mass.: Smithsonian Astrophysical
  Observatory, 1995., 23

\bibitem[{{Leaman}(2012)}]{leaman12}
{Leaman}, R. 2012, \aj, 144, 183

\bibitem[{{Lemasle} {et~al.}(2012){Lemasle}, {Hill}, {Tolstoy}, {Venn},
  {Shetrone}, {Irwin}, {de Boer}, {Starkenburg}, \& {Salvadori}}]{lemasle12}
{Lemasle}, B., {Hill}, V., {Tolstoy}, E.,  {et~al.} 2012, \aap, 538, A100

\bibitem[{{Lind} {et~al.}(2009){Lind}, {Asplund}, \& {Barklem}}]{lind09}
{Lind}, K., {Asplund}, M., \& {Barklem}, P.~S. 2009, \aap, 503, 541

\bibitem[{{Lind} {et~al.}(2012){Lind}, {Bergemann}, \& {Asplund}}]{lind12}
{Lind}, K., {Bergemann}, M., \& {Asplund}, M. 2012, \mnras, 427, 50

\bibitem[{{Lindg{\aa}rd} \& {Nielson}(1977)}]{lindgard77}
{Lindg{\aa}rd}, A., \& {Nielson}, S.~E. 1977, Atomic Data and Nuclear Data
  Tables, 19, 533

\bibitem[{{Lugaro} {et~al.}(2014){Lugaro}, {Tagliente}, {Karakas}, {Milazzo},
  {K{\"a}ppeler}, {Davis}, \& {Savina}}]{lugaro14}
{Lugaro}, M., {Tagliente}, G., {Karakas}, A.~I.,  {et~al.} 2014, \apj, 780, 95

\bibitem[{{Martell} {et~al.}(2011){Martell}, {Smolinski}, {Beers}, \&
  {Grebel}}]{martell11}
{Martell}, S.~L., {Smolinski}, J.~P., {Beers}, T.~C., \& {Grebel}, E.~K. 2011,
  \aap, 534, A136

\bibitem[{{Mashonkina} {et~al.}(2016){Mashonkina}, {Sitnova}, \&
  {Pakhomov}}]{mashonkina16}
{Mashonkina}, L.~I., {Sitnova}, T.~N., \& {Pakhomov}, Y.~V. 2016, Astronomy
  Letters, 42, 606

\bibitem[{{Mateo}(1998)}]{mateo98}
{Mateo}, M.~L. 1998, \araa, 36, 435

\bibitem[{{McWilliam}(1998)}]{mcwilliam98}
{McWilliam}, A. 1998, \aj, 115, 1640

\bibitem[{{McWilliam} {et~al.}(2013){McWilliam}, {Wallerstein}, \&
  {Mottini}}]{mcwilliam13}
{McWilliam}, A., {Wallerstein}, G., \& {Mottini}, M. 2013, \apj, 778, 149

\bibitem[{{Mighell}(1997)}]{mighell97}
{Mighell}, K.~J. 1997, \aj, 114, 1458

\bibitem[{{Monelli} {et~al.}(2014){Monelli}, {Milone}, {Fabrizio}, {Bono},
  {Stetson}, {Walker}, {Cassisi}, {Gallart}, {Nonino}, {Aparicio}, {Buonanno},
  {Dall'Ora}, {Ferraro}, {Iannicola}, {Pulone}, \& {Th{\'e}venin}}]{monelli14}
{Monelli}, M., {Milone}, A.~P., {Fabrizio}, M.,  {et~al.} 2014, \apj, 796, 90

\bibitem[{{Monelli} {et~al.}(2003){Monelli}, {Pulone}, {Corsi}, {Castellani},
  {Bono}, {Walker}, {Brocato}, {Buonanno}, {Caputo}, {Castellani}, {Dall'Ora},
  {Marconi}, {Nonino}, {Ripepi}, \& {Smith}}]{monelli03}
{Monelli}, M., {Pulone}, L., {Corsi}, C.~E.,  {et~al.} 2003, \aj, 126, 218

\bibitem[{{Nishimura} {et~al.}(2015){Nishimura}, {Takiwaki}, \&
  {Thielemann}}]{nishimura15}
{Nishimura}, N., {Takiwaki}, T., \& {Thielemann}, F.-K. 2015, \apj, 810, 109

\bibitem[{{Nomoto} {et~al.}(2013){Nomoto}, {Kobayashi}, \&
  {Tominaga}}]{nomoto13}
{Nomoto}, K., {Kobayashi}, C., \& {Tominaga}, N. 2013, \araa, 51, 457

\bibitem[{{Norris} \& {Da Costa}(1995)}]{norris95}
{Norris}, J.~E., \& {Da Costa}, G.~S. 1995, \apj, 447, 680

\bibitem[{{Norris} {et~al.}(1995){Norris}, {Da Costa}, \& {Tingay}}]{norris95a}
{Norris}, J.~E., {Da Costa}, G.~S., \& {Tingay}, S.~J. 1995, \apjs, 99, 637

\bibitem[{{Norris} {et~al.}(2001){Norris}, {Ryan}, \& {Beers}}]{norris01}
{Norris}, J.~E., {Ryan}, S.~G., \& {Beers}, T.~C. 2001, \apj, 561, 1034

\bibitem[{{Norris} {et~al.}(2010){Norris}, {Yong}, {Gilmore}, \&
  {Wyse}}]{norris10}
{Norris}, J.~E., {Yong}, D., {Gilmore}, G., \& {Wyse}, R.~F.~G. 2010, \apj,
  711, 350

\bibitem[{{Norris} {et~al.}(2017){Norris}, {Yong}, {Venn}, {Gilmore},
  {Casagrande}, \& {Dotter}}]{norris17a}
{Norris}, J.~E., {Yong}, D., {Venn}, K.~A.,  {et~al.} 2017, \apjs, submitted
  (Paper I)

\bibitem[{{Pasquini} {et~al.}(2002){Pasquini}, {Avila}, {Blecha}, {Cacciari},
  {Cayatte}, {Colless}, {Damiani}, {de Propris}, {Dekker}, {di Marcantonio},
  {Farrell}, {Gillingham}, {Guinouard}, {Hammer}, {Kaufer}, {Hill}, {Marteaud},
  {Modigliani}, {Mulas}, {North}, {Popovic}, {Rossetti}, {Royer}, {Santin},
  {Schmutzer}, {Simond}, {Vola}, {Waller}, \& {Zoccali}}]{pasquini02}
{Pasquini}, L., {Avila}, G., {Blecha}, A.,  {et~al.} 2002, The Messenger, 110,
  1

\bibitem[{{Perego} {et~al.}(2014){Perego}, {Rosswog}, {Cabez{\'o}n},
  {Korobkin}, {K{\"a}ppeli}, {Arcones}, \& {Liebend{\"o}rfer}}]{perego14}
{Perego}, A., {Rosswog}, S., {Cabez{\'o}n}, R.~M.,  {et~al.} 2014, \mnras, 443,
  3134

\bibitem[{{Pietrinferni} {et~al.}(2004){Pietrinferni}, {Cassisi}, {Salaris}, \&
  {Castelli}}]{pietrinferni04}
{Pietrinferni}, A., {Cassisi}, S., {Salaris}, M., \& {Castelli}, F. 2004, \apj,
  612, 168

\bibitem[{{Pietrinferni} {et~al.}(2006){Pietrinferni}, {Cassisi}, {Salaris}, \&
  {Castelli}}]{pietrinferni06}
---. 2006, \apj, 642, 797

\bibitem[{{Pignatari} {et~al.}(2016){Pignatari}, {Herwig}, {Hirschi},
  {Bennett}, {Rockefeller}, {Fryer}, {Timmes}, {Ritter}, {Heger}, {Jones},
  {Battino}, {Dotter}, {Trappitsch}, {Diehl}, {Frischknecht}, {Hungerford},
  {Magkotsios}, {Travaglio}, \& {Young}}]{pignatari16}
{Pignatari}, M., {Herwig}, F., {Hirschi}, R.,  {et~al.} 2016, \apjs, 225, 24

\bibitem[{{Preston} \& {Sneden}(2000)}]{preston00}
{Preston}, G.~W., \& {Sneden}, C. 2000, \aj, 120, 1014

\bibitem[{{Pryor} \& {Meylan}(1993)}]{pryor93}
{Pryor}, C., \& {Meylan}, G. 1993, in Astronomical Society of the Pacific
  Conference Series, Vol.~50, Structure and Dynamics of Globular Clusters, ed.
  S.~G. {Djorgovski} \& G.~{Meylan}, 357

\bibitem[{{Ram{\'{\i}}rez} \& {Mel{\'e}ndez}(2005)}]{ramirez05}
{Ram{\'{\i}}rez}, I., \& {Mel{\'e}ndez}, J. 2005, \apj, 626, 446

\bibitem[{{Ram{\'{\i}}rez} {et~al.}(2012){Ram{\'{\i}}rez}, {Mel{\'e}ndez}, \&
  {Chanam{\'e}}}]{ramirez12}
{Ram{\'{\i}}rez}, I., {Mel{\'e}ndez}, J., \& {Chanam{\'e}}, J. 2012, \apj, 757,
  164

\bibitem[{{Revaz} \& {Jablonka}(2012)}]{revaz12}
{Revaz}, Y., \& {Jablonka}, P. 2012, \aap, 538, A82

\bibitem[{{Sackmann} \& {Boothroyd}(1999)}]{sackmann99}
{Sackmann}, I.-J., \& {Boothroyd}, A.~I. 1999, \apj, 510, 217

\bibitem[{{Santana} {et~al.}(2016){Santana}, {Mu{\~n}oz}, {de Boer}, {Simon},
  {Geha}, {C{\^o}t{\'e}}, {Guzm{\'a}n}, {Stetson}, \& {Djorgovski}}]{santana16}
{Santana}, F.~A., {Mu{\~n}oz}, R.~R., {de Boer}, T.~J.~L.,  {et~al.} 2016,
  arXiv:1607.05312

\bibitem[{{Schlegel} {et~al.}(1998){Schlegel}, {Finkbeiner}, \&
  {Davis}}]{schlegel98}
{Schlegel}, D.~J., {Finkbeiner}, D.~P., \& {Davis}, M. 1998, \apj, 500, 525

\bibitem[{{Shetrone} {et~al.}(2003){Shetrone}, {Venn}, {Tolstoy}, {Primas},
  {Hill}, \& {Kaufer}}]{shetrone03}
{Shetrone}, M., {Venn}, K.~A., {Tolstoy}, E.,  {et~al.} 2003, \aj, 125, 684

\bibitem[{{Simon} {et~al.}(2011){Simon}, {Geha}, {Minor}, {Martinez}, {Kirby},
  {Bullock}, {Kaplinghat}, {Strigari}, {Willman}, {Choi}, {Tollerud}, \&
  {Wolf}}]{simon11}
{Simon}, J.~D., {Geha}, M., {Minor}, Q.~E.,  {et~al.} 2011, \apj, 733, 46

\bibitem[{{Smecker-Hane} {et~al.}(1994){Smecker-Hane}, {Stetson}, {Hesser}, \&
  {Lehnert}}]{smecker94}
{Smecker-Hane}, T.~A., {Stetson}, P.~B., {Hesser}, J.~E., \& {Lehnert}, M.~D.
  1994, \aj, 108, 507

\bibitem[{{Smecker-Hane} {et~al.}(1996){Smecker-Hane}, {Stetson}, {Hesser}, \&
  {Vandenberg}}]{smecker96}
{Smecker-Hane}, T.~A., {Stetson}, P.~B., {Hesser}, J.~E., \& {Vandenberg},
  D.~A. 1996, in Astronomical Society of the Pacific Conference Series,
  Vol.~98, From Stars to Galaxies: the Impact of Stellar Physics on Galaxy
  Evolution, ed. C.~{Leitherer}, U.~{Fritze-von-Alvensleben}, \& J.~{Huchra},
  328

\bibitem[{{Sneden}(1973)}]{sneden73}
{Sneden}, C. 1973, \apj, 184, 839

\bibitem[{{Sobeck} {et~al.}(2011){Sobeck}, {Kraft}, {Sneden}, {Preston},
  {Cowan}, {Smith}, {Thompson}, {Shectman}, \& {Burley}}]{sobeck11}
{Sobeck}, J.~S., {Kraft}, R.~P., {Sneden}, C.,  {et~al.} 2011, \aj, 141, 175

\bibitem[{{Starkenburg} {et~al.}(2010){Starkenburg}, {Hill}, {Tolstoy},
  {Gonz{\'a}lez Hern{\'a}ndez}, {Irwin}, {Helmi}, {Battaglia}, {Jablonka},
  {Tafelmeyer}, {Shetrone}, {Venn}, \& {de Boer}}]{starkenburg10}
{Starkenburg}, E., {Hill}, V., {Tolstoy}, E.,  {et~al.} 2010, \aap, 513, A34

\bibitem[{{Stetson}(2005)}]{stetson05}
{Stetson}, P.~B. 2005, \pasp, 117, 563

\bibitem[{{Stetson} {et~al.}(2011){Stetson}, {Monelli}, {Fabrizio}, {Walker},
  {Bono}, {Buonanno}, {Caputo}, {Cassisi}, {Corsi}, {Dall'Ora},
  {Degl'Innocenti}, {Fran{\c c}ois}, {Ferraro}, {Gilmozzi}, {Iannicola},
  {Merle}, {Nonino}, {Pietrinferni}, {Moroni}, {Pulone}, {Romaniello}, \&
  {Th{\'e}venin}}]{stetson11}
{Stetson}, P.~B., {Monelli}, M., {Fabrizio}, M.,  {et~al.} 2011, The Messenger,
  144, 32

\bibitem[{{Stetson} \& {Pancino}(2008)}]{stetson08}
{Stetson}, P.~B., \& {Pancino}, E. 2008, \pasp, 120, 1332

\bibitem[{{Tolstoy} {et~al.}(2009){Tolstoy}, {Hill}, \& {Tosi}}]{tolstoy09}
{Tolstoy}, E., {Hill}, V., \& {Tosi}, M. 2009, \araa, 47, 371

\bibitem[{{Tolstoy} {et~al.}(2003){Tolstoy}, {Venn}, {Shetrone}, {Primas},
  {Hill}, {Kaufer}, \& {Szeifert}}]{tolstoy03}
{Tolstoy}, E., {Venn}, K.~A., {Shetrone}, M.,  {et~al.} 2003, \aj, 125, 707

\bibitem[{{Tominaga} {et~al.}(2014){Tominaga}, {Iwamoto}, \&
  {Nomoto}}]{tominaga14}
{Tominaga}, N., {Iwamoto}, N., \& {Nomoto}, K. 2014, \apj, 785, 98

\bibitem[{{Tsujimoto} \& {Nishimura}(2015)}]{tsujimoto15}
{Tsujimoto}, T., \& {Nishimura}, N. 2015, \apjl, 811, L10

\bibitem[{{VandenBerg} {et~al.}(2006){VandenBerg}, {Bergbusch}, \&
  {Dowler}}]{vandenberg06}
{VandenBerg}, D.~A., {Bergbusch}, P.~A., \& {Dowler}, P.~D. 2006, \apjs, 162,
  375

\bibitem[{{VandenBerg} {et~al.}(2015){VandenBerg}, {Stetson}, \&
  {Brown}}]{vandenberg15}
{VandenBerg}, D.~A., {Stetson}, P.~B., \& {Brown}, T.~M. 2015, \apj, 805, 103

\bibitem[{{Venn} {et~al.}(2012){Venn}, {Shetrone}, {Irwin}, {Hill}, {Jablonka},
  {Tolstoy}, {Lemasle}, {Divell}, {Starkenburg}, {Letarte}, {Baldner},
  {Battaglia}, {Helmi}, {Kaufer}, \& {Primas}}]{venn12}
{Venn}, K.~A., {Shetrone}, M.~D., {Irwin}, M.~J.,  {et~al.} 2012, \apj, 751,
  102

\bibitem[{{Walker} {et~al.}(2009{\natexlab{a}}){Walker}, {Mateo}, {Olszewski},
  {Pe{\~n}arrubia}, {Wyn Evans}, \& {Gilmore}}]{walker09b}
{Walker}, M.~G., {Mateo}, M., {Olszewski}, E.~W.,  {et~al.} 2009{\natexlab{a}},
  \apj, 704, 1274

\bibitem[{{Walker} {et~al.}(2009{\natexlab{b}}){Walker}, {Mateo}, \&
  {Olszewski}}]{walker09}
{Walker}, M.~G., {Mateo}, M., \& {Olszewski}, E.~W. 2009{\natexlab{b}}, \aj,
  137, 3100

\bibitem[{{Wallerstein} \& {Sneden}(1982)}]{wallerstein82}
{Wallerstein}, G., \& {Sneden}, C. 1982, \apj, 255, 577

\bibitem[{{Weisz} {et~al.}(2014){Weisz}, {Dolphin}, {Skillman}, {Holtzman},
  {Gilbert}, {Dalcanton}, \& {Williams}}]{weisz14}
{Weisz}, D.~R., {Dolphin}, A.~E., {Skillman}, E.~D.,  {et~al.} 2014, \apj, 789,
  148

\bibitem[{{Wise} {et~al.}(2012){Wise}, {Turk}, {Norman}, \& {Abel}}]{wise12}
{Wise}, J.~H., {Turk}, M.~J., {Norman}, M.~L., \& {Abel}, T. 2012, \apj, 745,
  50

\bibitem[{{Woosley} \& {Weaver}(1995)}]{woosley95}
{Woosley}, S.~E., \& {Weaver}, T.~A. 1995, \apjs, 101, 181

\bibitem[{{Worley} {et~al.}(2013){Worley}, {Hill}, {Sobeck}, \&
  {Carretta}}]{worley13}
{Worley}, C.~C., {Hill}, V., {Sobeck}, J., \& {Carretta}, E. 2013, \aap, 553,
  A47

\bibitem[{{Yong} {et~al.}(2008){Yong}, {Lambert}, {Paulson}, \&
  {Carney}}]{yong08}
{Yong}, D., {Lambert}, D.~L., {Paulson}, D.~B., \& {Carney}, B.~W. 2008, \apj,
  673, 854

\bibitem[{{Yong} {et~al.}(2013){Yong}, {Norris}, {Bessell}, {Christlieb},
  {Asplund}, {Beers}, {Barklem}, {Frebel}, \& {Ryan}}]{yong13a}
{Yong}, D., {Norris}, J.~E., {Bessell}, M.~S.,  {et~al.} 2013, \apj, 762, 26

\end{thebibliography}
\end{document}